\documentclass{emulateapj}

\usepackage{graphicx}
\usepackage{apjfonts}
\usepackage{verbatim}
\usepackage{morefloats}
\usepackage{color}
\usepackage{bmpsize}
\usepackage{rotating}
\newcommand{\redpen}[1]{{\textcolor{black}{#1}}}

\newcommand{\mytilde}{\raise.17ex\hbox{$\scriptstyle\mathtt{\sim}$}}
\providecommand{\e}[1]{\ensuremath{\times 10^{#1}}}

\shorttitle{High-z  3C Survey}
\shortauthors{Hilbert et al.}

\slugcomment{Accepted for Publication in ApJS}

\begin{document}


\title{Powerful Activity in the Bright Ages. I. A Visible/IR Survey of High Redshift 3C Radio Galaxies and Quasars}


\author{B. Hilbert\altaffilmark{1}, M. Chiaberge\altaffilmark{1,2},J.~P. Kotyla\altaffilmark{1}, G.~R. Tremblay\altaffilmark{3}, C. Stanghellini\altaffilmark{4}, W.~B. Sparks\altaffilmark{1}, 
S. Baum\altaffilmark{5,6}, A. Capetti\altaffilmark{7}, F.~D. Macchetto\altaffilmark{1}, 
G.~K. Miley\altaffilmark{8}, C.~P. O'Dea\altaffilmark{5,6}, E.~S. Perlman\altaffilmark{9}, A.~C. Quillen\altaffilmark{6}}

\altaffiltext{1}{Space Telescope Science Institute, 3700 San Martin Drive, Baltimore, MD 21218}
\altaffiltext{2}{Center for Astrophysical Sciences, Johns Hopkins University, 3400 N. Charles Street, Baltimore, MD 21218, USA}
\altaffiltext{3}{Yale University, Department of Astronomy and Yale Center for Astronomy and Astrophysics, 52 Hillhouse Ave, New Haven, CT 06511}
\altaffiltext{4}{INAF - Istituto di Radioastronomia, Via P. Gobetti, 101 40129 Bologna, Italy}
\altaffiltext{5}{University of Manitoba, Dept of Physics and Astronomy, 66 Chancellors Cir., Winnipeg, MB R3T 2N2 Canada}
\altaffiltext{6}{Rochester Institute of Technology, School of Physics \& Astronomy, 84 Lomb Memorial Dr., Rochester, NY, 14623, USA}
\altaffiltext{7}{Osservatorio Astronomico de Torino, Corso Savona, 10024 Moncalieri TO, Italy}
\altaffiltext{8}{Universiteit Leiden, Rapenburg 70, 2311 EZ Leiden, Netherlands}
\altaffiltext{9}{Florida Institute of Technology, 150 W University Blvd, Melbourne, FL 32901}



\begin{abstract}
We present new rest frame UV and visible observations of 22 high-$z$ ($1 < z < 2.5$) 3C 
radio galaxies and QSOs obtained with the Hubble Space Telescope's (HST) Wide Field 
Camera 3 (WFC3) instrument. Using a custom data reduction strategy in order 
to assure the removal of cosmic rays, persistence signal, and other data artifacts, 
we have produced high-quality science-ready images of the targets and their 
local environments. We observe targets with regions of UV emission suggestive of 
active star formation. In addition, several targets exhibit highly distorted host 
galaxy morphologies in the rest frame visible images. Photometric analyses reveals 
that brighter QSOs tend to be generally redder than their dimmer counterparts.
Using emission line fluxes from the literature, we estimate that emission line contamination is relatively 
small in the rest frame UV images for the QSOs. Using archival VLA data, we have also created radio 
map overlays for each of our targets, allowing for analysis of the optical and radio axes alignment.

\end{abstract}




\section{Introduction}

In the epoch between redshifts 1 and 2, the Universe was in a period of rampant star formation,
AGN phenomena were 1000 times more common than at the present time, and there were many powerful, massive quasars 
and radio galaxies. There are many questions about the behavior and evolution of these dynamic 
objects and processes during these "Bright Ages". What fraction of these objects are in clusters 
or active mergers? How does the AGN affect star formation in its local environment? Do radio synchrotron 
jets quench or trigger star formation?

In order to provide a large database for studies aimed at addressing all of these questions, 
 we have used the WFC3 instrument to obtain multiband optical and IR images of a set of 
1$<z<$2.5 radio galaxies and QSOs in the 3CR catalog \citep{spinrad85}. The 3CR catalog is unbiased 
with respect to orientation effects and contains 
the most powerful radio sources at any given redshift, allowing us a view of
the behaviors of the progenitors of some of the most massive, dominant cluster elliptical galaxies.
From these data, we can begin to address many of the fundamental questions of 
galaxy and cluster behavior at this exciting time in the Universe's history, including details 
of star formation rates and cluster environments and the prevalence of mergers.

Here we present our calibrated data, along with qualitative descriptions 
of the features in the observations, as well as initial photometric and 
emission line analyses of the objects. 

Combined with data from previous SNAPSHOTS of 3C objects at lower redshifts (e.g. \citet{madrid06,tremblay09})
we can look for evolutionary trends over long timescales, including AGN and host
galaxy interaction from z\mytilde2 to today. Scientific results from these data are presented in 
a series of papers e.g. \citet{marco15}, Kotyla et al. (2015, in preparation), and Chaiberge et al. (2016, in preparation).


\section{The observed sample}


We used WFC3 on board HST to obtain visible and near-IR 
images of our 3C targets under SNAPSHOT proposal 13023. These data were collected between December of 2012 and May 2013.

The sample definition is the revised 3C catalog as defined by \citep{bennett62a}, \citep{bennett62b}, \citep{spinrad85}. The 3CR catalog includes all radio 
sources with flux density at 178MHz $S_{178} > 9$Jy that are located at a latitude $|b|>10^\circ$. From these, we restricted our proposed target list 
to the 58 sources with $z>1.0$.  Over the course of Cycle 20, we obtained data for 22 
of these objects, representing 38\% of our proposed sample.  
Of these 22 objects, 12 are high-excitation narrow line radio galaxies (RGs) and 10 are QSOs. Observed targets are listed in table \ref{tab:objinfo} 
along with some of their basic properties. These objects cover a range in redshift from 1.055 to 1.825, with the exception of 3C 257 at $z=2.474$.


\subsection{Statistical properties of the observed sample}

The average completion rate for HST/WFC3 SNAPSHOT programs is currently (Cycle 20-21)  $\sim 30\%$ \citep{work14}. 
This is most likely due to the STScI policy that lowers the scheduling priority of targets belonging to programs 
that reach the $35\%$ completion rate limit. As a result, our program was efficiently scheduled in the first part 
of the HST Cycle 20. Unfortunately, as soon as we reached the $35\%$ limit, the 
observations stopped, and no other targets were scheduled after June 2013.

\begin{figure*}
\epsscale{0.8}
\plottwo{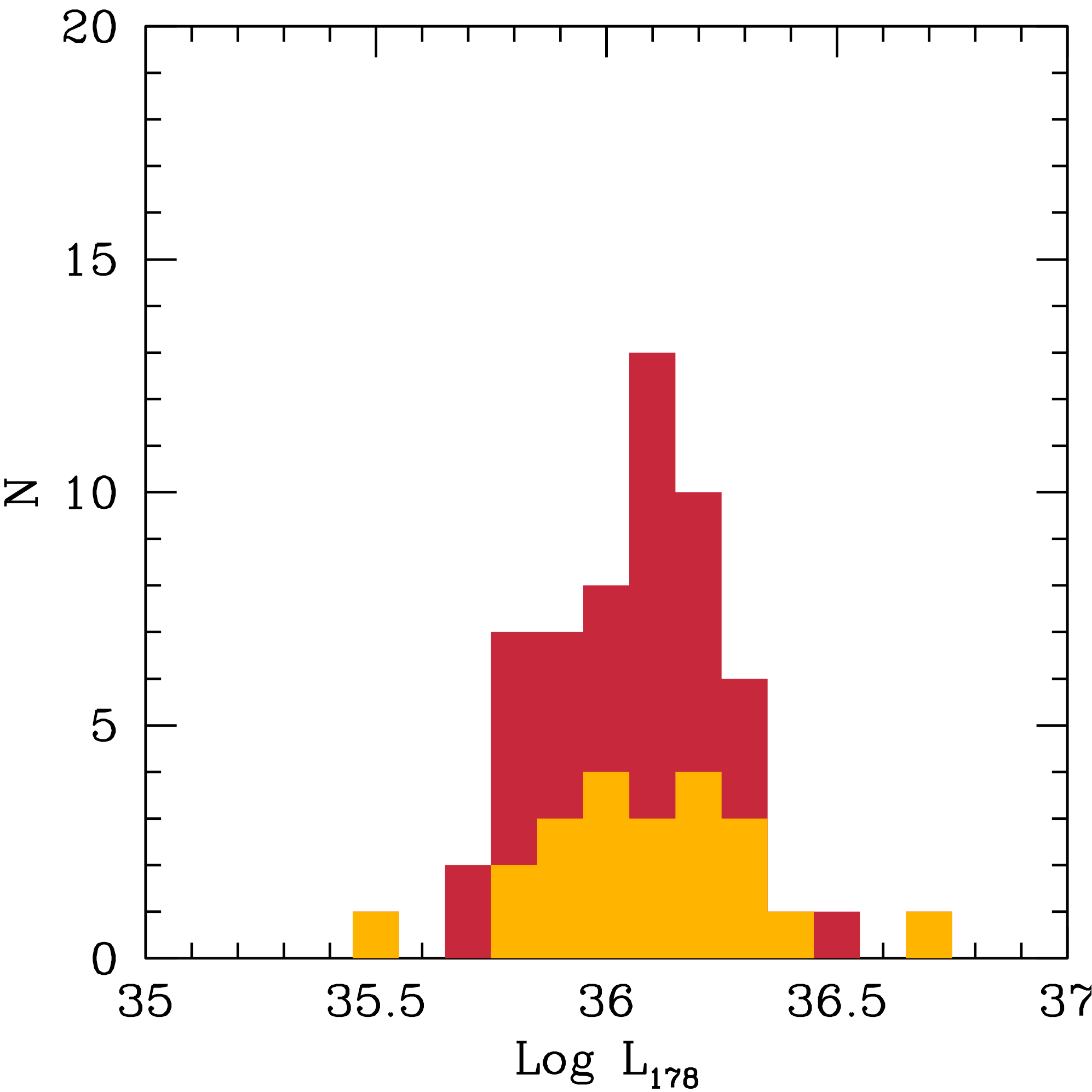}{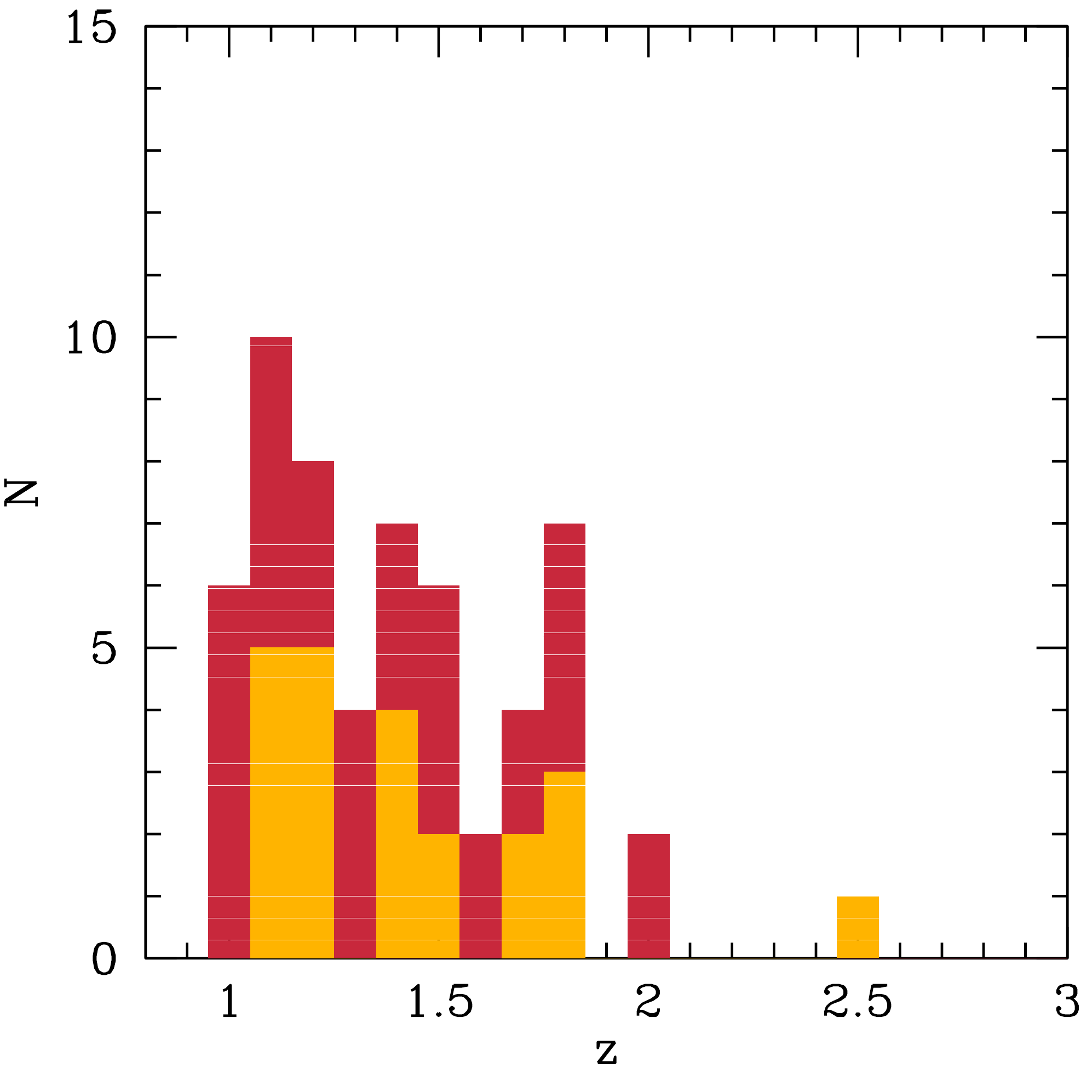}
\caption{The distribution in total radio power per unit frequency (at 178MHz, left panel) and redshift (right panel) for the whole 3CR sample with $z>1$ (in red) and for the observed sub-sample (yellow). \label{histo}}
\end{figure*}

Our total of 22 observed objects represents 
a rather small fraction of the complete catalog of sources. This is particularly significant for statistical work, for example to firmly
establish the fraction of objects that show specific properties and to derive important information on the nature of these radio sources.
The STScI procedure for selecting a subset of sources from a snapshot proposal should not result in any bias in terms of source properties. 
Nevertheless, it is very important to test that the observed sample is representative of the
entire population of objects belonging to the 3CR. In Fig.~\ref{histo} we show the distribution of total radio power at 178MHz (left panel) and
redshift (right panel) for the whole 3CR sample with $z>1$ (in red) and for our observed sub-sample (in yellow). 
Visually, the yellow and red histograms appear to be similar. In fact, the Kolmogorov-Smirnov test shows that we cannot
reject the null hypothesis that the observed and the full sample follow the same redshift and radio power distributions. 
The p-values are $p=0.998$ and $p=0.995$ for the
radio power and redshift distributions, respectively. 
The statistical test is performed using the $R$ function {\it ks.boot} included in the {\it Matching} package \citep{sekhon11}. This function performs a
bootstrap version of the K-S test that is useful when statistical ties are present in the samples, as it is in our case. 
However, an identical result is returned by the classical KS test ({\it ks.test}).


While this analysis does not prove that the observed and original distributions
are the same, it provides support to the idea that our sample of 22 objects is a good representation of the original 3CR catalog at $z>1$.
However,
note that with only 22 observed objects the statistical accuracy is significantly reduced (by a factor of $\sim 2$, assuming the binomial 
statistics and depending on 
the resulting object fraction) compared to the original sample. For example, if we observed that 70\% of our objects display a specific property
(e.g. the source is a member of a cluster of galaxies, the source is associated with a host galaxy brighter than a certain magnitude, etc.), 
we would not not be able to statistically establish whether our sample is different from any 
(larger) comparison sample that shows only 50\% of objects associated with clusters. Instead, this would be possible if at least 77\% of the full 
sample was observed. A higher level of statistical accuracy is clearly very important for a number of tasks, e.g. testing unification scenarios, 
comparing radio-loud and radio-quiet AGNs, establishing differences between inactive and active galaxies, etc.

\subsection{HST Observation Strategy}

Table \ref{tab:objinfo} lists details of the HST data that have been collected for this analysis.  
We imaged each object with both the UVIS and IR channels of WFC3, through the filters which 
cover the highest DQE spectral regions of those channels. See the 
WFC3 Instrument Handbook \citep{dressel14} for more technical details on the two channels within 
WFC3.

UVIS observations were made using the F606W filter, which has a pivot wavelength of 588.7 nm and a 
width of 218.2 nm.  The field of view in the UVIS channel is 162" x 162" at a resolution of 0.04" per pixel. 
For each object, we collected 2 images with equal exposure times and a 2.4" pointing offset between one 
another, in order to enable cosmic ray rejection and to cover the 1.2" gap between the two UVIS CCDs.  
For QSO targets, we also collected a 30-second image, which will be used to facilitate PSF subtraction. 
This work will be described in a future paper.


The IR channel observing strategy was similar. Two observations of each object were made through the 
F140W filter (pivot wavelength 1392.2 nm, width 384 nm). These observations covered a 123" x 136" field 
of view, at a resolution of 0.13" per pixel.  For each observation, we used the SPARS50 sample sequence 
with 11 non-destructive detector readouts, translating into an exposure time of 249 seconds.  We also imposed 
a ~1.25" dither between the two IR channel observations of each object.

\begin{deluxetable*}{lrrlcccc}
\tablecolumns{8}
\tabletypesize{\scriptsize}
\tablecaption{Basic characteristics and exposure times of observed objects.}
\tablewidth{0pt}
\tablehead{
\colhead{3C Num} & \colhead{RA} & \colhead{Dec} &    \colhead{z} & \colhead{F606W}            & \colhead{F140W}          &    \colhead{$S_{178 MHz}$}   & \colhead{Log $L_{178 MHz}$} \\
                 &              &               &                & \colhead{Exp Time (sec)}   & \colhead{Exp Time (sec)} &    \colhead{(Jy)}           & \colhead{(erg/sec/Hz)}     \\
          }   
\startdata
\cutinhead{Radio Galaxies}
3C 210      & 8:58:10.0  & +27:50:52 	& 1.169   & 2 x 500       & 2 x 249.23 & 9.5       	   &  35.85  	\\
3C 230      & 9:51:58.8  & -00:01:27    & 1.487   & 2 x 510       & 2 x 249.23 & 19.2  	           &  36.37 	\\
3C 255      & 11:19:25.2 & -03:02:52    & 1.355   & 2 x 500       & 2 x 249.23 & 13.3  	           &  36.13  	\\
3C 257      & 11:23:09.2 & +05:30.19    & 2.474   & 2 x 520       & 2 x 249.23 & 9.7  	           &  36.30	\\
3C 297      & 14:17:24.0 & -04:00:48 	& 1.406   & 2 x 500       & 2 x 249.23 & 10.3\tablenotemark{a} &  36.05 \\
3C 300.1    & 14:28:31.3 & -01:24:08 	& 1.159   & 2 x 510       & 2 x 249.23 & 10.1	           &  35.87  	\\
3C 305.1    & 14:47:09.5 & +76:56:22 	& 1.132   & 2 x 520       & 2 x 249.23 & 4.6  	           &  35.50  	\\
3C 322      & 15:35:01.2 & +55:36:53 	& 1.168   & 2 x 530       & 2 x 249.23 & 10.2  	           &  36.19  	\\
3C 324      & 15:49:48.9 & +21:25:38 	& 1.206   & 2 x 490       & 2 x 249.23 & 13.6  	           &  36.04  	\\
3C 326.1    & 15:56:10.1 & +20:04:20 	& 1.825   & 2 x 500       & 2 x 249.23 & 9.0  	           &  36.19  	\\
3C 356      & 17:24:19.0 & +50:57:40 	& 1.079   & 2 x 348 + 664 & 2 x 249.23 & 11.3  	           &  35.85  	\\
3C 454.1    & 22:50:32.9 & +71:29:19 	& 1.841   & 2 x 500       & 2 x 249.23 & 10.2  	           &  36.25     \\
\cutinhead{QSOs}                                                                                        
3C 68.1     & 02:32:28.9 & +34:23:47 	& 1.238   & 2 x 550 + 30  & 2 x 249.23 & 12.1	           &  36.01     \\
3C 186      & 07:44:17.4 & +37:53:17 	& 1.069   & 2 x 550 + 30  & 2 x 249.23 & 13.0  	           &  35.90  	\\
3C 208      & 08:53:08.6 & +13:52:55 	& 1.112   & 2 x 550 + 30  & 2 x 249.23 & 17.0                  &  36.06 \\
3C 220.2    & 09:30:33.5 & +36:01:24 	& 1.157   & 2 x 550 + 30  & 2 x 249.23 & 8.6	           &  35.80      \\
3C 268.4    & 12:09:13.6 & +43:39:21 	& 1.402   & 2 x 550 + 30  & 2 x 249.23 & 9.5                   &  36.01 \\
3C 270.1    & 12:20:33.9 & +33:43:12 	& 1.528   & 2 x 550 + 30  & 2 x 249.23 & 12.7  	           &  36.21     \\
3C 287      & 13:30:37.7 & +25:09:11 	& 1.055   & 2 x 550 + 30  & 2 x 249.23 & 16.0                  &  35.98 \\
3C 298      & 14:19:08.2 & +06:28:35 	& 1.438   & 2 x 550 + 30  & 2 x 249.23 & 47.1	           &  36.73     \\
3C 418      & 20:38:37.0 & +51:19:13 	& 1.686   & 2 x 500       & 2 x 249.23 & 11.9  	           &  36.26 	\\
3C 432      & 21:22:46.3 & +17:04:38 	& 1.785   & 2 x 550 + 30  & 2 x 249.23 & 12.5  	           &  36.32  	\\
\enddata
\tablenotetext{a}{Radio flux value from \citep{keller69}.}
\label{tab:objinfo}
\end{deluxetable*}

\section{Data Reduction}

In this section, we focus on the data reduction strategies and techniques we use to produce the highest-quality visible and 
near-IR images for each object. We customize the data reduction and image combination process, rather than using the 
drizzled images produced by the standard HST pipeline and available in the Mikulski Archive for Space Telescopes (MAST). 
Our calibrated and drizzled data are available at http://hz3c.stsci.edu, and will also be ingested into MAST as High Level 
Science Products (HLSP) available for download at http://archive.stsci.edu/hlsp/index.html.

For both the UVIS and IR channel data, we begin our custom data reduction routine with with the {\it flt} files output by the 
standard {\it calwf3} data reduction pipeline. The pipeline performs basic calibration steps on the raw data, including bias 
and dark current subtraction, linearity correction, flat fielding, and bad pixel masking. Additionally, for IR data {\it calwf3} 
identifies and removes cosmic ray effects. It also fits a line to the multiple signal values for each pixel, and outputs the 
pixel's signal rate in the {\it flt} file. For more details on the calibration steps performed by {\it calwf3}, see the WFC3 Data 
Handbook \citep{deustua16}. We also check to be sure that {\it calwf3} is run with the best available reference files for our data.

\subsection{UVIS Data}

Beginning with the {\it calwf3}-output {\it flt} images, we first correct charge transfer efficiency (CTE) effects in the 
files by running them through the CTE correction algorithm \citep{anderson10}. This correction models the 
details of how charge traps in the UVIS detector grab and release charge during the readout process. From this, the 
correction script is able to identify charge which was caught in charge traps during the detector readout, and return it 
to the proper pixels. This greatly reduces the magnitude of the CTE-induced charge trails in the original {\it flt} images. The
resulting data are saved in {\it flc} files.

After the CTE correction is performed we are able to focus on cosmic ray rejection. Astrodrizzle, which 
is part of the Drizzlepac software package \citep{gonzaga12}, allows for identification and removal cosmic rays from WFC3 data, as well as
the removal of the geometric distortion from the images. It also then combines the individual images into a final, clean image. 
Astrodrizzle eliminates cosmic rays by placing multiple images of the same section 
of sky onto the same pixel grid and then comparing the measured signals in each pixel through the stack of observations. If, for a 
given pixel, one image in the stack shows an anomalous signal compared to the mean of all the signals, Astrodrizzle will flag the 
anomalous signal as a cosmic ray hit.

However, with our data composed of only two dithered images per object, we only have two measures of the signal at any given 
point on the sky. Any location impacted by a cosmic ray in one image therefore leaves us with one good measure of the signal for 
that point on the sky, along with one bad measurement. In this situation, Astrodrizzle has no knowledge of which of the two measures 
is good and which is bad.  The traditional way to deal with this situation when combining images with Astrodrizzle is simply to assume 
that the lower of the two values is the uncontaminated signal, and to use that value in the final combined image. This limits the accuracy
of the counts measured in the resulting combined image.

Instead, we attempt to recover some of the sky values in CR-contaminated pixels prior to combining images with Astrodrizzle. We use the 
Python version of LA Cosmic \citep{vandokkum01}, cosmics.py, to identify and remove cosmic rays in our images. LA Cosmic identifies cosmic rays 
by using a variation of Laplacian edge detection, and can distinguish between cosmic rays and undersampled PSFs. 
For more details, see \citet{vandokkum01}. While this method is effective at locating cosmic rays within the data, we note that it 
replaces the signal values in the cosmic ray-impacted pixels with extrapolated counts from the surrounding pixels. It therefore introduces 
larger uncertainties in the recovered pixels compared to the use of a method such as Astrodrizzle for cosmic ray correction.

We run LA Cosmic on 
each individual {\it flc} file, using a set of conservative parameters to 
ensure that no signal is mistakenly removed from the astronomical sources. These parameters include a sigclip
value of 5, sigfrac of 0.85, and objlim of 5. This run of LA Cosmic (hereafter Run 1) identifies the most obvious cosmic rays 
in each image, and replaces each impacted pixel with the median value of the surrounding good pixels. 

We then run LA Cosmic a second time (Run 2) on the original {\it flc} file using more stringent parameters: sigclip
value of 4.5, sigfrac of 0.65, and objlim of 2.  The motivation for this
is to improve the data quality in the areas of sky covered by the 
inter-chip gap in each of the original input images. The width of the inter-chip gap is approximately 1.2$\arcsec$, while the dither between 
our two observations is roughly twice as large. With the inter-chip gap falling on a different area of the sky in each of the two exposures, 
when combining the images we create two 1.2$\arcsec$ wide strips where the sky was only imaged once. Any cosmic rays impacting within 
these strips leave us with no uncontaminated measurements of the signal in that location. 

We extract these regions 
from the clean {\it flc} obtained after Run 2 and insert them in the appropriate position of the {\it flc}
file obtained after Run 1. Correcting a significant fraction of the 
cosmic ray population with LA Cosmic then allows Astrodrizzle to use 2 measures of the signal during image combination for many of the cosmic 
ray-impacted pixels.

After these initial cosmic ray correction steps, we use Astrodrizzle to remove residual cosmic rays and geometric distortion, to align, and combine the two individual images for 
each object into a final image.  Final images were rotated into a north-up and east-left orientation.

Even after the previous cosmic ray correction steps, some residual cosmic rays in the drizzled images are still present. This is mainly due to the fact 
that two dither-points are not always sufficient to
provide a fully clean image. This happens for two reasons: i) if two cosmic rays cover the same area of sky in both images, and 
ii) for particularly bright cosmic rays, the rejection algorithm in Astrodrizzle may not work perfectly, for reasons that are not 
completely understood but that 
most likely depend on the specific settings of that task and on the image noise level. 
We remove these residual cosmic rays using a custom procedure. First, we make a mask that includes pixels showing significant flux excess compared to
the surrounding pixels. These are identified by a simple algorithm that compares each drizzled image with both the difference and the ratio of the original two images. 
The marked pixels in the mask are then grown using a Gaussian kernel of appropriate FWHM (generally $\sim 1$ pixel), in order to fix a slightly larger area. 
This is important because these residual cosmic rays appear much more smoothed in the drizzled image than in the original FLT files, and while the brightest pixels are
easily identified with this procedure, the adjacent fainter external pixels might not be correctly removed.

\subsection{IR Data}

The data reduction process for the IR channel data is relatively simpler than that for the UVIS data, due to the fact that the 
multiple non-destructive detector readouts within each observation allowed for easy cosmic ray identification and removal 
within the {\it calwf3} data reduction pipeline.  

The first effect we deal with is persistence. This is an after-image observed in pixels which in previous exposures were subjected to 
high flux levels. See \citet{long11} for more details on persistence. Our goal is to remove 
any persistence signal present in our data, in order to avoid contaminating later photometry.

\begin{figure*}[h]
\epsscale{0.65}
\plotone{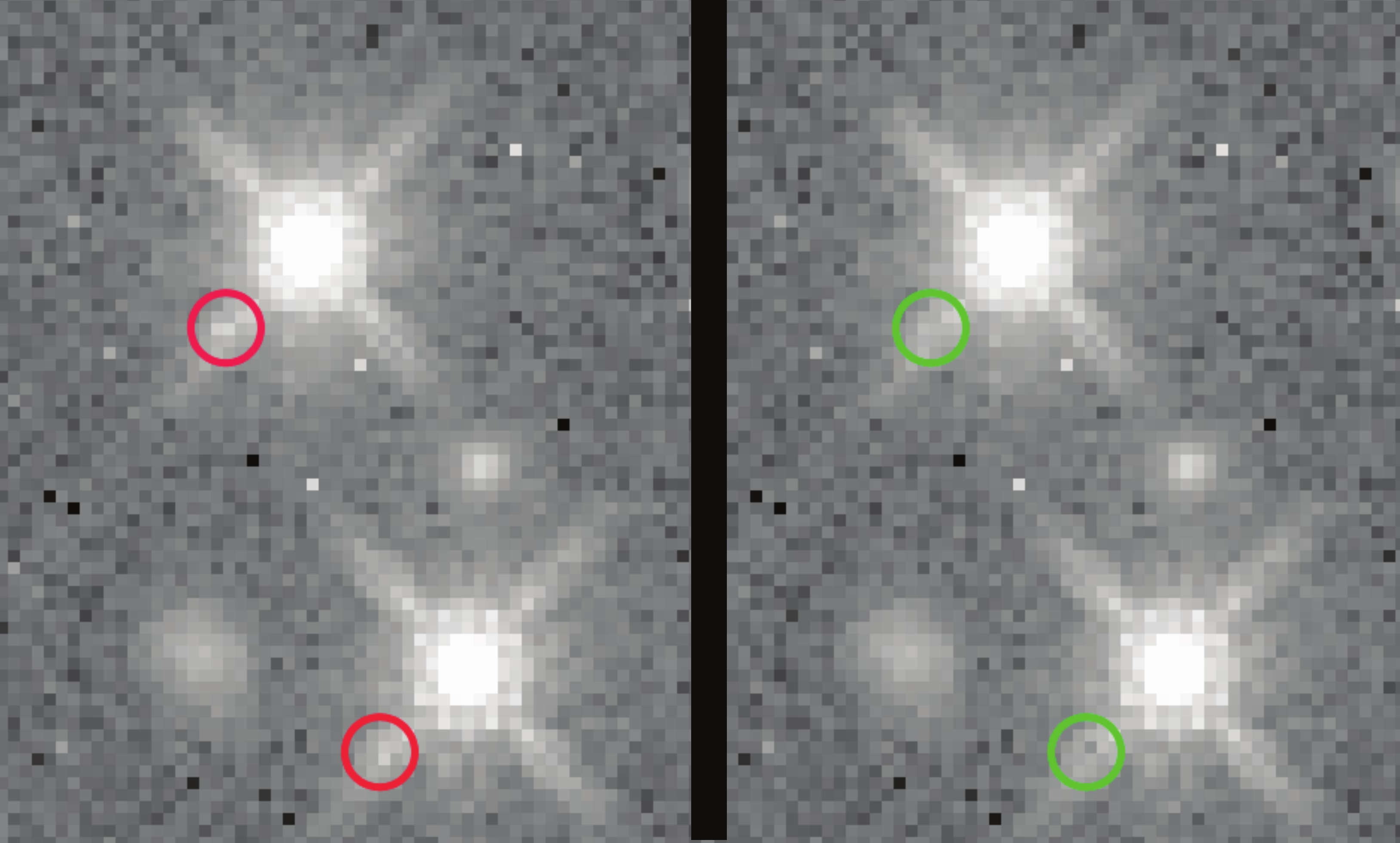}
\caption{Example of self-persistence, seen in before and after versions of our second image of 3C 287. On the left side, persistence signal is visible inside the red circles. 
This persistence is from the cores of these same objects in our first 3C 287 image. The persistence is offset from the cores in the second image 
due to our 1.25$\arcsec$ dither between images.  On the right, we show the persistence-cleaned version of the image. Note that in the green 
circles, the persistence is no longer visible. \label{persist}}
\end{figure*}

We begin by retrieving the persistence masks and persistence-corrected {\it flt} files of our observations from MAST. These 
files have had the persistence signal modeled and subtracted from them, following the model described in section 8.3 of version 4.0 of the 
WFC3 Data Handbook \citep{deustua16}, and are therefore different from the standard {\it flt} files available in the archive. Examination of these 
persistence masks reveals that for most objects there is no appreciable persistence contamination from prior observations. 
However, in some cases we do fall victim to self-persistence, where the bright sources in our initial exposure induced persistence 
in the following image.  While the 1.25$\arcsec$ (approximately 10-pixel) dither between observations is large enough that self-persistence 
from the central object is shifted outside of the object in the second image, there are cases where the shift results in this persistence signal 
appearing close to surrounding objects of interest. An example is shown in figure \ref{persist}, where the panel on the left shows signal due to 
self-persistence inside the red circles. In the panel on the right, the persistence has been removed using the persistence model.


After careful examination of the persistence-corrected {\it flt} files from MAST, we chose to use these files for our subsequent analyses. 

The next step in our data reduction is to remove the geometric distortion from all of the {\it flt} files, and to combine the two images 
for each object into a final IR image. We use Astrodrizzle to accomplish both of these steps.

We then use Tweakreg (also part of the Drizzlepac software package) to align this final image to the same world coordinate system 
present in the corresponding UVIS-channel drizzled image. 
 
At this point it is possible to overlay the UVIS and IR image for a particular object and compare the morphology and brightness in 
the two observation bands.

\section{Photometry} \label{photsection}

Photometric analysis of the objects in our images is performed using SExtractor \citep{bertin96}. Our strategy is to use the IR
images for object identification and aperture definition. The resulting catalogs are then used for the photometry on both the IR and UVIS images. The 
motivation for this strategy lies in the fact that our UVIS data, which have an effective wavelength of roughly 6000 \AA, are capturing UV rest frame photons
for our targets at $z > 1$. These data are therefore biased towards showing irregularly shaped star forming regions rather than the entire galaxies, whose rest frame 
optical emission has been shifted into the wavelength range covered by our IR data. This means that 
many sources present in the IR data appear at a very low signal level, not at all, or with a significantly different
morphology in the UVIS data. This renders the UVIS images unsuitable for aperture definition.

\begin{figure*}[h]
\epsscale{0.70}
\plotone{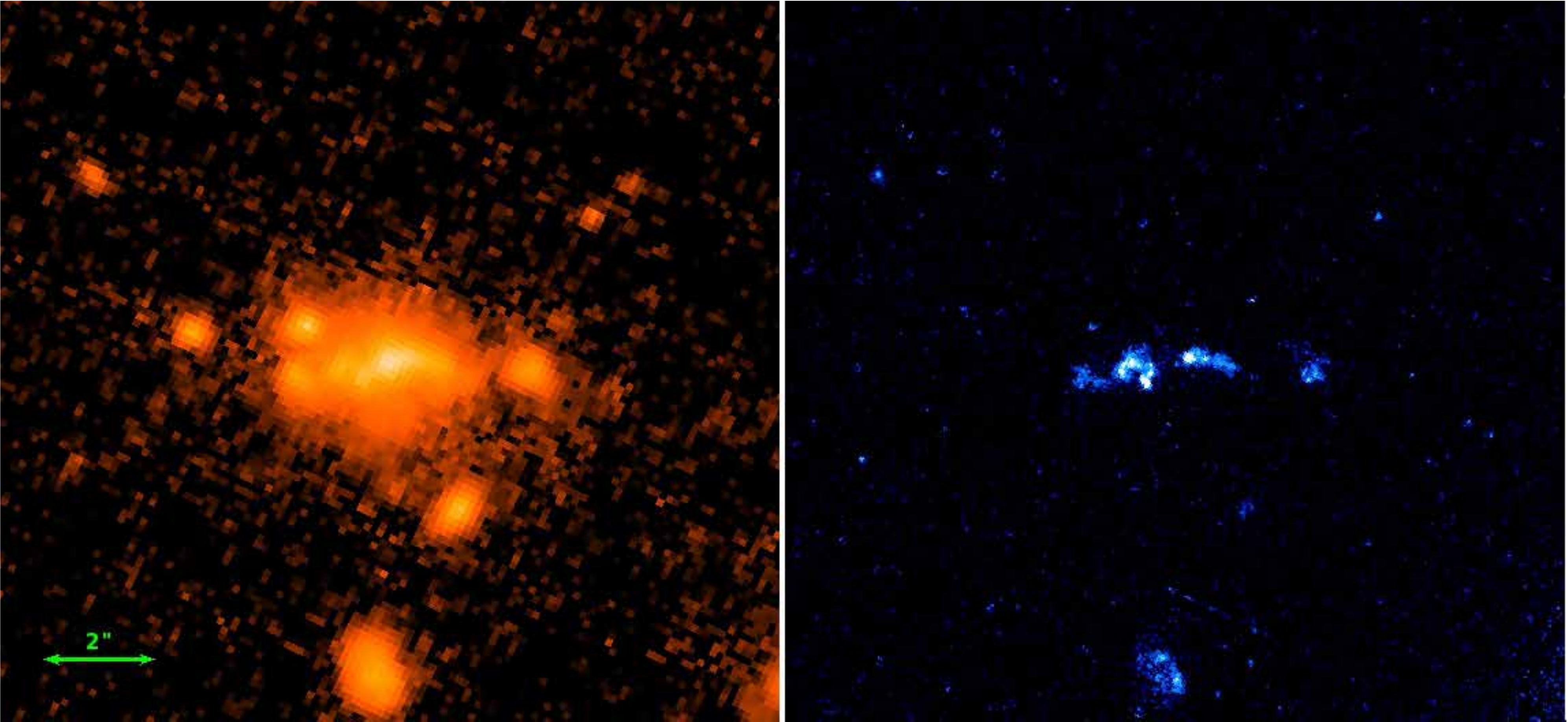}
\caption[]{3C 324 IR and UVIS images. We use the photometric apertures defined in the IR image for photometry in the UVIS data to avoid counting
each small star forming region as a separate source.}
\label{fig:3c324_zoom}
\end{figure*}

For example, the left panel of fig \ref{fig:3c324_zoom} shows that 3C 324 appears as an extended source stretching primarily east-west. However the corresponding
UVIS image, seen in the right panel, reveals appreciable signal only in several distinct star forming regions within the host galaxy. 

By basing the aperture used in
the UVIS image on that defined within the IR image, we are sure to measure the total UVIS signal associated with 3C 324, rather than having SExtractor
erroneously treat each star forming region as a separate object.


We therefore perform our photometry in two steps. First, using SExtractor in MAG\_BEST mode, we obtain measurements of each target's flux in the IR data. In this mode, SExtractor determines the best elliptical aperture to use, and measures the flux inside that aperture. As an ancillary output, we have SExtractor calculate $R_{.9}$, the radius of a circle which would encompass 90\% of the target's light.

We then use $R_{.9}$ as the basis of the UVIS photometry for the target. Testing on real and simulated sources of varying morphologies reveals that using a circular aperture which is 20\% larger than $R_{.9}$ results in the full recovery of all of the object's flux in the UVIS image.

The photometry results are then corrected for galactic absorption and converted into AB magnitudes using the zeropoints provided by the WFC3 team.

Table \ref{tab:abmags} lists the resulting AB magnitudes and uncertainties for all of our targets. Note that in this paper we focus only on the results for our target 3CR sources rather than all of the 
objects in each image.

\begin{deluxetable}{lcc}[h]
\tabletypesize{\scriptsize}
\tablecaption{AB magnitudes for all objects, corrected for galactic absorption.}
\tablewidth{0pt}
\tablehead{
\colhead{Target} & \colhead{F606W Mag} & \colhead{F140W Mag} 
}
\startdata

\cutinhead{Radio Galaxies}
3C210	&	21.770 $\pm$ 0.035	&	19.790 $\pm$ 0.005\\
3C230	&	22.386 $\pm$ 0.040	&	19.971 $\pm$ 0.016\\ 
3C255	&	23.356 $\pm$ 0.082	&	20.755 $\pm$ 0.009\\
3C257	&	24.422 $\pm$ 0.221	&	21.404 $\pm$ 0.018\\
3C297	&	21.855 $\pm$ 0.033	&	19.377 $\pm$ 0.003\\
3C300.1	&	22.508 $\pm$ 0.072	&	19.837 $\pm$ 0.005\\
3C305.1	&	21.330 $\pm$ 0.023	&	19.446 $\pm$ 0.003\\
3C322	&	23.491 $\pm$ 0.074	&	20.584 $\pm$ 0.006\\
3C324	&	22.233 $\pm$ 0.040	&	19.688 $\pm$ 0.004\\
3C326.1	&	25.140 $\pm$ 0.251	&	22.197 $\pm$ 0.033\\
3C356	&	22.032 $\pm$ 0.030	&	19.762 $\pm$ 0.004\\
3C454.1	&	22.497 $\pm$ 0.124	&	19.577 $\pm$ 0.003\\
\cutinhead{QSOs}
3C68.1	&	19.474 $\pm$ 0.002	&	17.299 $\pm$ 0.001\\
3C186	&	17.747 $\pm$ 0.001	&	17.367 $\pm$ 0.001\\
3C208	&	17.829 $\pm$ 0.001	&	17.027 $\pm$ 0.001\\
3C220.2	&	18.451 $\pm$ 0.001	&	17.705 $\pm$ 0.001\\
3C268.4	&	17.915 $\pm$ 0.001	&	16.548 $\pm$ 0.001\\
3C270.1	&	18.531 $\pm$ 0.001	&	17.795 $\pm$ 0.001\\
3C287	&	18.176 $\pm$ 0.001	&	17.818 $\pm$ 0.001\\
3C298	&	17.276 $\pm$ 0.001	&	15.981 $\pm$ 0.001\\
3C418	&	17.798 $\pm$ 0.003	&	17.224 $\pm$ 0.001\\
3C432	&	18.163 $\pm$ 0.001	&	18.137 $\pm$ 0.001\\

\enddata
\label{tab:abmags}
\end{deluxetable}


\subsection{Emission Line Contamination}
\label{emissionlines}

\begin{figure*}[]
\begin{center}
\centering \includegraphics[width=15cm]{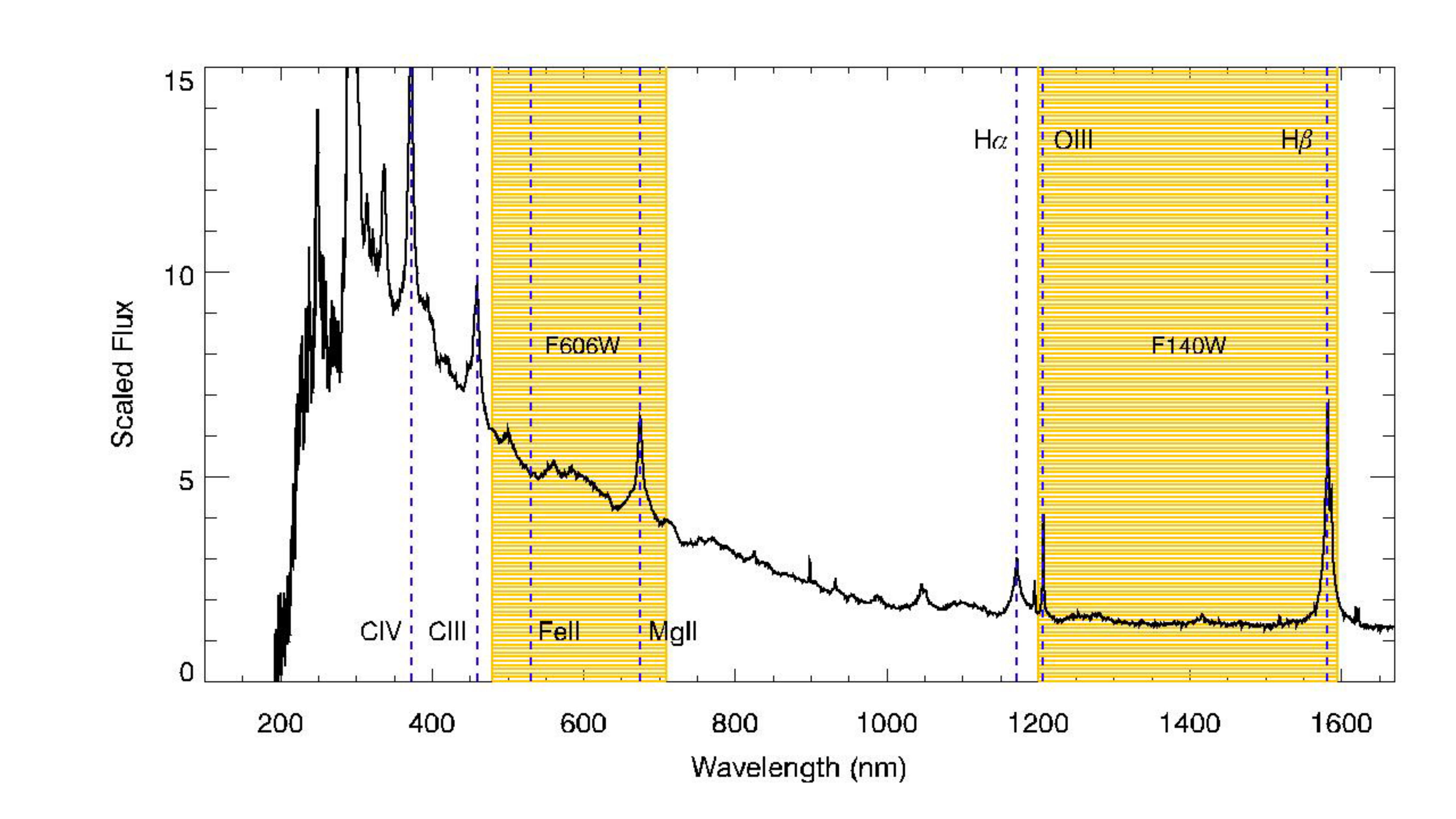}
\caption{This figure shows an example QSO spectrum (from \citet{vanden01}) redshifted to z=1.41 (roughly matching 3C 268.4), 
plotted on top of the bandpasses for the F606W (left) and F140W (right) filters, shown in yellow. In this case, two emission lines are present within each of the filter bandpasses.\label{fig:emissionlines}}
\end{center}
\end{figure*}

In addition to the broadband photometry results above, we are also interested in the amount of emission line contamination 
in the measured fluxes of our objects. Depending on the redshift of a particular target, it is possible to have flux from 
any of a series of emission lines shifted into 
the bandpasses of the F606W and F140W filters used to collect our data. As an example, Figure \ref{fig:emissionlines} shows a median composite QSO 
spectrum from \citet{vanden01} which we have redshifted to {\it z}=1.41 (e.g. similar to our target 
3C~268.4) for illustrative purposes, with our filter bandpasses overlaid. In this case, we see that the FeII and MgII emission lines are present within the 
F606W bandpass, and that the [OIII] and H$\beta$ lines are present at opposite ends of the F140W bandpass. This suggests that flux from these four emission 
lines may be contributing to the total fluxes we calculated in our photometry calculations. Table \ref{tab:possibleemlines} lists the emission lines which fall
within our two filter bandpasses for all of our targets. We base these lists on the redshift of each target, as well as the rest frame wavelength of each
emission line feature.

For each of our targets, we search the literature for published fluxes for emission lines which fall within the F140W and F606W bandpasses. We input 
each flux value into the WFC3 Exposure Time Calculator (ETC) to estimate the measured count rate that emission line would produce on the relevant WFC3 detector.
We then compare these predicted emission line signals with our broadband photometry results to obtain a measure of the emission line contamination.

\begin{deluxetable}{ccc}[h]
\tabletypesize{\scriptsize}
\tablecaption{Emission lines in our target images.}
\tablewidth{0pt}
\tablehead{
\colhead{3C Num}  & \colhead{Em. Lines within} & \colhead{Em. Lines within}  \\ 
                  & \colhead{UVIS bandpass}    & \colhead{IR bandpass}        
          }
\startdata
\cutinhead{Radio Galaxies}
3C 210    & MgII               & H$\alpha$          \\
3C 230    & MgII               & H$\beta$, [OIII]   \\
3C 255    & MgII               & H$\alpha$          \\
3C 257    & CIII]             & none               \\
3C 297    & MgII               & H$\alpha$, [OIII]  \\
3C 300.1  & MgII               & H$\alpha$          \\
3C 305.1  & MgII               & H$\alpha$          \\
3C 322    & MgII               & H$\alpha$          \\
3C 324    & MgII               & H$\alpha$          \\
3C 326.1  & CIII]             & H$\beta$, [OIII]   \\
3C 356    & MgII               & H$\alpha$          \\
3C 454.1  & CIII]             & H$\beta$, [OIII]   \\
\cutinhead{QSOs} 
3C 68.1   & FeII, MgII         & H$\alpha$         \\ 
3C 186    & MgII               & H$\alpha$          \\
3C 208    & MgII               & H$\alpha$          \\
3C 220.2  & MgII               & H$\alpha$          \\
3C 268.4  & FeII, MgII         & H$\alpha$, [OIII]  \\
3C 270.1  & CIII], FeII, MgII & H$\beta$, [OIII]   \\
3C 287    & MgII               & H$\alpha$          \\
3C 298    & FeII, MgII         & [OIII]             \\
3C 418    & CIII], FeII       & H$\beta$, [OIII]   \\
3C 432    & CIII], FeII       & H$\beta$, [OIII]   \\
\enddata
\tablecomments{Here we list the emission lines which fall within the filter bandpasses for all targets. This information is based on the redshifts of the targets
as well as the rest frame wavelengths of the emission lines. The rest frame wavelengths of the emission lines are: MgII 2798\AA, CIII] 1909\AA, FeII 2200-2800\AA,
[OIII] 5007\AA, H$\alpha$ 6563\AA, H$\beta$ 4861\AA. }
\label{tab:possibleemlines}
\end{deluxetable}

\begin{deluxetable*}{ccccccc}
\tabletypesize{\scriptsize}
\tablecaption{Emission line information for our targets}
\tablewidth{0pt}
\tablehead{
\colhead{3C Num}  & Filter & \colhead{Em. Lines}      & \colhead{Em. Line Flux}   & \colhead{Em. Line}      & \colhead{Em. Line}   & \colhead{Other Em. Lines}\\ 
                  &        & \colhead{within bandpass}& \colhead{$erg/s/cm^{2}$}  & \colhead{Ref}            & \colhead{Contamination} & \colhead{Potentially Present}\\
                  &        &                  & \colhead{$\e{-15}$}       &                          &                        & \\
          }
\startdata
\cutinhead{Radio Galaxies}
\it{3C 297} & \it{F140W} & \it{[OIII]5007}        & $10.3$&\tablenotemark{a}& 0.03\%  & $H\alpha$  \\
3C 356    & F606W & MgII2798               & $17$       &\tablenotemark{b}   &     1.6\%      &  -  \\                   
\cutinhead{QSOs}
3C 186    & F606W & MgII2798               & $12.6$  & \tablenotemark{c}     &    2.4\%  & -\\
3C 208    & F606W & MgII2798               & $13.4$  &\tablenotemark{c}      &     2.6\% & -\\
3C 220.2  & F606W & MgII2798               & $6.9$   &\tablenotemark{c}      &    2.4\%  & -\\
3C 268.4  & F606W & MgII2798               & $9.6$   &\tablenotemark{c}      &    2.1\%  & FeII2200-2800\\
3C 270.1  & F606W & MgII2798               & $6.0$   & \tablenotemark{c}     &     1.7\% & FeII2200-2800\\
3C 270.1  & F606W & CIII]1909              & $5.4$   & \tablenotemark{c}      &     1.3\% & FeII2200-2800 \\
3C 287    & F606W & MgII2798               & $3.2$    & \tablenotemark{c}     &     0.8\% & -\\
3C 298    & F606W & MgII2798               & $19.8$     & \tablenotemark{c}   &   2.5\%  & FeII2200-2800 \\
\enddata
\tablenotetext{a}{Flux used to estimate emission line signals from \citep{jr97}. The stated uncertainty in this flux is 15-25\%, which has a small effect on our calculated contamination percentage.}
\tablenotetext{b}{Flux used to estimate emission line contamination from \citet{lacy94}. We also note a similar flux of 13$\e{-15} erg/s/cm^{-2}$ from \citet{cimatti97}.}
\tablenotetext{c}{Emission line flux from SDSS \citep{ahn12}.}
\tablecomments{The second column lists the filter bandpass within which the emission line appears, and the third column lists the emission line and corresponding rest wavelength. The fluxes listed in the fourth column are measured from SDSS  \citep{ahn12}
with the exception of the [OIII] line in 3C 297 and the MgII line in 3C 356. The sixth column lists the percentage of signal in our observations which could come from these emission lines. The rightmost column lists other emission lines which are present within the bandpasses of our observations, but 
for which we do not have measured flux values, and therefore no contamination estimates. The only F140W emission line for which we have a measured flux is the [OIII] line for 3C 297, indicated with italics. All other emission line fluxes are for contamination in the F606W observations.}
\label{tab:exptimes}
\end{deluxetable*}


Table \ref{tab:exptimes} shows the results, where each row of the table contains the contamination estimate for a single emission line in one of our targets. We find published 
emission line fluxes for a limited set of targets and emission lines. Most fluxes are from data contained in the Sloan Digital Sky Survey's 
Data Release 9 \citep{ahn12}, and are for emission lines present within the F606W bandpass for the QSO targets. We find fluxes for only two emission lines in 
radio galaxy targets. The first is the flux of the MgII2798 emission line within the F606W bandpass for 3C 356 from \citet{lacy94}. The only measured flux for 
an emission line in the F140W bandpass comes from \citet{jr97}, for the [OIII]5007 line in 3C 297.

All emission line fluxes from the literature were measured 
using 3$\arcsec$ apertures, with the exception of the 2$\arcsec$ aperture used by \citet{lacy94} for the 3C 356 observations. 
Examining a single exposure taken with a 3$\arcsec$ aperture, 
\citet{lacy94} find that the only emission line for which
the signal extends beyond their 2$\arcsec$ aperture is the [OII] 3727 line.

While these apertures do not exactly match our broadband photometry 
apertures, we are confident that our apertures capture all of the 
broadband QSO and RG flux. In addition, we believe that the 3$\arcsec$ emission line apertures captured all of the emission line flux.

For the QSOs the 3$\arcsec$ apertures should have captured all of the emission line flux because the dominant part of the MgII line is 
from the broad component, which is emitted within the point source. For the radio galaxy emission line fluxes, in agreement with the findings
of \citet{lacy94}, we find that 3C 356 is smaller than 1$\arcsec$ in the rest frame
UV image, indicating that a 
3$\arcsec$ aperture should capture all of the flux. The galaxy associated with 3C 297 is larger than 3$\arcsec$. However, at the distance of 3C 297, 3$\arcsec$ corresponds to
about 25 kpc. The typical size of the narrow line region (NLR) in radio galaxies is of order 10kpc (e.g. \citep{baum89}). Therefore we are confident that a 
3$\arcsec$ aperture should contain most if not all of the emission line flux for this case as well.

Using these signals as inputs to the ETC, we find that no
single emission line contributes more than 2.6\% of an object's total signal.

We attempt to estimate contaminations for a larger set of emission lines, using typical emission line ratios (e.g. \citep{netzer90}) to estimate fluxes 
for emission lines without published fluxes. However, this method does not account for the dust obscuration of our targets, and leads to highly uncertain results.


\subsection{Radio Observations}
In addition to the visible/IR observations, we also make use of archival VLA radio observations of our objects. The combination of the HST and VLA data allows us to examine the relative positions of radio lobes and star forming regions. 

Over the last few decades, 3CR radio sources have been target sources for many VLA projects, 
and the NRAO VLA archive contains many useful data sets. 

Radio data from the NRAO VLA archive have been selected on the basis of the following criteria:
a) longer integration time to increase general sensitivity; b) frequency and array configuration to match the angular resolution of the visible/IR images as much as possible. Different frequencies on the same target have been selected, when possible, to obtain the spectral index information.
Table \ref{tab:radioobs} lists the basic information on the radio data selected.

The selected VLA archive data were retrieved from the online database in FITS format and loaded into the NRAO Astronomical Image
Processing System (AIPS). Data reduction followed 
the standard procedure. 
The absolute flux density scale have been determined with the observations of a primary calibrator (3C286, 3C48 or 3C147).
AIPS includes values of the flux densities of the primary calibrators taken during several years. We used the values closer in time to
the target source observations (task SETJY). 
The flux density scale used at the VLA up to 15 GHz, for these archival data, is based on the flux density of 3C295 determined by
\citet{baars77}.  At frequencies above 15 GHz the flux density scale is based
on observations and emission models for the planet Mars. 
Phase calibration has been performed using the phase calibrators available for each archive data set, closer to the targets.

Errors on the flux density due to the calibration procedure, unless differently indicated, are estimated to be 3$\%$ at 5/8.4GHz, and 5$\%$ at 15/22GHz.

After manual editing of bad fringe visibilities, iterations of phase self-calibration were performed to correct for 
antenna based errors, until the process converged to a stable solution (task CALIB), followed by a final step of
amplitude and phase self-calibration, if the improvement of the r.m.s. on the final image was significant.

Imaging have been through a clean and restore algorithm (task IMAGR), with a suitable choice of fringe visibility weighting 
(ROBUST/UVBOX parameters) to get the best balance between angular resolution, sensitivity to the extended emission, r.m.s. on the images, then optimizing the comparison with the visible/IR images.

\begin{deluxetable*}{lccccrcr}[]
\tabletypesize{\scriptsize}
\tablecaption{VLA archive radio data}
\tablewidth{0pt}
\tablehead{
\colhead{Target} & \colhead{VLA code} & \colhead{freq.(GHz)} &
\colhead{VLA conf.} & \colhead{obs. date} & \colhead{\# vis.}
& \colhead{FWHM($"$)} & \colhead{rms(uJy)}
}
\startdata

\cutinhead{Radio Galaxies}
3C210    &    AO230    &4.86 &B &19-Apr-2009 &39754
&1.21x1.02@-17$^\circ$&100    \\
3C230    &    AK403    &8.44 &A &27-Jul-1995
&13845&0.32x0.23@-38$^\circ$&70    \\
3C255    &    AV157    &8.44 &A &22-Dec-1988
&32643&0.29x0.23@20$^\circ$&30    \\
3C257    &    AV164    & 8.44 & A & 11-May-1990 &
62429&0.41x0.23@43$^\circ$&30 \\
3C297    &    AV164    & 8.44 & A & 10-May-1990 &
62452&0.44x0.29@37$^\circ$&20    \\
3C300.1    &AK403& 8.44 & A & 27-Jul-1995 &
37364&0.29x0.24@38$^\circ$&50    \\
3C305.1    &AM141&14.94 & A & 18-Feb-1985 &
16987&0.17x0.10@-33$^\circ$&250    \\
3C322    &        AV133& 4.86 & A & 16-Apr-1986
&111530&0.49x0.29@85$^\circ$&80    \\
3C324    &        AB755& 8.21 & A & 11-Jul-1995 &472658&0.20x0.20&15    \\
3C326.1    &AV133&14.94 & A & 16-Apr-1986
&73118&0.17x0.12@-69$^\circ$&150    \\
3C356    &    AF186&  4.85&  A & 22-Apr-1990 &334997&0.35x0.35&30    \\
3C454.1&AM213&4.86        &  A & 17-Aug-1987
&80633&0.36x0.34@-90$^\circ$&100    \\
3C454.1    &AM213        &14.94&A&  17-Aug-1987
&119846&0.12x0.11@35$^\circ$&150\\
\cutinhead{QSOs}
3C68.1    &AB369 &4.86&A&29-Mar-1986&678366&0.29x0.26@-3$^\circ$&30 \\
3C186    &AA129 &8.44&A&14-Sep-1991&8296&0.22x0.20@30$^\circ$&100 \\
3C208    &    AL280 &8.44&A&13-Dec-1992&109850&0.20x0.20&40 \\
3C220.2    &    AK403    & 8.44 & A & 27-Jul-1995 & 12493 & 0.25x0.25& 40 \\
3C268.4    &    AB796    & 8.46 & A & 07-Nov-1996 &
457566&0.18x0.17@-8$^\circ$&30    \\
3C270.1    &    AB796    & 8.46 & A & 07-Nov-1996 &554306
&0.18x0.18&40    \\
3C287    &AP263 &8.44&A&02-Mar-1994&28434&0.26x0.21@75$^\circ$&400 \\
3C298    &AA149 &22.46&A&10-Nov-1992&15021&0.093x0.075@-33$^\circ$&400 \\
3C418    &AM299 &4.86&A&14-May-1990&247857&0.54x0.33@55$^\circ$&700 \\
3C432    &AL280        &8.44&A&
13-Dec-1992&140375&0.23x0.20@-85$^\circ$&40    \\
\enddata
\tablecomments{VLA archive radio data. The second column from the left lists the project identification number of the observations. The third column from the right lists the number of fringe visibilities of the final dataset after flagging and calibration. Each visibility is a single data point of 2 antennas of the interferometer and has a 10 second integration time.}
\label{tab:radioobs}
\end{deluxetable*}

\section{Color Magnitude Diagram}

\begin{figure*}[h]
\begin{center}
\centering \includegraphics[width=12cm]{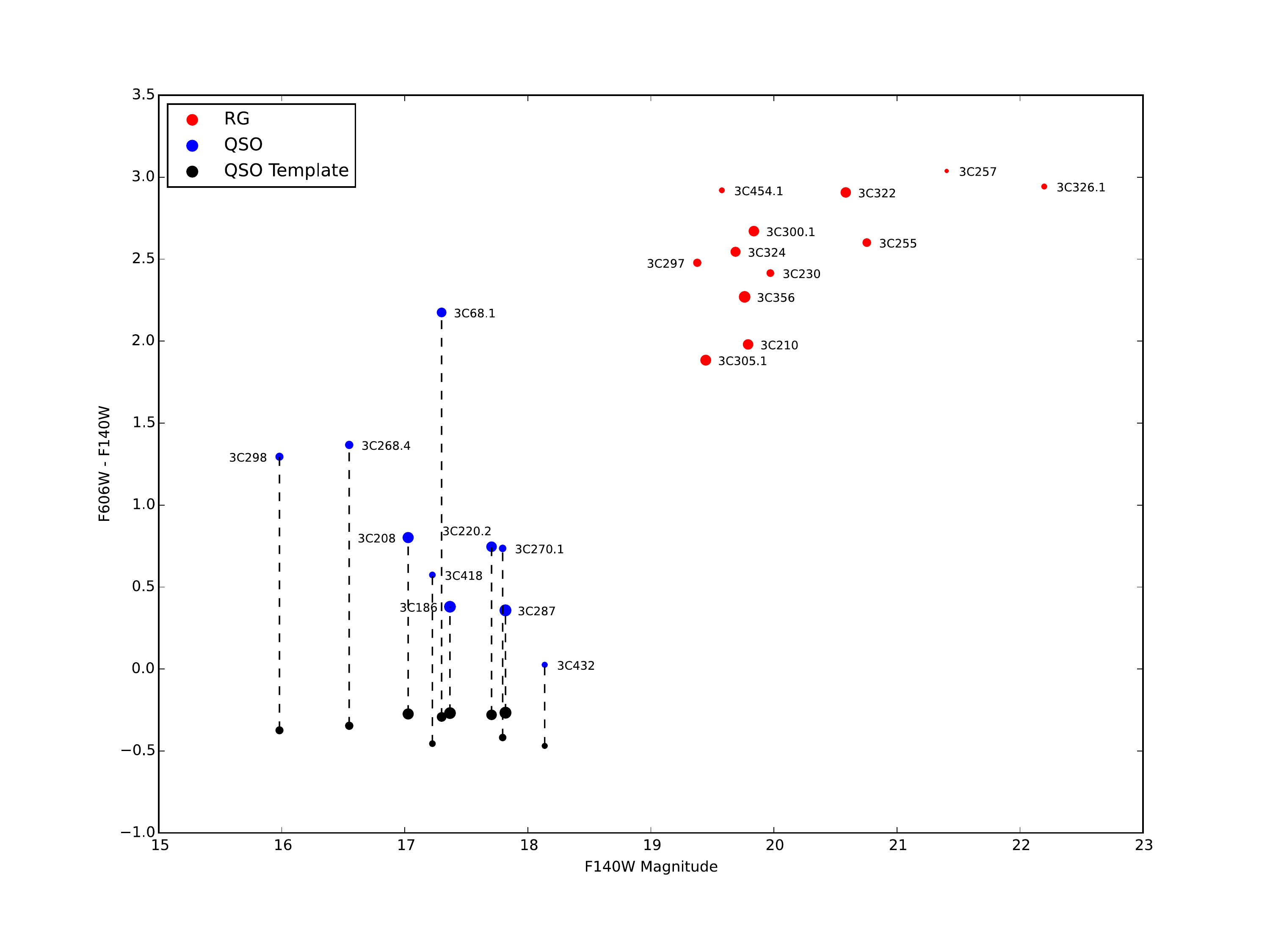}
\caption[]{CMD for all of our targets. Red points show results for the radio galaxies, and blue points represent QSOs. The size of the data points is inversely proportional to the targets' redshift values. This shows that two of our farthest targets (3C 257 and 3C 326.1) are also the reddest. The black points show calculated colors from a template QSO spectrum that has been redshifted and renormalized to match the F140W magnitudes of each of our targets.}
\label{fig:colormag}
\end{center}
\end{figure*}

Figure \ref{fig:colormag} shows the color-magnitude diagram (CMD) produced from our photometry. The blue points show the results for the QSO targets, while the red points represent the RG targets. 

We see that the QSOs are generally more blue than the RGs. This is expected as the bluer AGN spectrum dominates over the redder host galaxy spectrum in the QSOs.

3C 68.1 is the exception to this rule, with a color comparable to many of the RGs. \citet{w13} measures an absorbing 
column density of $9.0\e{22} cm^{-2}$ for 3C 68.1. \citet{brotherton98} showed that the object is characterized by reddened polarized scattered light from the nucleus.
Without reddening, the location of the 3C 68.1 point would move down and to the left (i.e. bluer and brighter), closer to the values for the other QSOs.

Among the RG targets, two of the three farthest targets, 3C 257 and 3C 326.1, are the reddest. Beyond that, the RGs show no clear color
trends. Conversely, as a group the QSO targets appear to become more red as their brightness increases. In order to assure ourselves that this trend was not the result of the different redshifts of the sources, we produce a set of simulated QSO magnitudes and colors for comparison. We use the QSO1 template spectrum from \citep{polletta}, which was created by calculating the average spectrum from a set of 35 SDSS quasar spectra and rest frame IR data. We redshift and renormalize the template spectrum to match the redshift and F140W magnitude of each of our target QSOs. We then calculate the F606W magnitude for each object, and plot the calculated color versus the F140W magnitude as the black points in figure \ref{fig:colormag}. Vertical dashed black lines connect each observed QSO color (in blue), with the model color (in black) for each object. 

The color differences between the blue and black points may imply that the spectra of our targets do not match the template spectrum we used to estimate the 
location of our QSOs in the CMD. A possible effect that might contribute to redden our sources is obscuration, as in the case of 3C68.1.
However, even if reddening affects the flux measured in F606W more significantly than that in F140W, we would still expect to see a correlation between 
reddening and F140W magnitude (i.e. the redder the source, the fainter). Unless the brighter QSOs in the CMD are intrinsically brighter at a level sufficient to counter the obscuration effect in F140W, we do not believe that the presence of reddening is the correct interpretation. Another possible effect is the presence of a stronger emission line contamination in the redder filter in sources that are intrinsically brighter. 
We currently do not have enough information to distinguish which of these situations is the true cause. However, since the black dots span a narrow range in color, 
we can conclude that redshift alone has a very small effect in determining the location of the QSOs in the CMD plot.

\section{Notes on individual objects} \label{objsumm}
\begin{figure*}
\begin{minipage}[b]{0.45\linewidth}
\centering
\includegraphics[width=\textwidth]{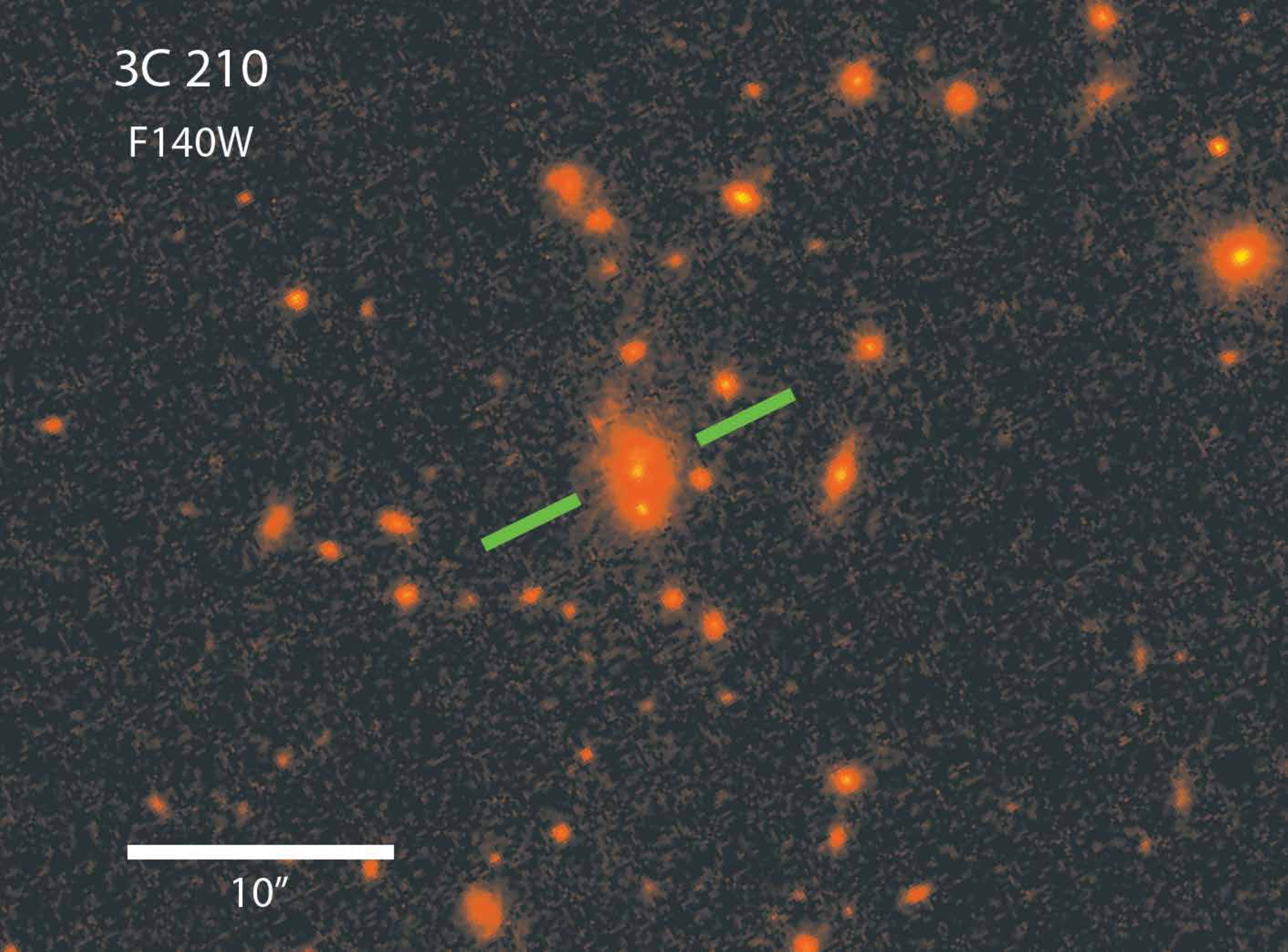}
\caption{Central $35\arcsec \times 50\arcsec$ of the IR image of RG 3C 210. \redpen{The image has been rotated so 
that North is up and East to the left. Green lines are placed on 
either side of the target to help identify its location.}}
\label{fig:3c210_ir}
\end{minipage}
\hspace{0.5cm}
\begin{minipage}[b]{0.45\linewidth}
\centering
\includegraphics[width=\textwidth]{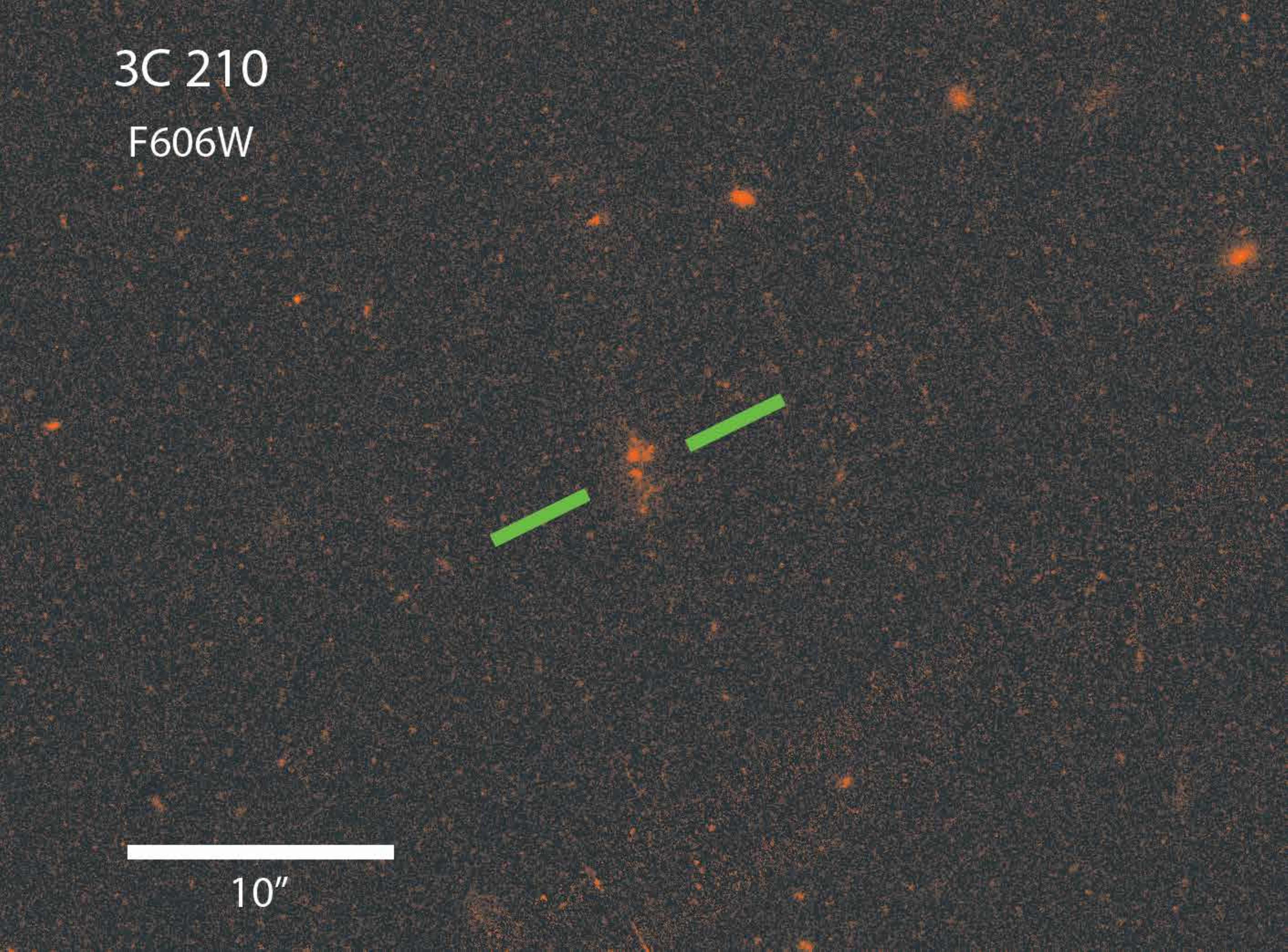}
\caption{Central $35\arcsec \times 50\arcsec$ of the UVIS image of RG 3C 210. \redpen{The image has been rotated
so that North is up and East to the left. Green lines are placed on 
either side of the target to help identify its location.}}
\label{fig:3c210_uv}
\end{minipage}
\end{figure*}

\begin{figure*}[htp]
\begin{minipage}[b]{0.45\linewidth}
\centering
\includegraphics[width=\textwidth]{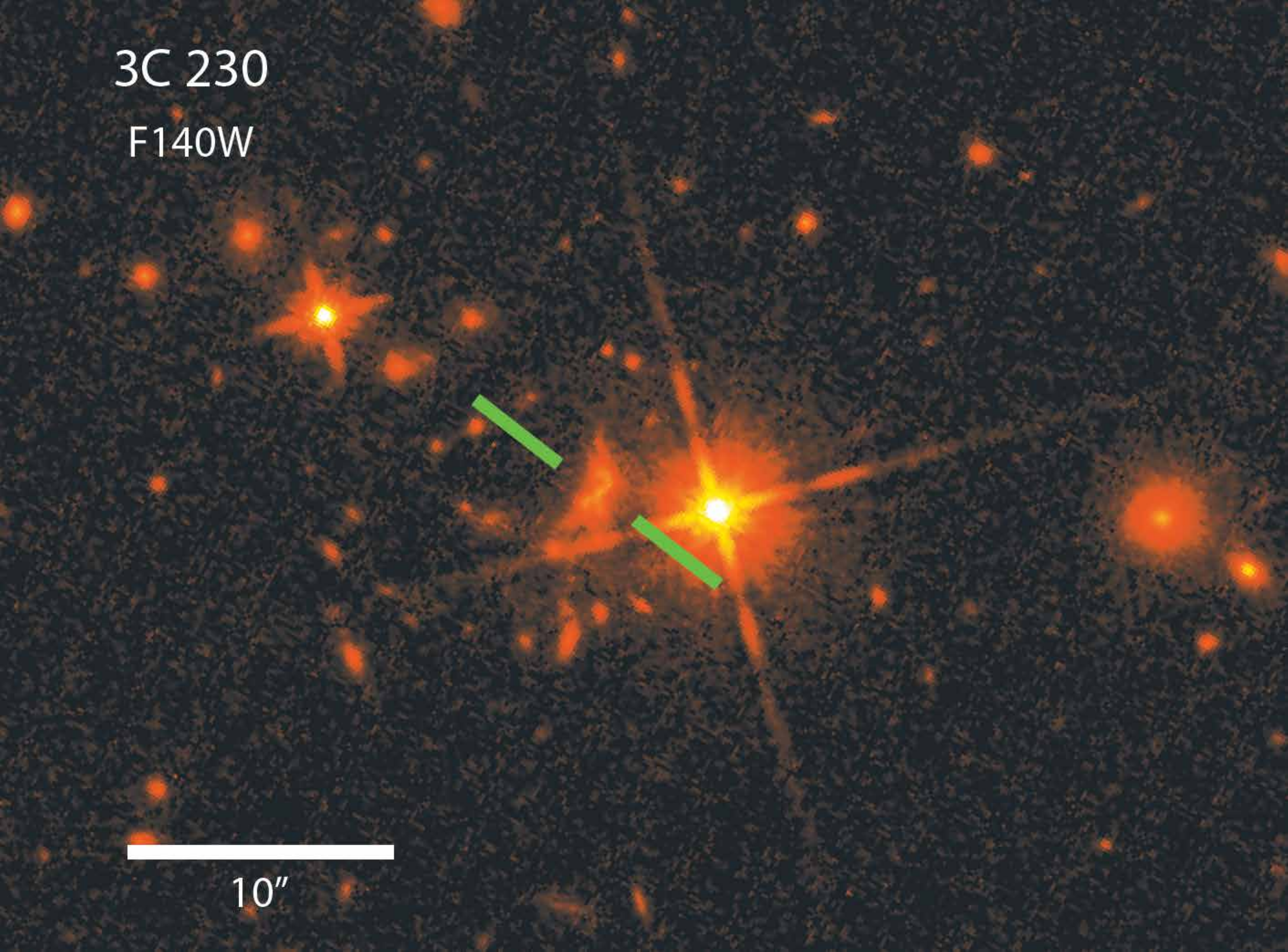}
\caption{Central $35\arcsec \times 50\arcsec$ of the IR image of RG 230. \redpen{The image has been rotated so 
that North is up and East to the left. Green lines are placed on 
either side of the target to help identify its location.}}
\label{fig:3c230_ir}
\end{minipage}
\hspace{0.5cm}
\begin{minipage}[b]{0.45\linewidth}
\centering
\includegraphics[width=\textwidth]{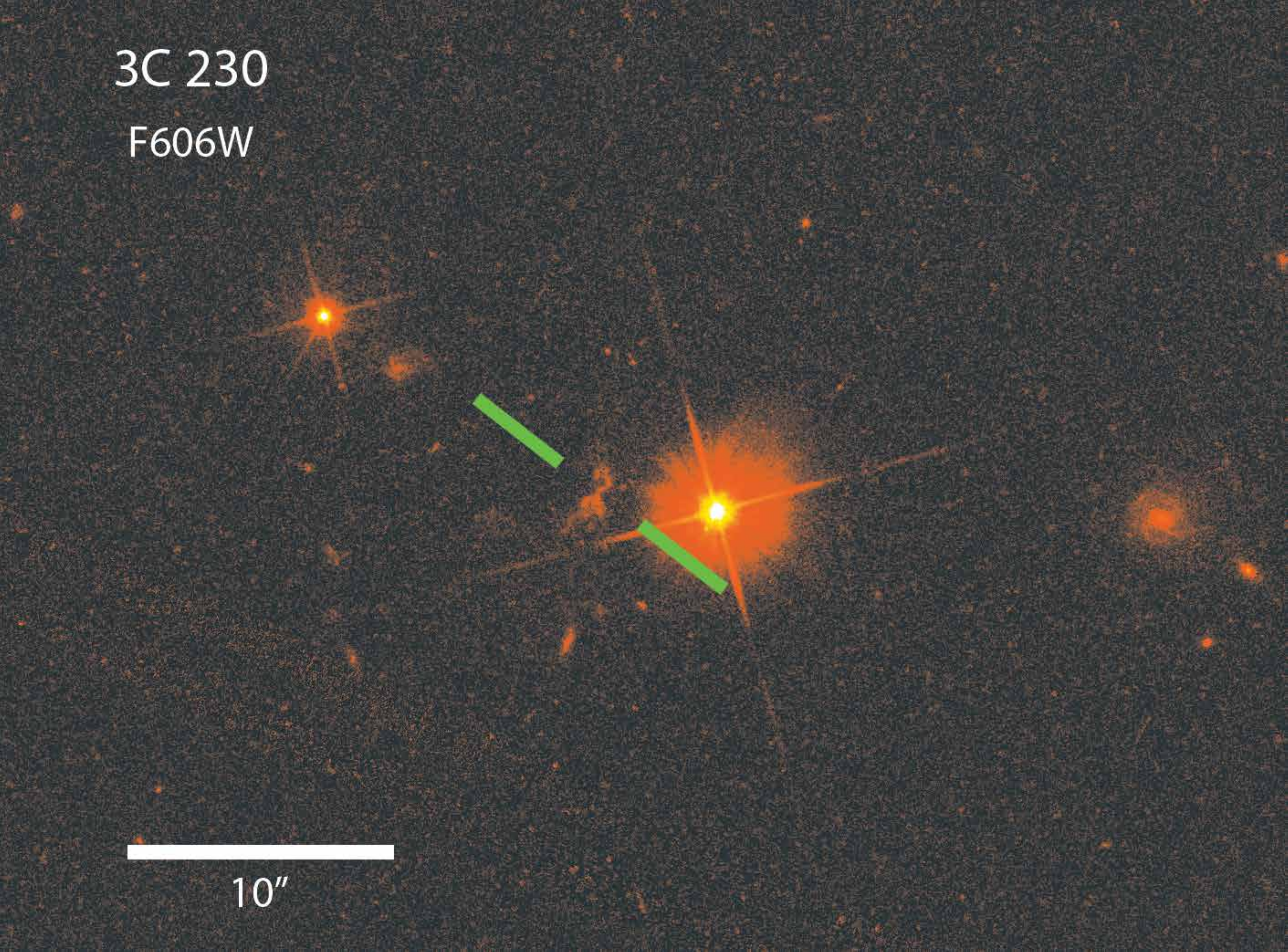}
\caption{Central $35\arcsec \times 50\arcsec$ of the IR image of RG 3C 230. \redpen{The image has been rotated
so that North is up and East to the left. Green lines are placed on 
either side of the target to help identify its location.}}
\label{fig:3c230_uv}
\end{minipage}
\end{figure*}

\begin{figure*}[htp]
\begin{minipage}[b]{0.45\linewidth}
\centering
\includegraphics[width=\textwidth]{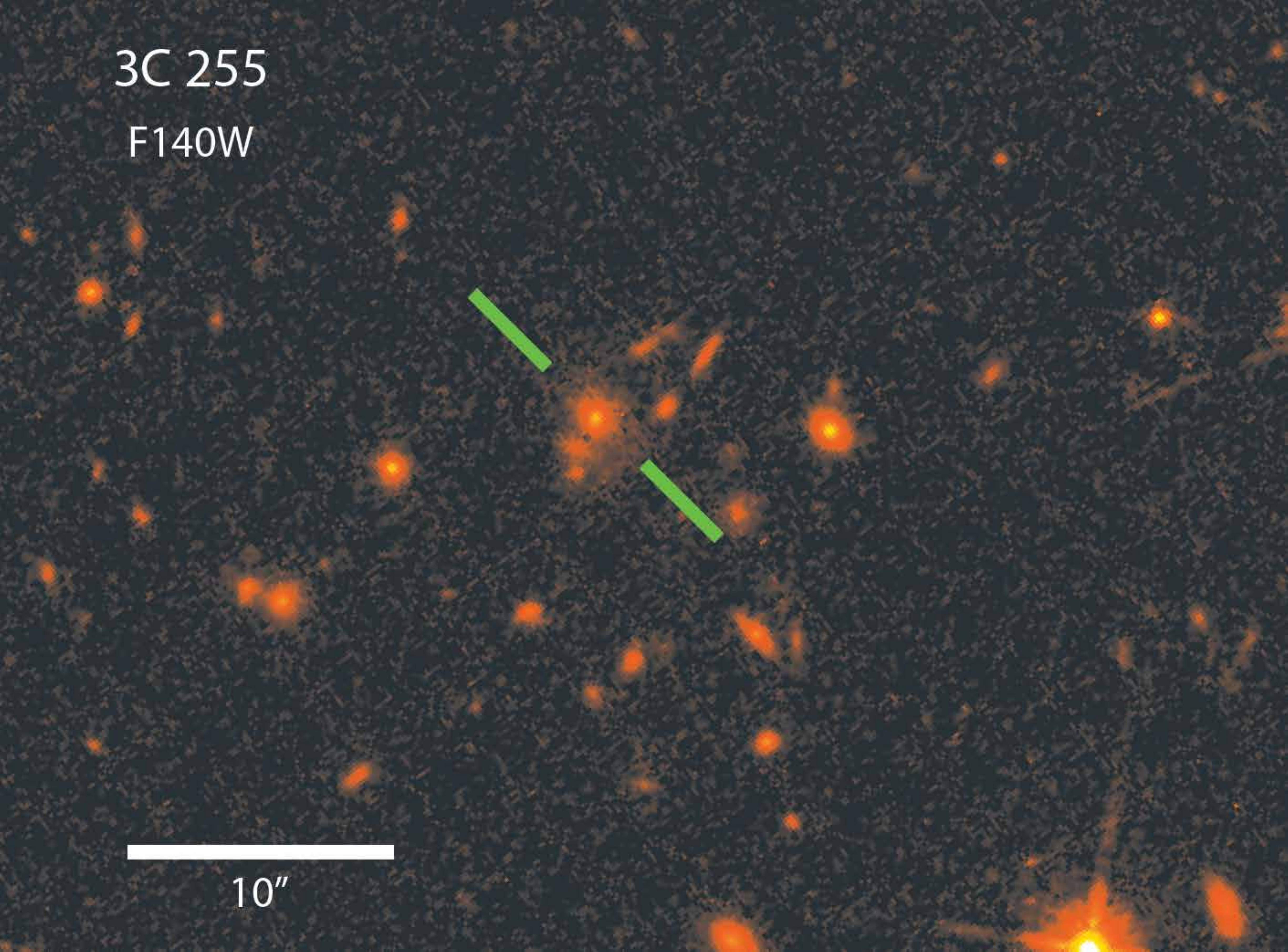}
\caption{Central $35\arcsec \times 50\arcsec$ of the IR image of RG 3C 255. \redpen{The image has been rotated so 
that North is up and East to the left. Green lines are placed on 
either side of the target to help identify its location.}}
\label{fig:3c255_ir}
\end{minipage}
\hspace{0.5cm}
\begin{minipage}[b]{0.45\linewidth}
\centering
\includegraphics[width=\textwidth]{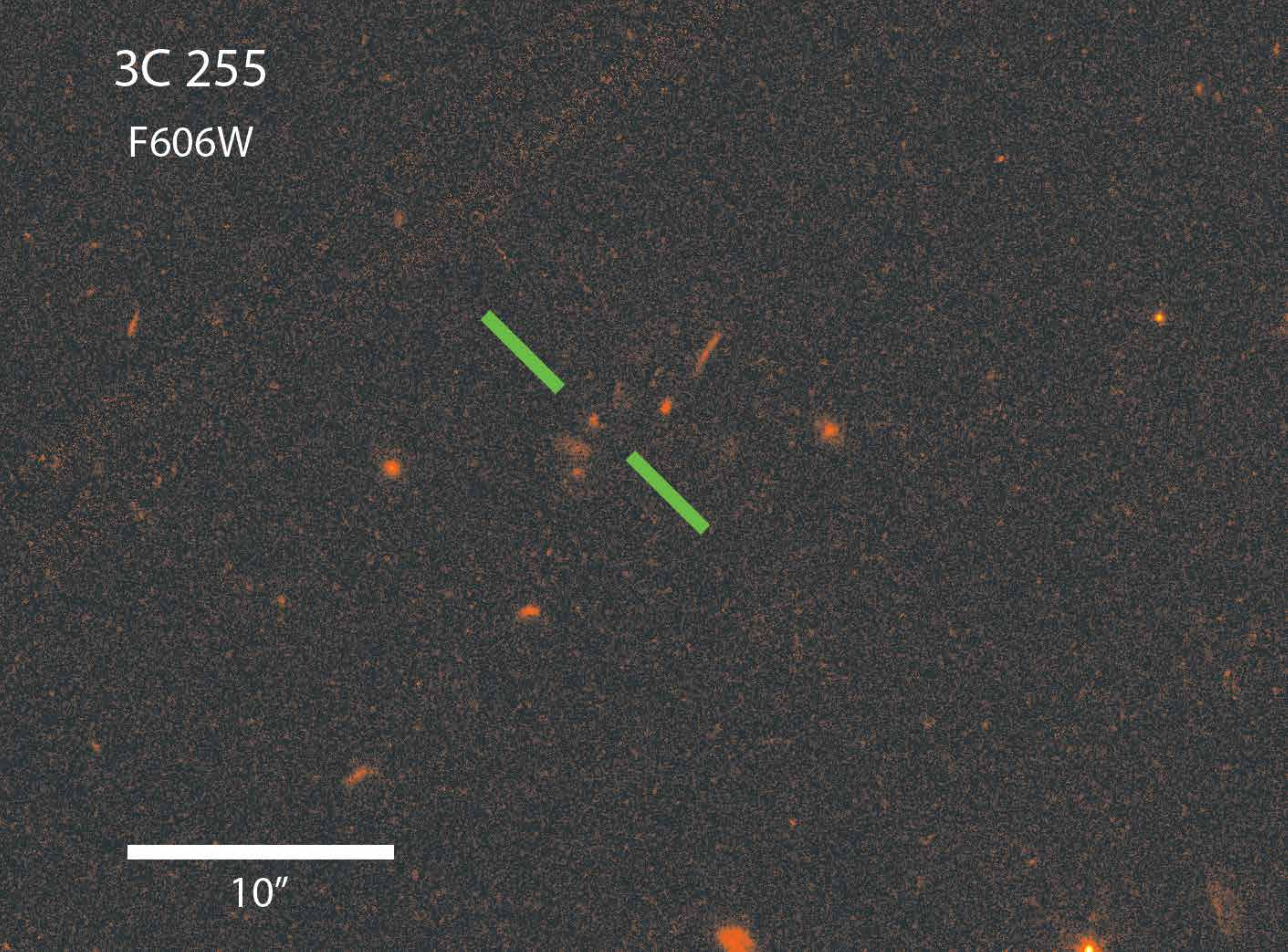}
\caption{Central $35\arcsec \times 50\arcsec$ of the UVIS image of RG 3C 255. \redpen{The image has been rotated
so that North is up and East to the left. Green lines are placed on 
either side of the target to help identify its location.}}
\label{fig:3c255_uv}
\end{minipage}
\end{figure*}

\begin{figure*}[htp]
\begin{minipage}[b]{0.45\linewidth}
\centering
\includegraphics[width=\textwidth]{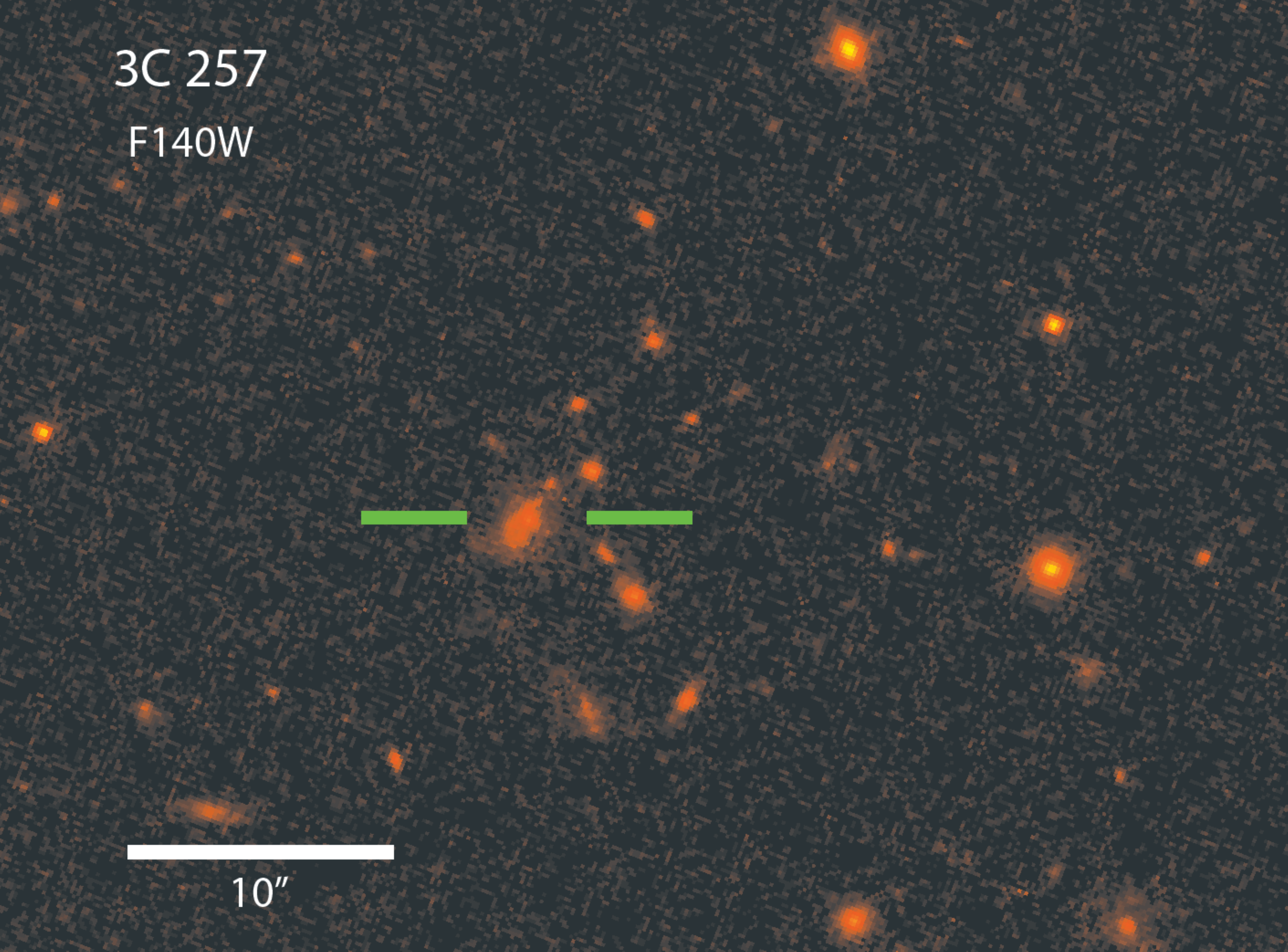}
\caption{Central $35\arcsec \times 50\arcsec$ of the IR image of RG 3C 257. \redpen{The image has been rotated so 
that North is up and East to the left. Green lines are placed on 
either side of the target to help identify its location.}}
\label{fig:3c257_ir}
\end{minipage}
\hspace{0.5cm}
\begin{minipage}[b]{0.45\linewidth}
\centering
\includegraphics[width=\textwidth]{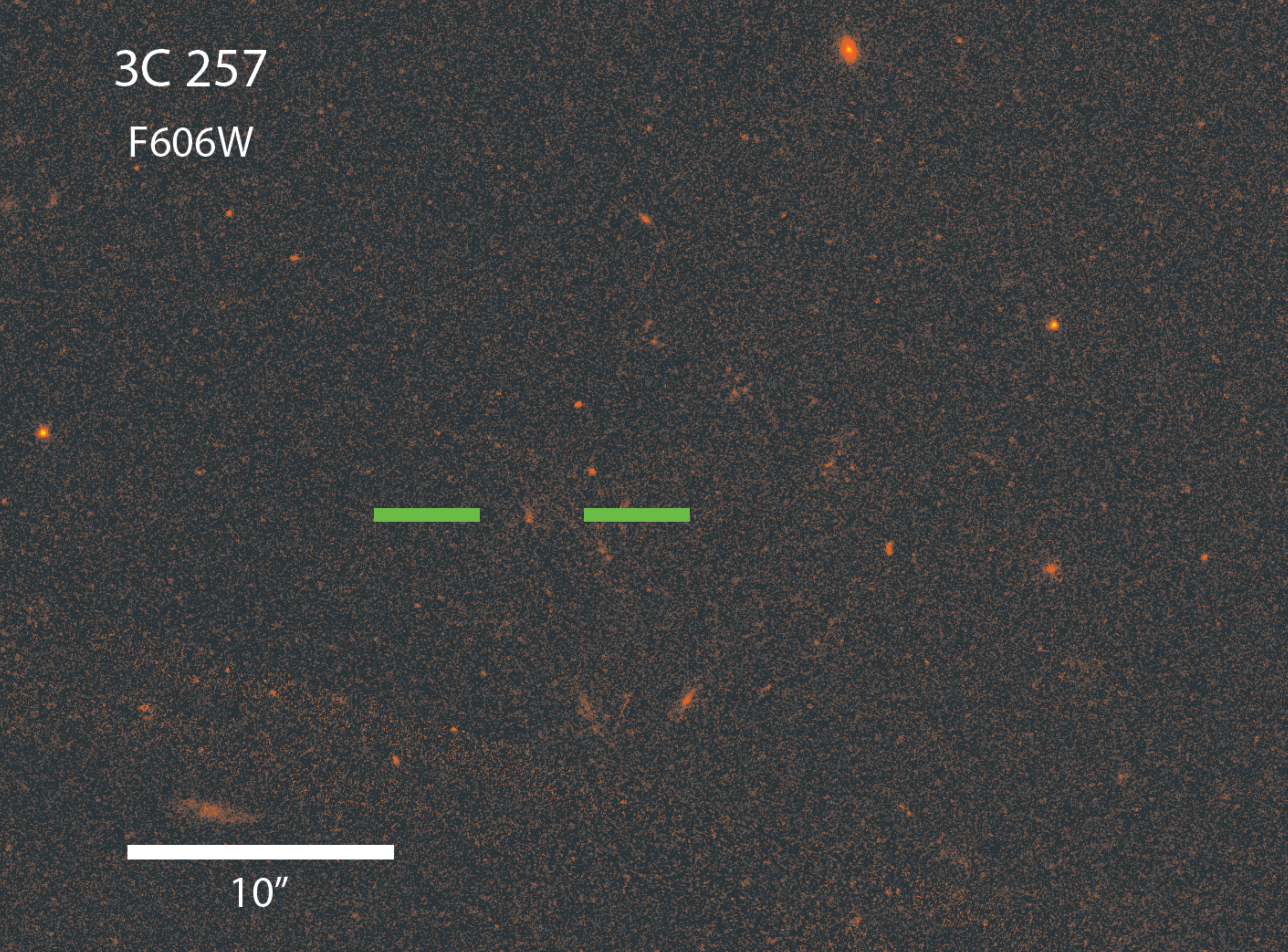}
\caption{Central $35\arcsec \times 50\arcsec$ of the UVIS image of RG 3C 257. \redpen{The image has been rotated
so that North is up and East to the left. Green lines are placed on 
either side of the target to help identify its location.}}
\label{fig:3c257_uv}
\end{minipage}
\end{figure*}

\begin{figure*}[htp]
\begin{minipage}[b]{0.45\linewidth}
\centering
\includegraphics[width=\textwidth]{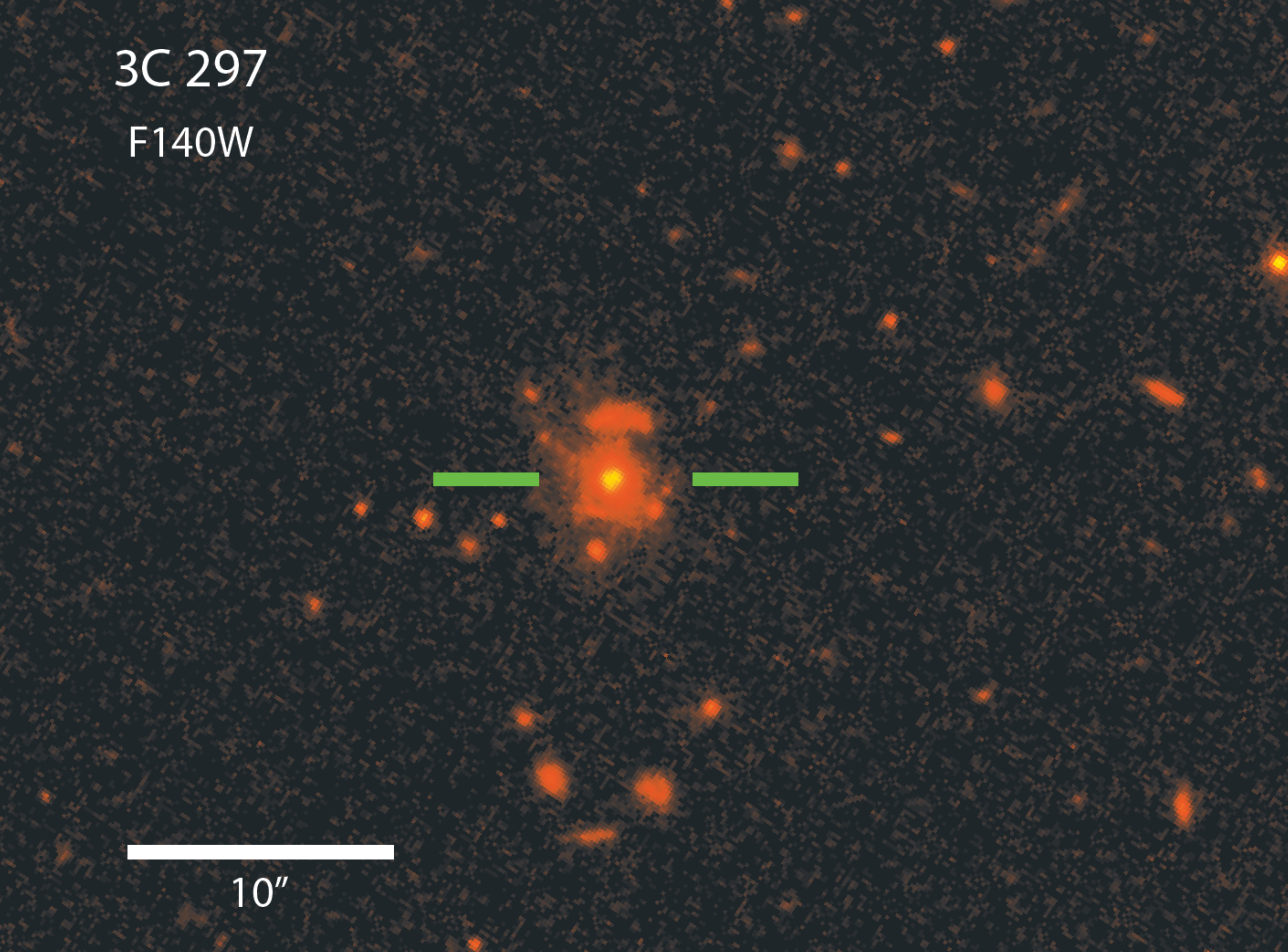}
\caption{Central $35\arcsec \times 50\arcsec$ of the IR image of RG 3C 297. \redpen{The image has been rotated so 
that North is up and East to the left. Green lines are placed on 
either side of the target to help identify its location.}}
\label{fig:3c297_ir}
\end{minipage}
\hspace{0.5cm}
\begin{minipage}[b]{0.45\linewidth}
\centering
\includegraphics[width=\textwidth]{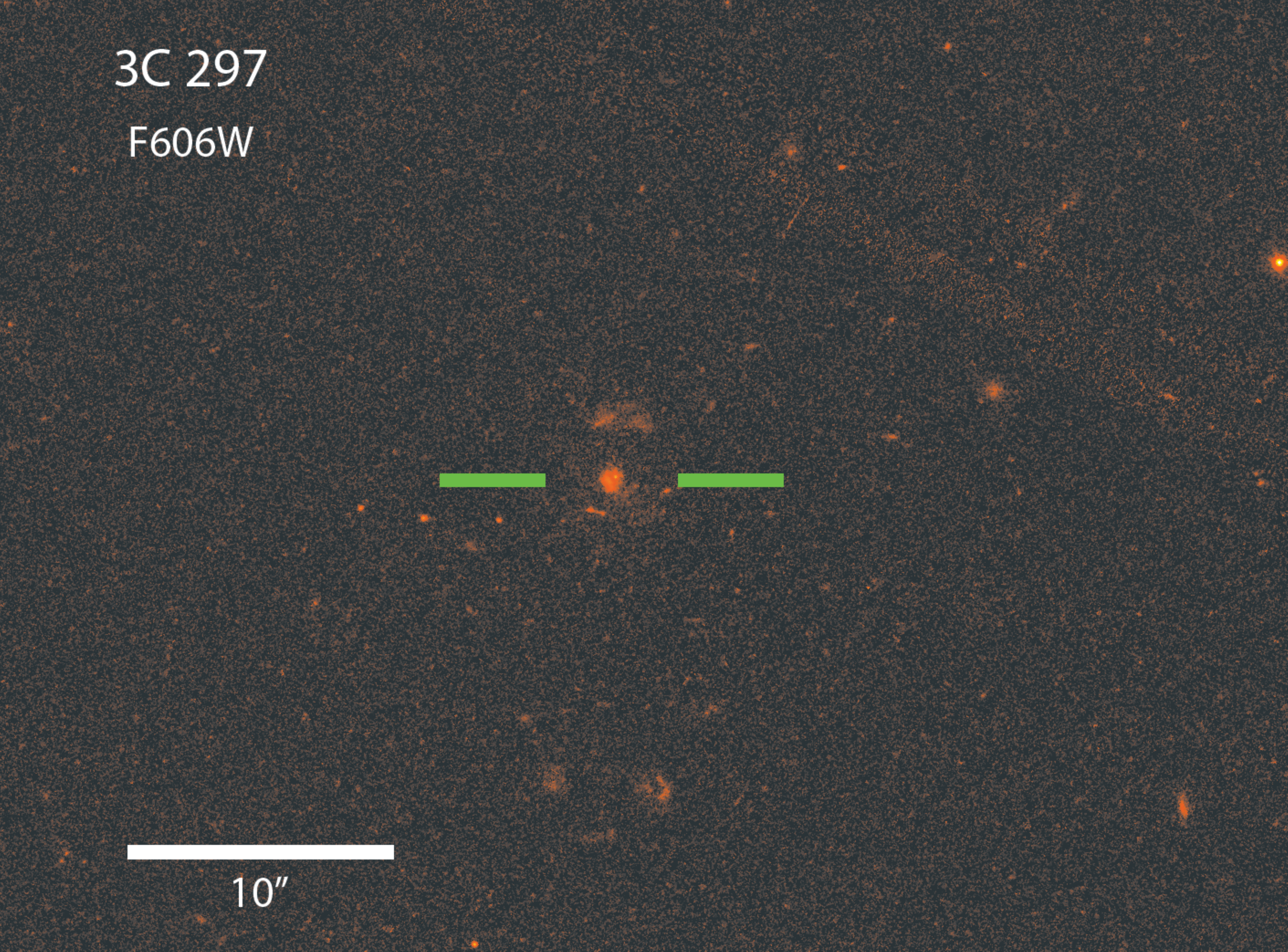}
\caption{Central $35\arcsec \times 50\arcsec$ of the UVIS image of RG 3C 297. \redpen{The image has been rotated
so that North is up and East to the left. Green lines are placed on 
either side of the target to help identify its location.}}
\label{fig:3c297_uv}
\end{minipage}
\end{figure*}

\begin{figure*}[htp]
\begin{minipage}[b]{0.45\linewidth}
\centering
\includegraphics[width=\textwidth]{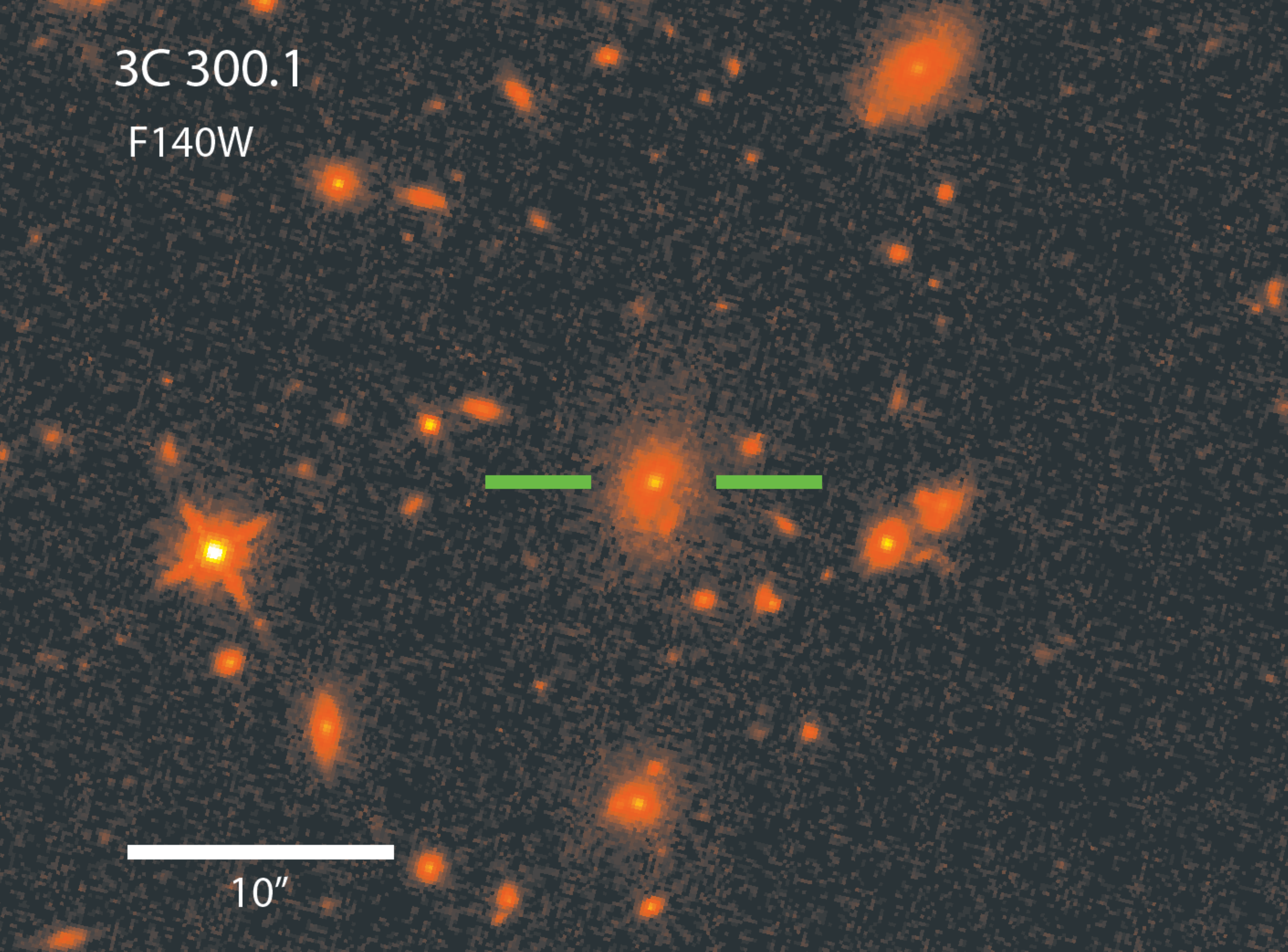}
\caption{Central $35\arcsec \times 50\arcsec$ of the IR image of RG 3C 300.1. \redpen{The image has been rotated so 
that North is up and East to the left. Green lines are placed on 
either side of the target to help identify its location.}}
\label{fig:3c300p1_ir}
\end{minipage}
\hspace{0.5cm}
\begin{minipage}[b]{0.45\linewidth}
\centering
\includegraphics[width=\textwidth]{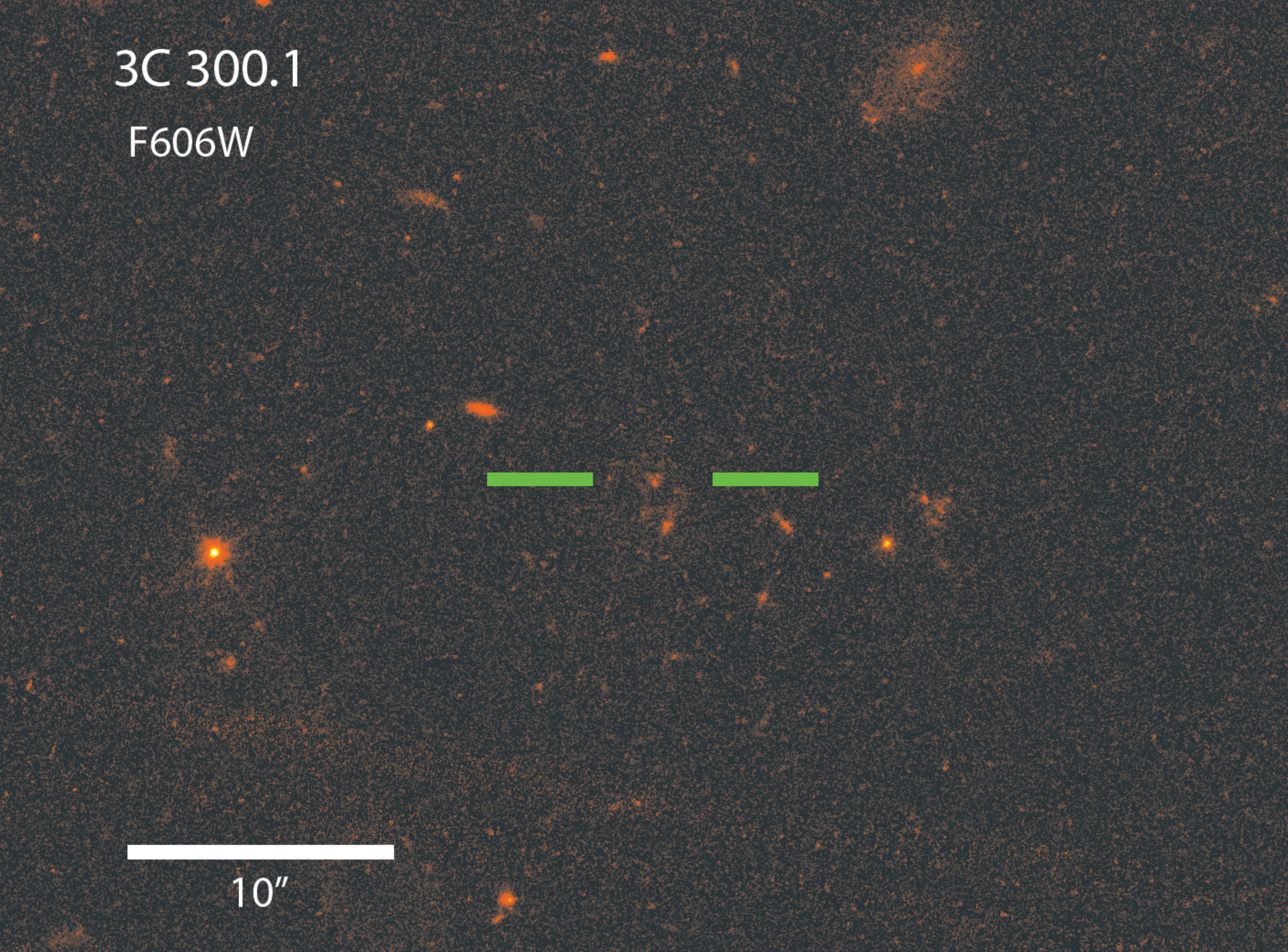}
\caption{Central $35\arcsec \times 50\arcsec$ of the UVIS image of RG 3C 300.1. \redpen{The image has been rotated
so that North is up and East to the left. Green lines are placed on 
either side of the target to help identify its location.}}
\label{fig:3c300p1_uv}
\end{minipage}
\end{figure*}

\begin{figure*}[htp]
\begin{minipage}[b]{0.45\linewidth}
\centering
\includegraphics[width=\textwidth]{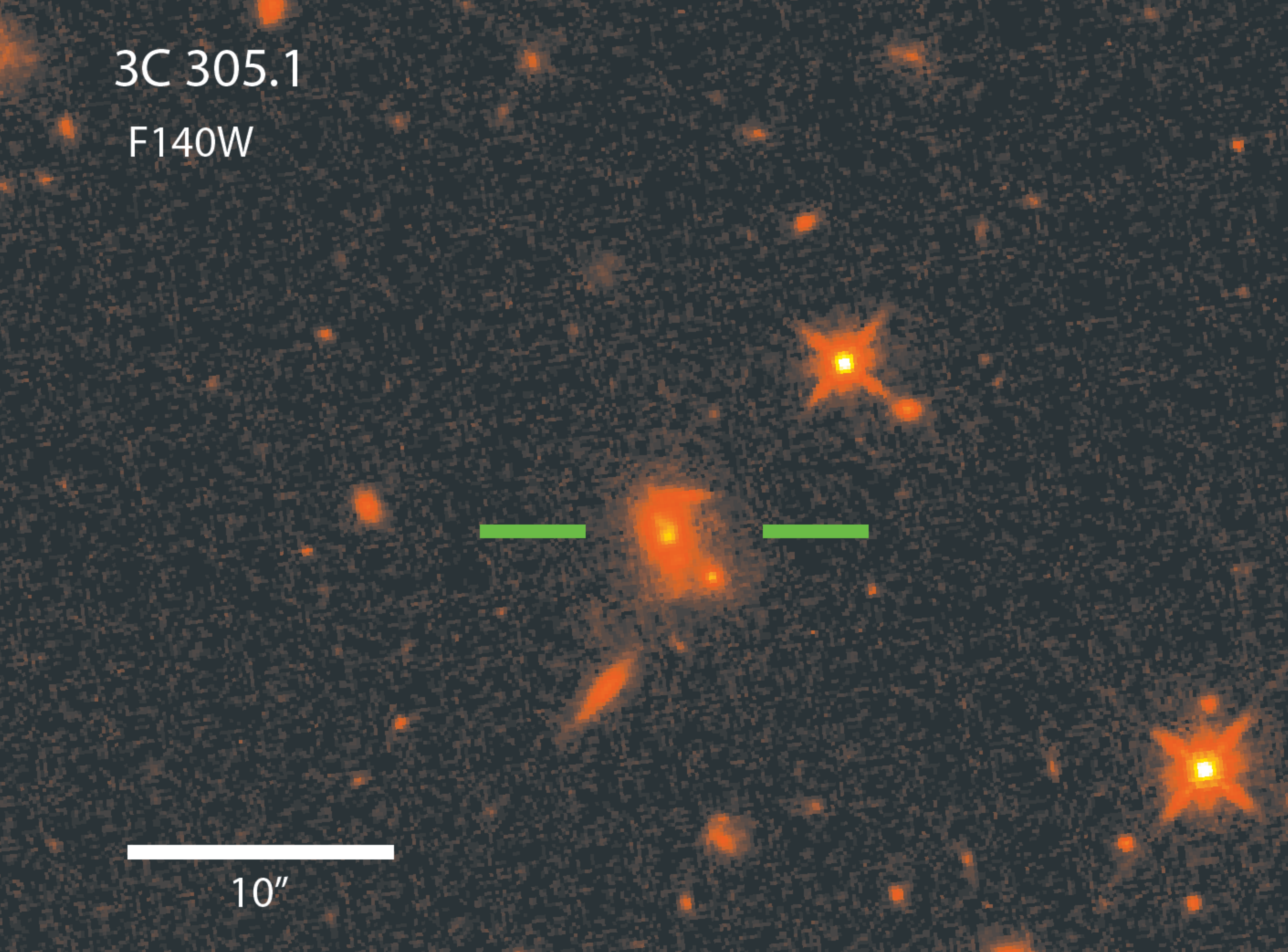}
\caption{Central $35\arcsec \times 50\arcsec$ of the IR image of RG 3C 305.1. \redpen{The image has been rotated so 
that North is up and East to the left. Green lines are placed on 
either side of the target to help identify its location.}}
\label{fig:3c305p1_ir}
\end{minipage}
\hspace{0.5cm}
\begin{minipage}[b]{0.45\linewidth}
\centering
\includegraphics[width=\textwidth]{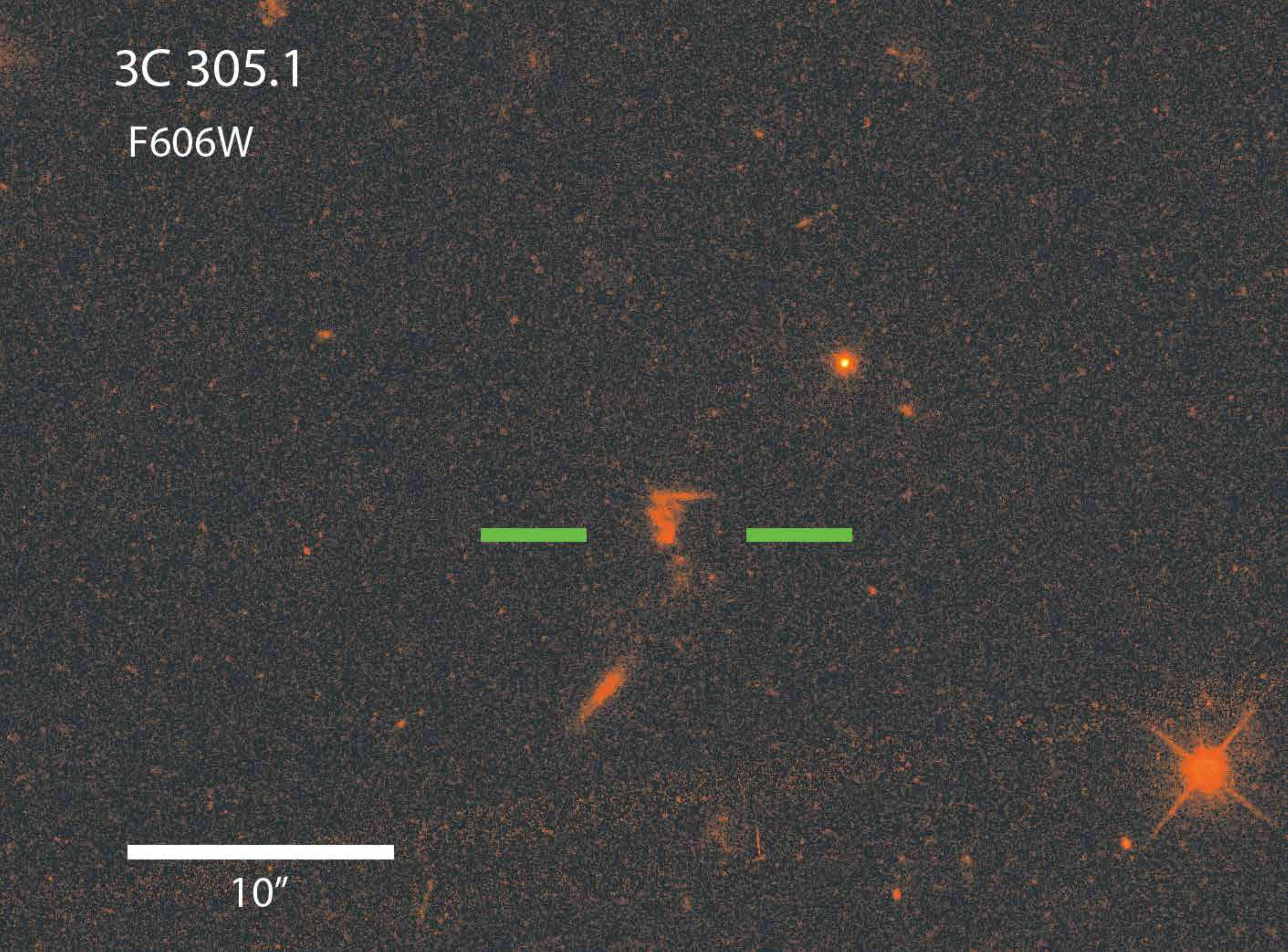}
\caption{Central $35\arcsec \times 50\arcsec$ of the UVIS image of RG 3C 305.1. \redpen{The image has been rotated
so that North is up and East to the left. Green lines are placed on 
either side of the target to help identify its location.}}
\label{fig:3c305p1_uv}
\end{minipage}
\end{figure*}

\begin{figure*}[htp]
\begin{minipage}[b]{0.45\linewidth}
\centering
\includegraphics[width=\textwidth]{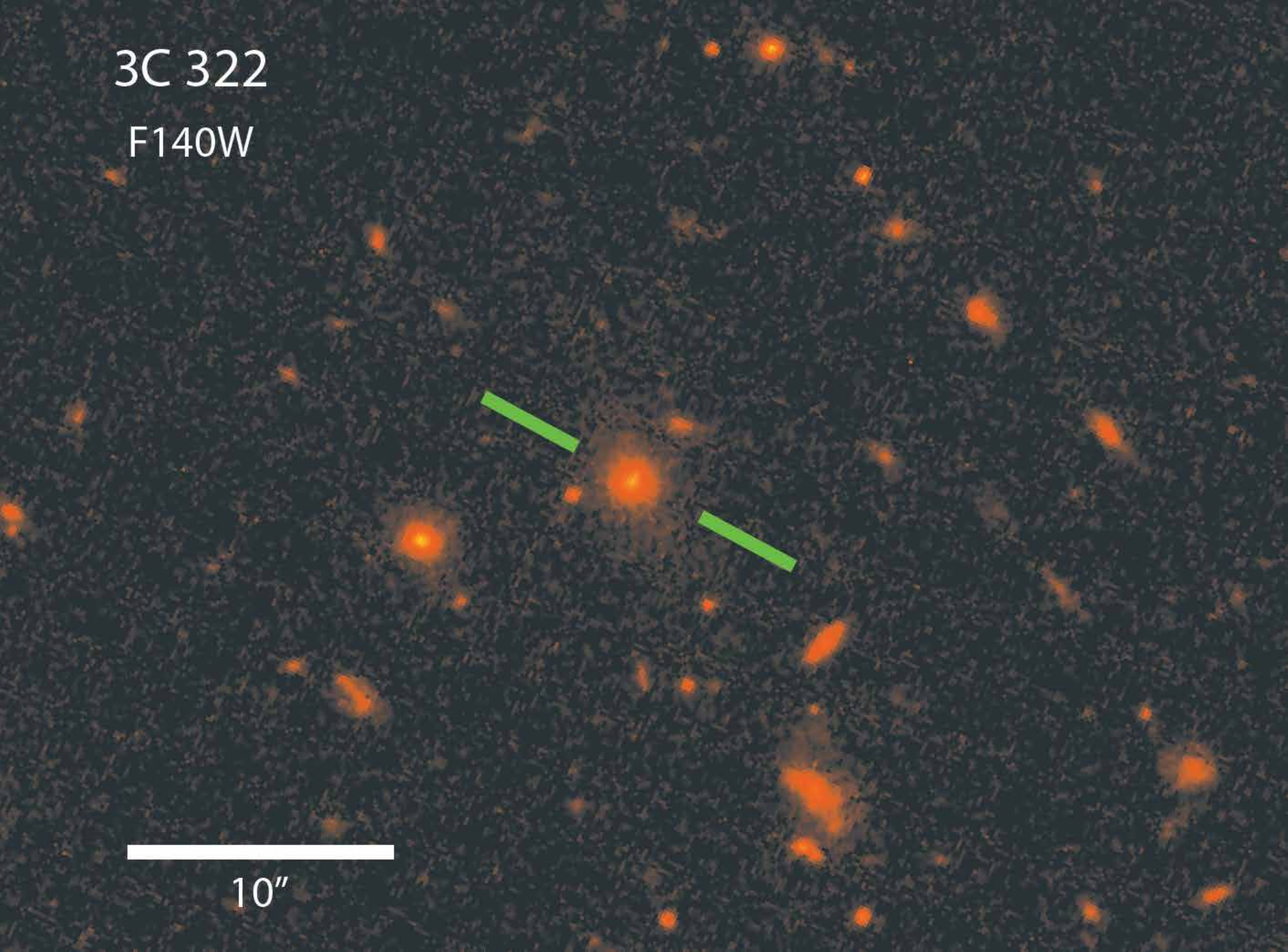}
\caption{Central $35\arcsec \times 50\arcsec$ of the IR image of RG 3C 322. \redpen{The image has been rotated so 
that North is up and East to the left. Green lines are placed on 
either side of the target to help identify its location.}}
\label{fig:3c322_ir}
\end{minipage}
\hspace{0.5cm}
\begin{minipage}[b]{0.45\linewidth}
\centering
\includegraphics[width=\textwidth]{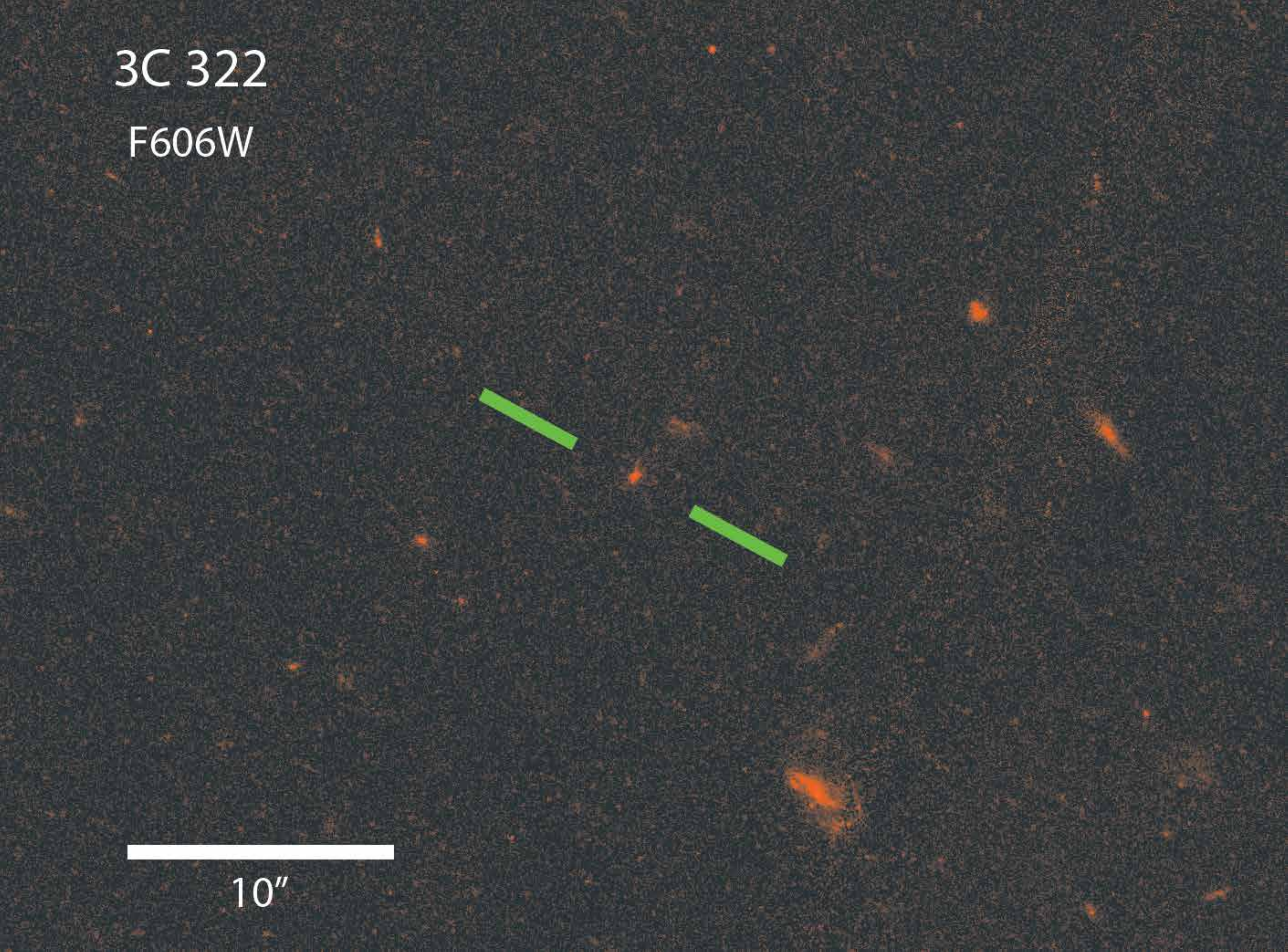}
\caption{Central $35\arcsec \times 50\arcsec$ of the UVIS image of RG 3C 322. \redpen{The image has been rotated
so that North is up and East to the left. Green lines are placed on 
either side of the target to help identify its location.}}
\label{fig:3c322_uv}
\end{minipage}
\end{figure*}

\begin{figure*}[htp]
\begin{minipage}[b]{0.45\linewidth}
\centering
\includegraphics[width=\textwidth]{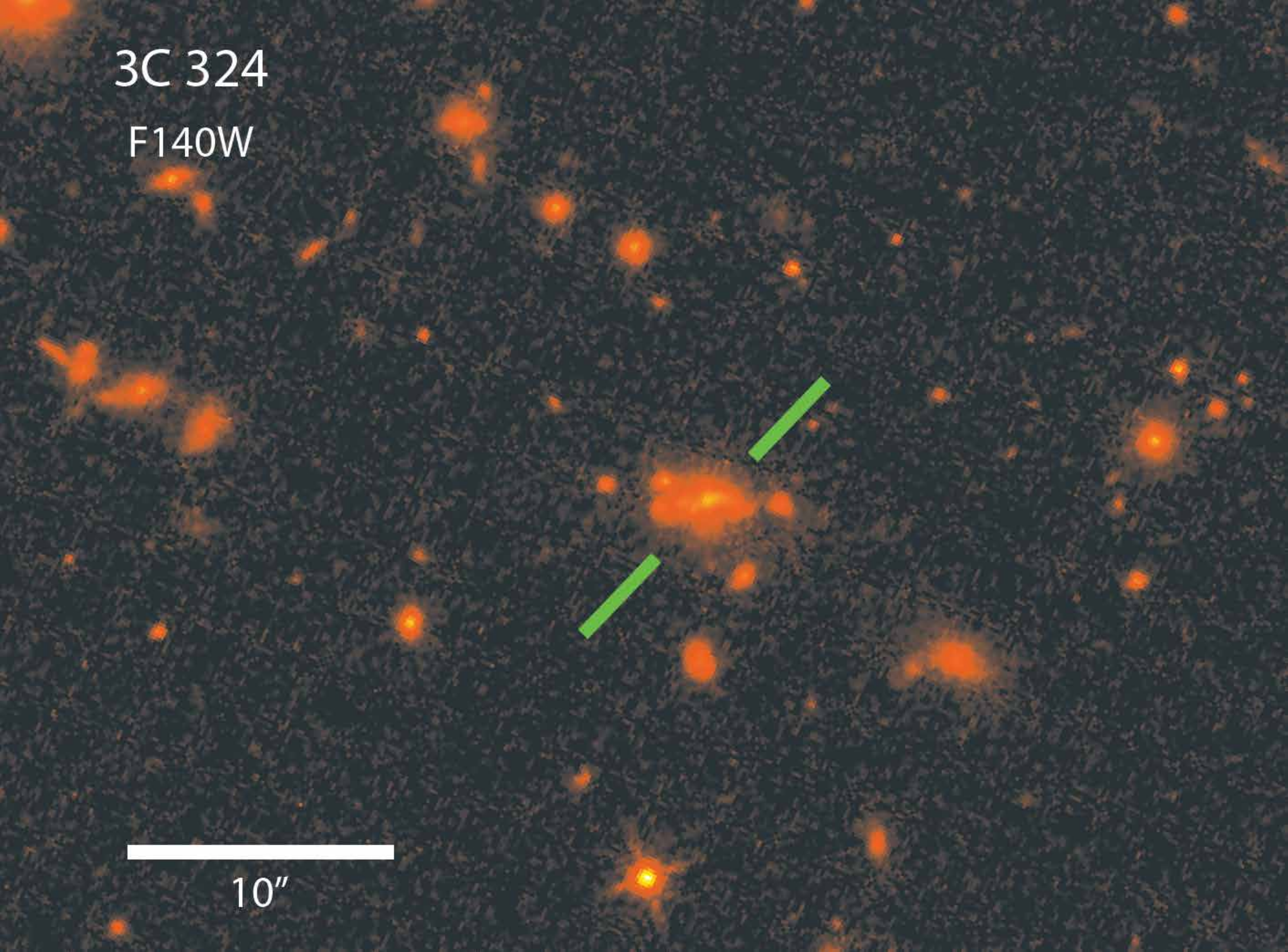}
\caption{Central $35\arcsec \times 50\arcsec$ of the IR image of RG 3C 324. \redpen{The image has been rotated so 
that North is up and East to the left. Green lines are placed on 
either side of the target to help identify its location.}}
\label{fig:3c324_ir}
\end{minipage}
\hspace{0.5cm}
\begin{minipage}[b]{0.45\linewidth}
\centering
\includegraphics[width=\textwidth]{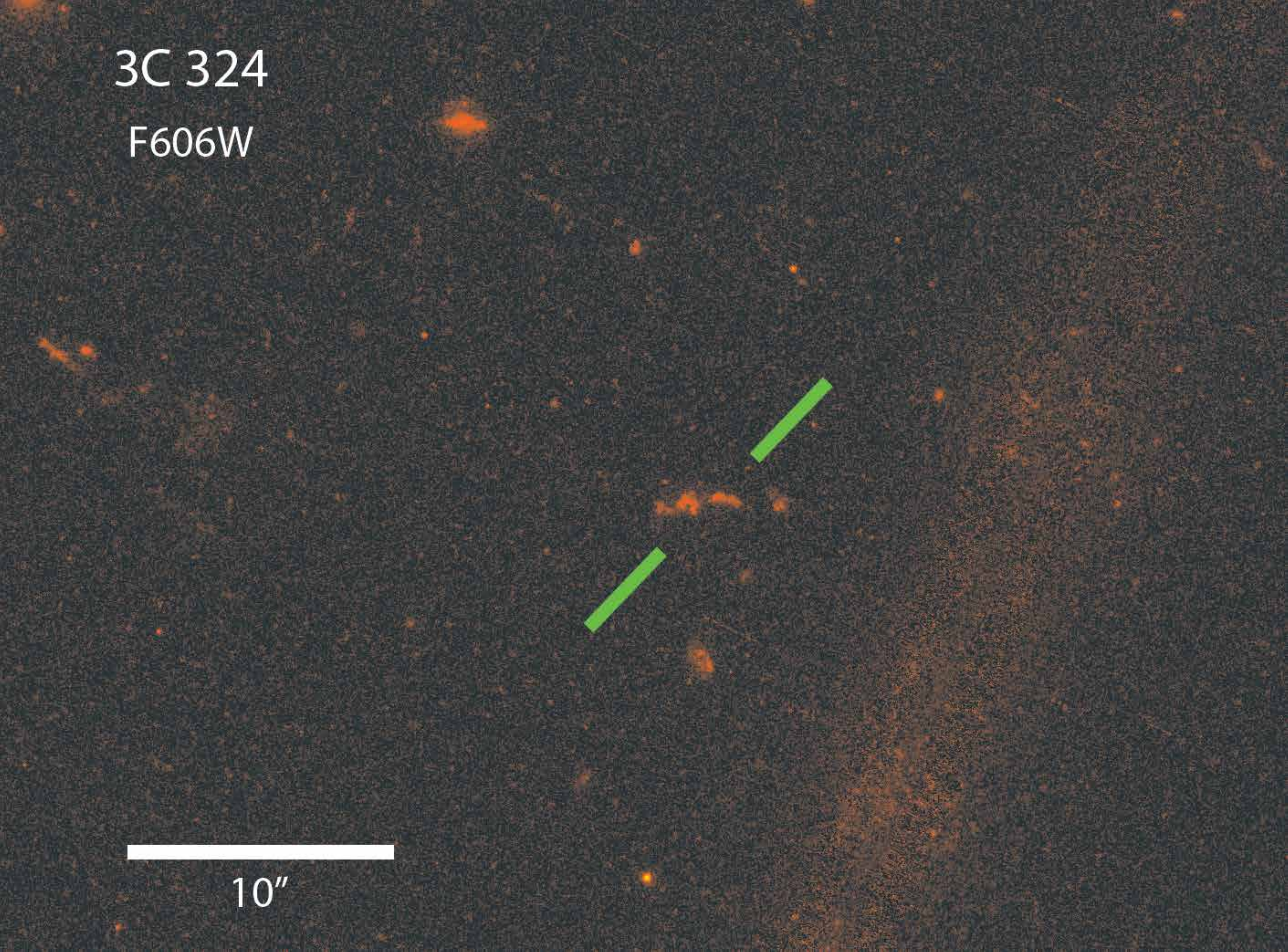}
\caption{Central $35\arcsec \times 50\arcsec$ of the UVIS image of RG 3C 324. \redpen{The image has been rotated
so that North is up and East to the left. Green lines are placed on 
either side of the target to help identify its location.}}
\label{fig:3c324_uv}
\end{minipage}
\end{figure*}

\begin{figure*}[htp]
\begin{minipage}[b]{0.45\linewidth}
\centering
\includegraphics[width=\textwidth]{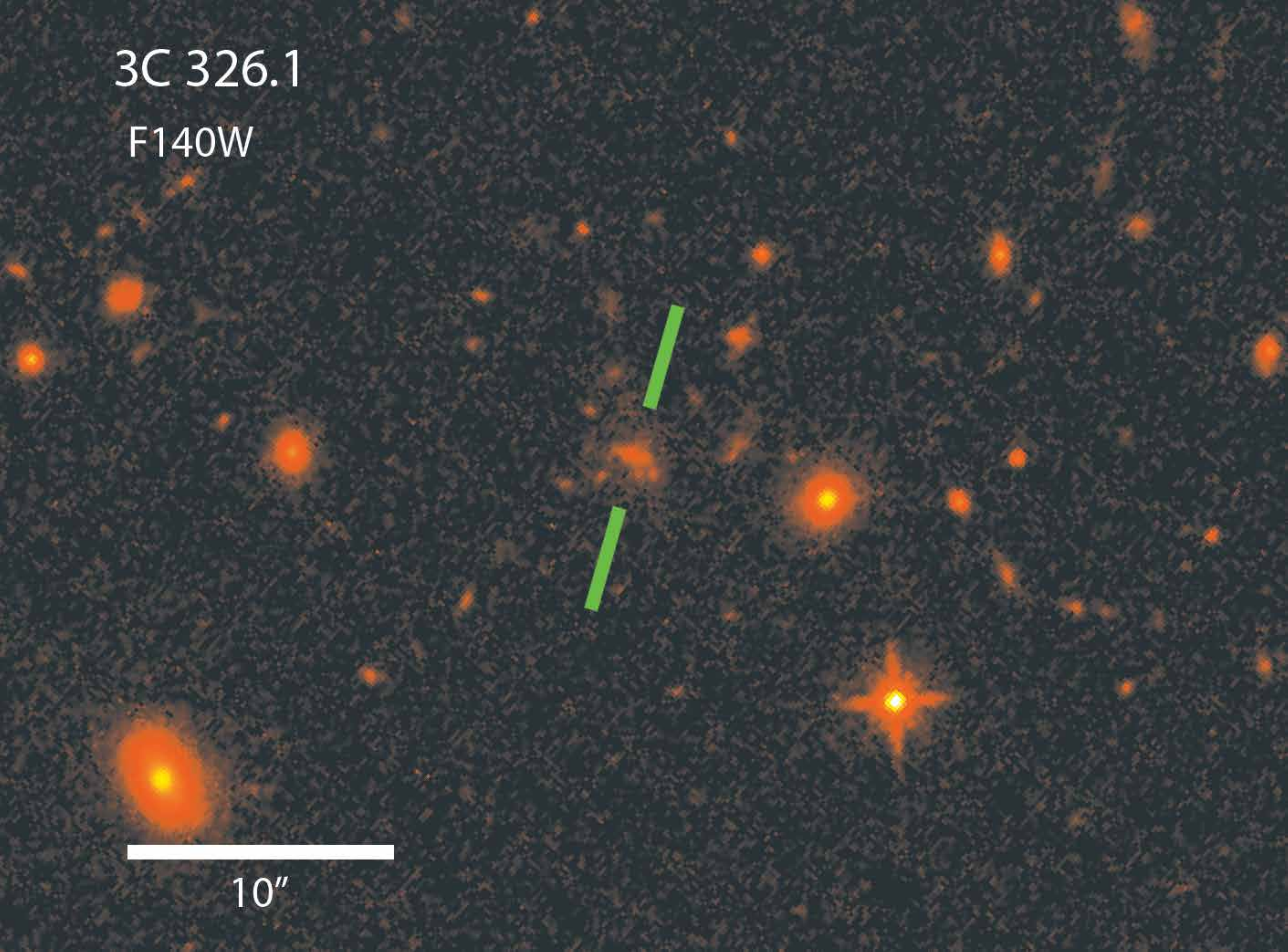}
\caption{Central $35\arcsec \times 50\arcsec$ of the IR image of RG 3C 326.1. \redpen{The image has been rotated so 
that North is up and East to the left. Green lines are placed on 
either side of the target to help identify its location.}}
\label{fig:3c326p1_ir}
\end{minipage}
\hspace{0.5cm}
\begin{minipage}[b]{0.45\linewidth}
\centering
\includegraphics[width=\textwidth]{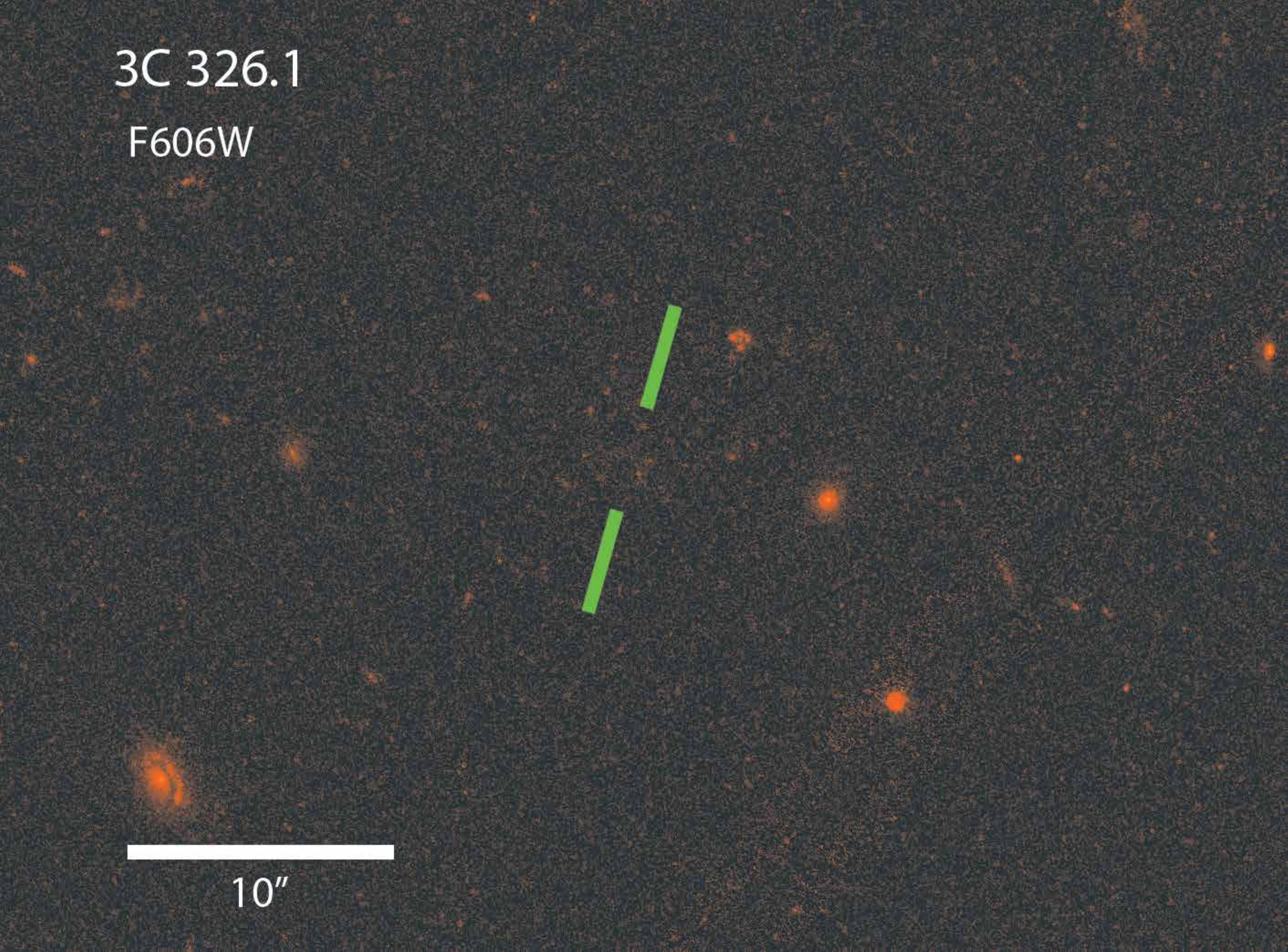}
\caption{Central $35\arcsec \times 50\arcsec$ of the UVIS image of RG 3C 326.1. \redpen{The image has been rotated
so that North is up and East to the left. Green lines are placed on 
either side of the target to help identify its location.}}
\label{fig:3c326p1_uv}
\end{minipage}
\end{figure*}

\begin{figure*}[htp]
\begin{minipage}[b]{0.45\linewidth}
\centering
\includegraphics[width=\textwidth]{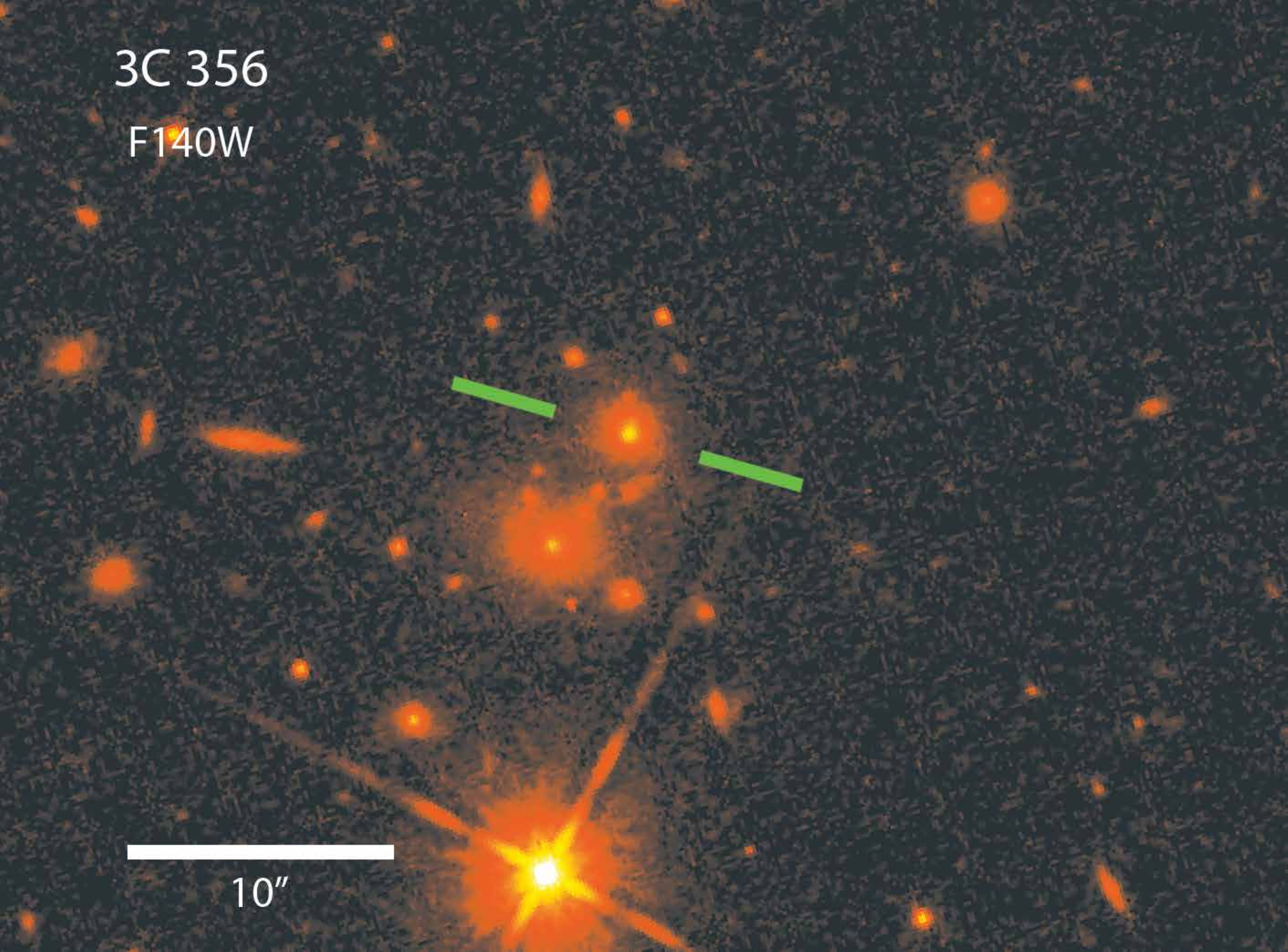}
\caption{Central $35\arcsec \times 50\arcsec$ of the IR image of RG 3C 356. \redpen{The image has been rotated so 
that North is up and East to the left. Green lines are placed on 
either side of the target to help identify its location.}}
\label{fig:3c356_ir}
\end{minipage}
\hspace{0.5cm}
\begin{minipage}[b]{0.45\linewidth}
\centering
\includegraphics[width=\textwidth]{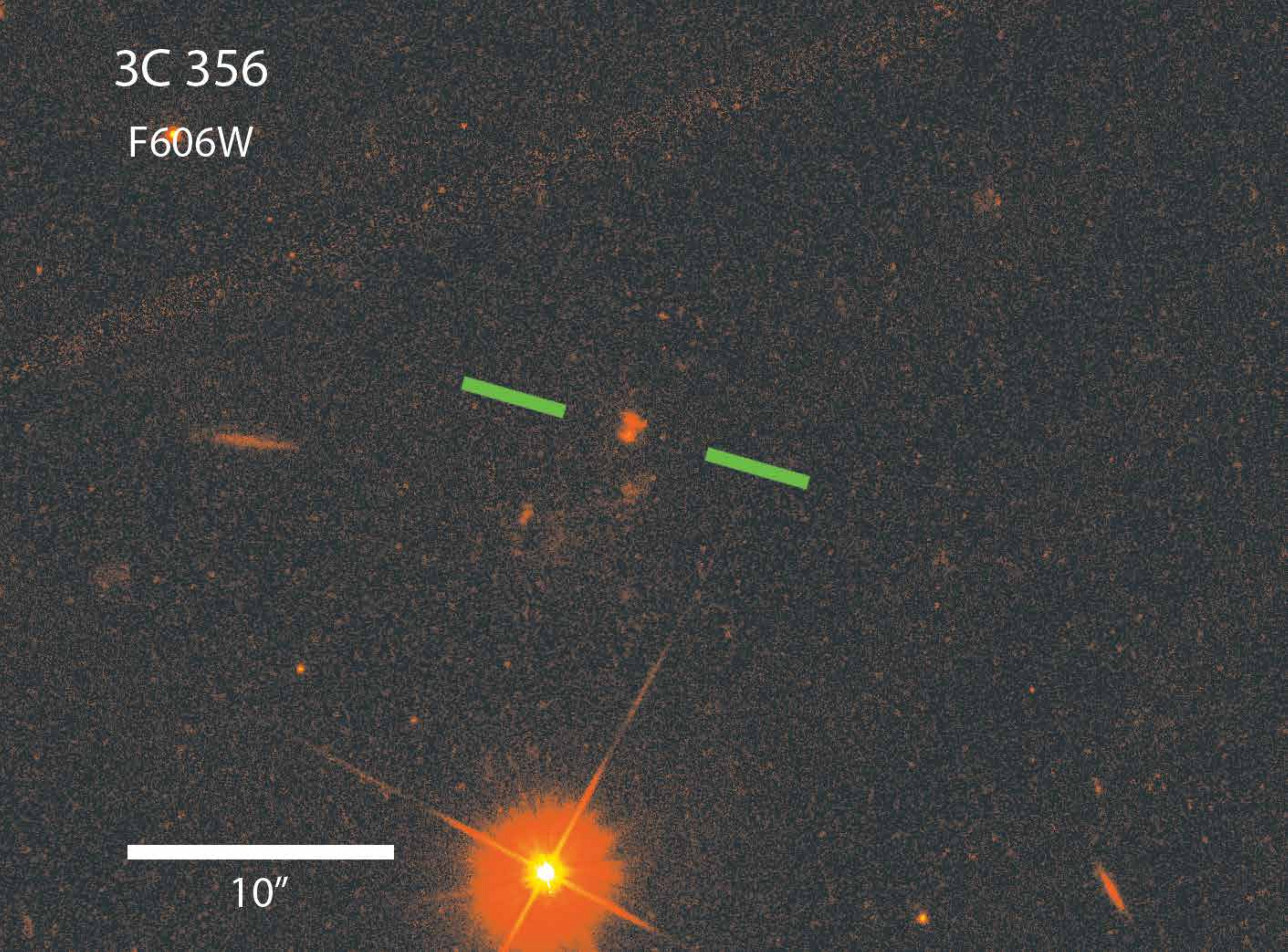}
\caption{Central $35\arcsec \times 50\arcsec$ of the UVIS image of RG 3C 356. \redpen{The image has been rotated
so that North is up and East to the left. Green lines are placed on 
either side of the target to help identify its location.}}
\label{fig:3c356_uv}
\end{minipage}
\end{figure*}

\begin{figure*}[htp]
\begin{minipage}[b]{0.45\linewidth}
\centering
\includegraphics[width=\textwidth]{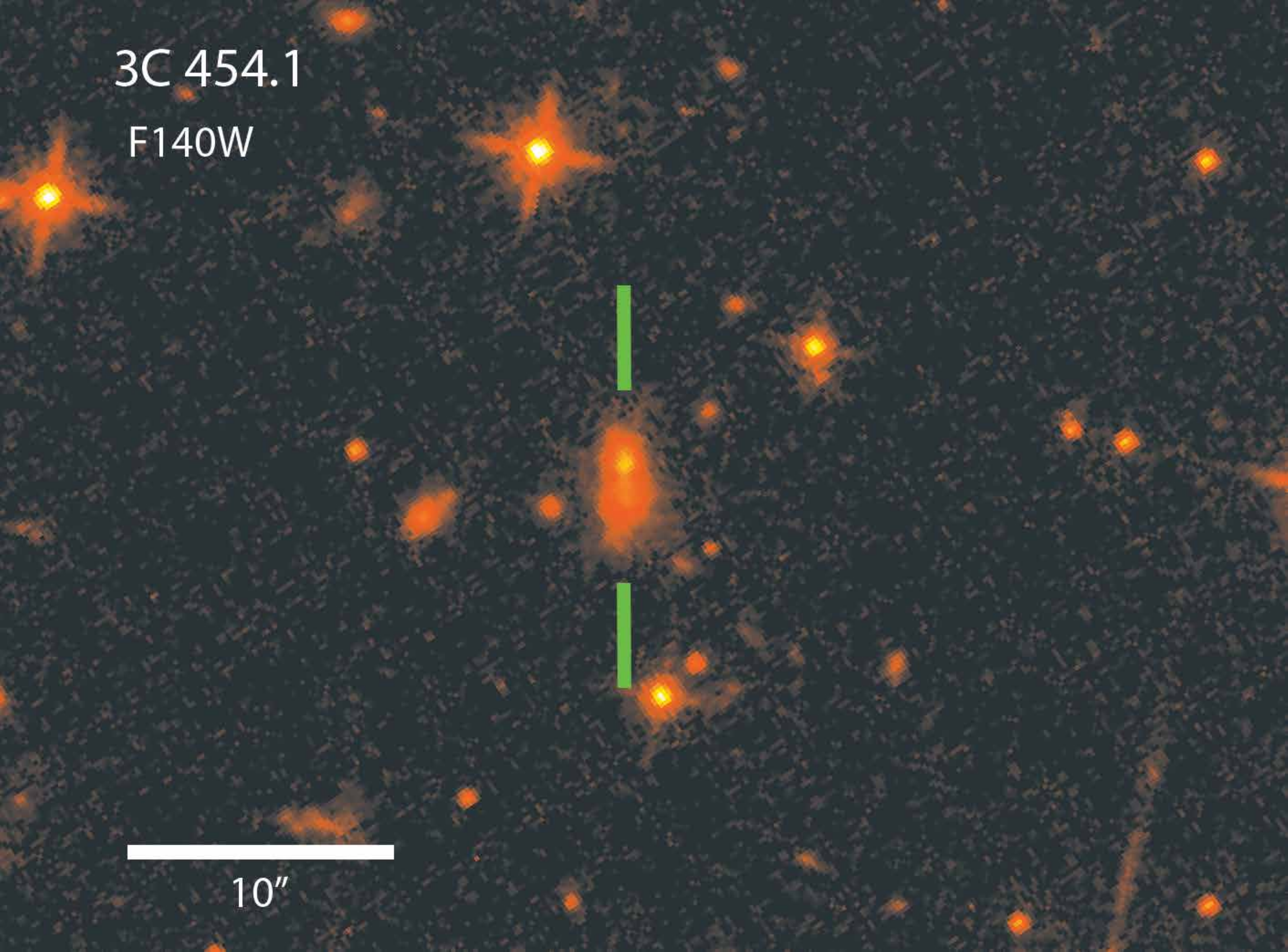}
\caption{Central $35\arcsec \times 50\arcsec$ of the IR image of RG 3C 454.1. \redpen{The image has been rotated so 
that North is up and East to the left. Green lines are placed on 
either side of the target to help identify its location.}}
\label{fig:3c454p1_ir}
\end{minipage}
\hspace{0.5cm}
\begin{minipage}[b]{0.45\linewidth}
\centering
\includegraphics[width=\textwidth]{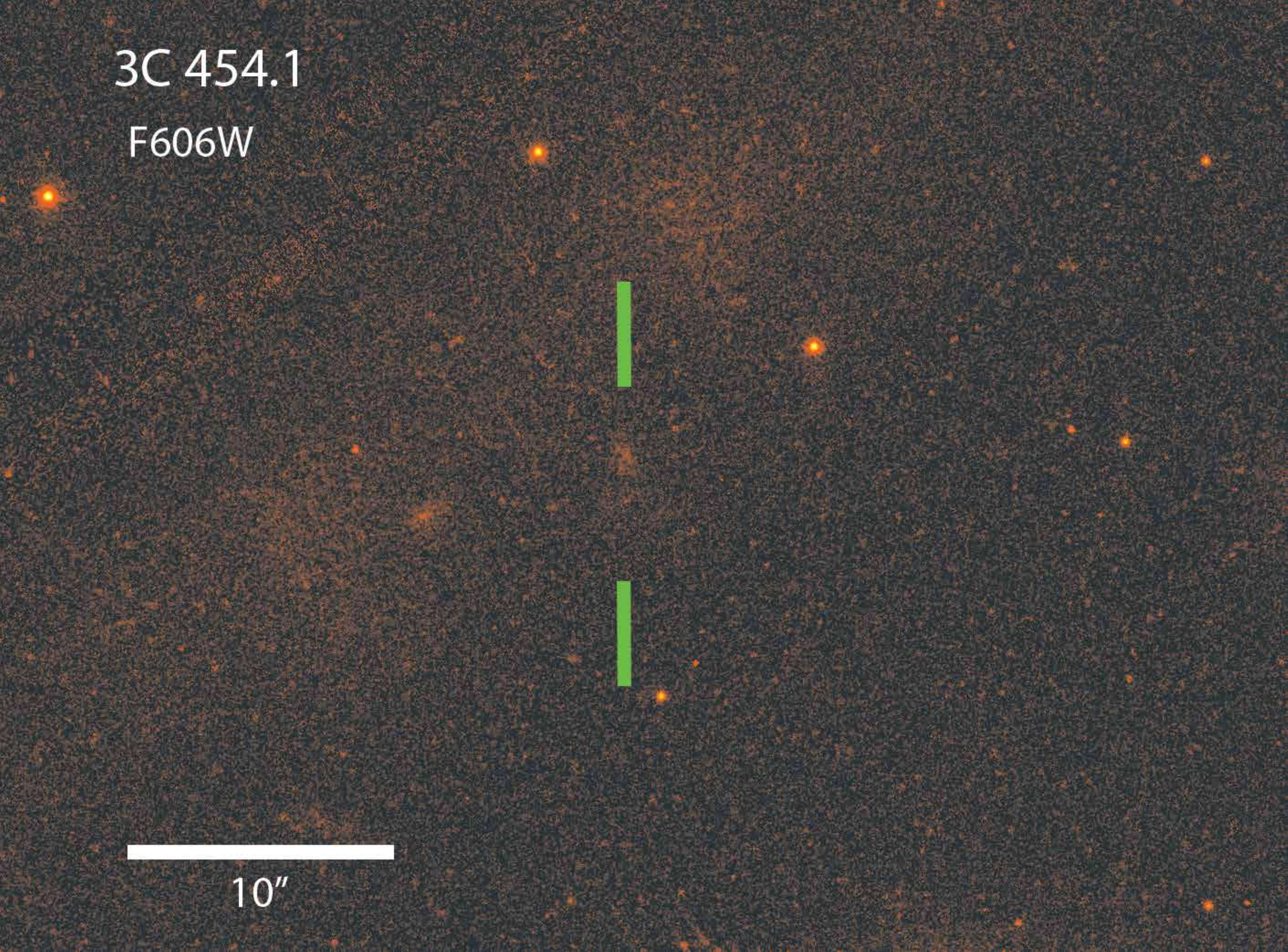}
\caption{Central $35\arcsec \times 50\arcsec$ of the UVIS image of RG 3C 454.1. \redpen{The image has been rotated
so that North is up and East to the left. Green lines are placed on 
either side of the target to help identify its location.}}
\label{fig:3c454p1_uv}
\end{minipage}
\end{figure*}

\begin{figure*}[ht]
\begin{minipage}[b]{0.45\linewidth}
\centering
\includegraphics[width=\textwidth]{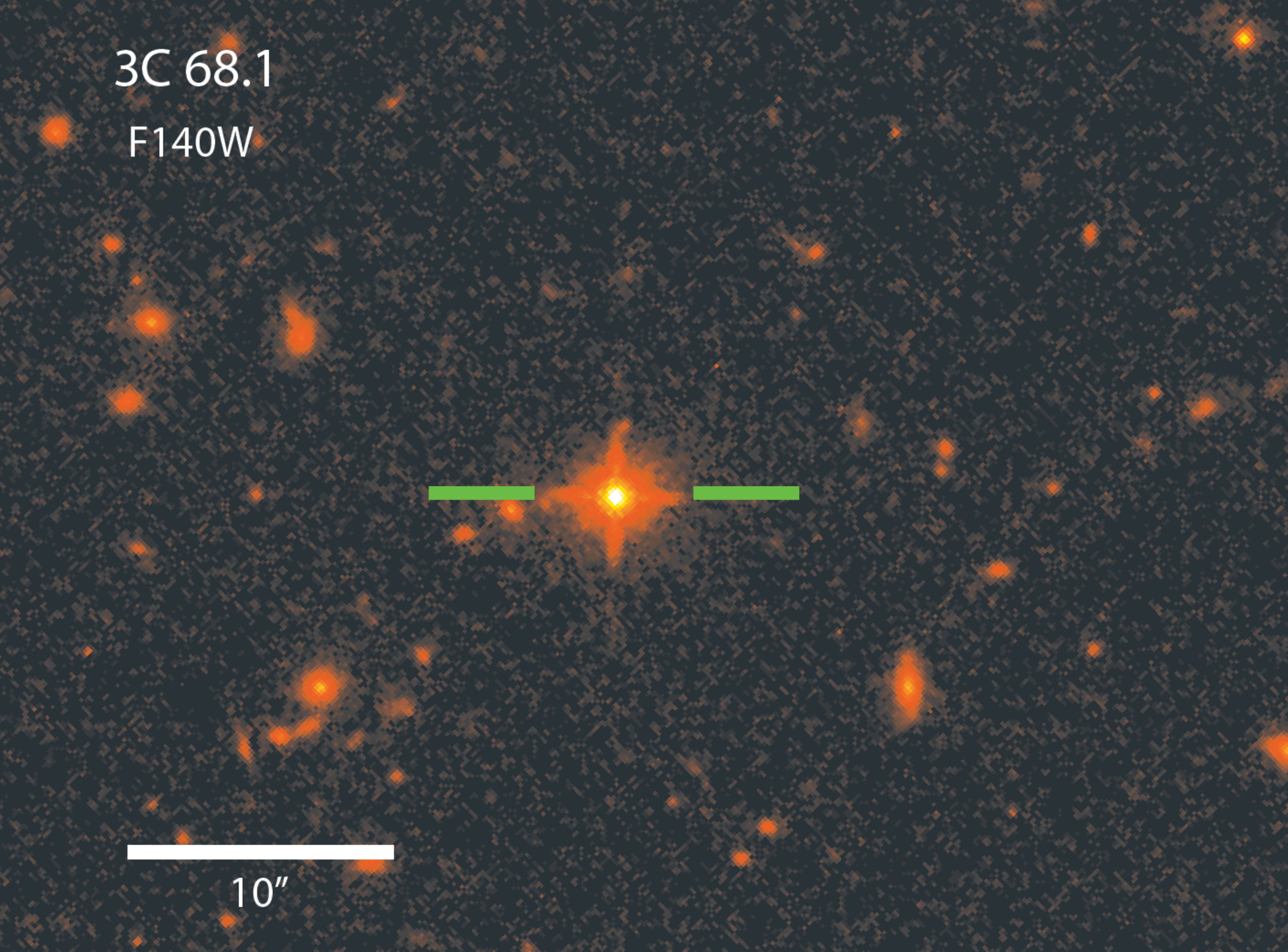}
\caption{Central $35\arcsec \times 50\arcsec$ of the IR image of QSO 3C 68.1. \redpen{The image has been rotated so 
that North is up and East to the left. Green lines are placed on 
either side of the target to help identify its location.}}
\label{fig:3c68p1_ir}
\end{minipage}
\hspace{0.5cm}
\begin{minipage}[b]{0.45\linewidth}
\centering
\includegraphics[width=\textwidth]{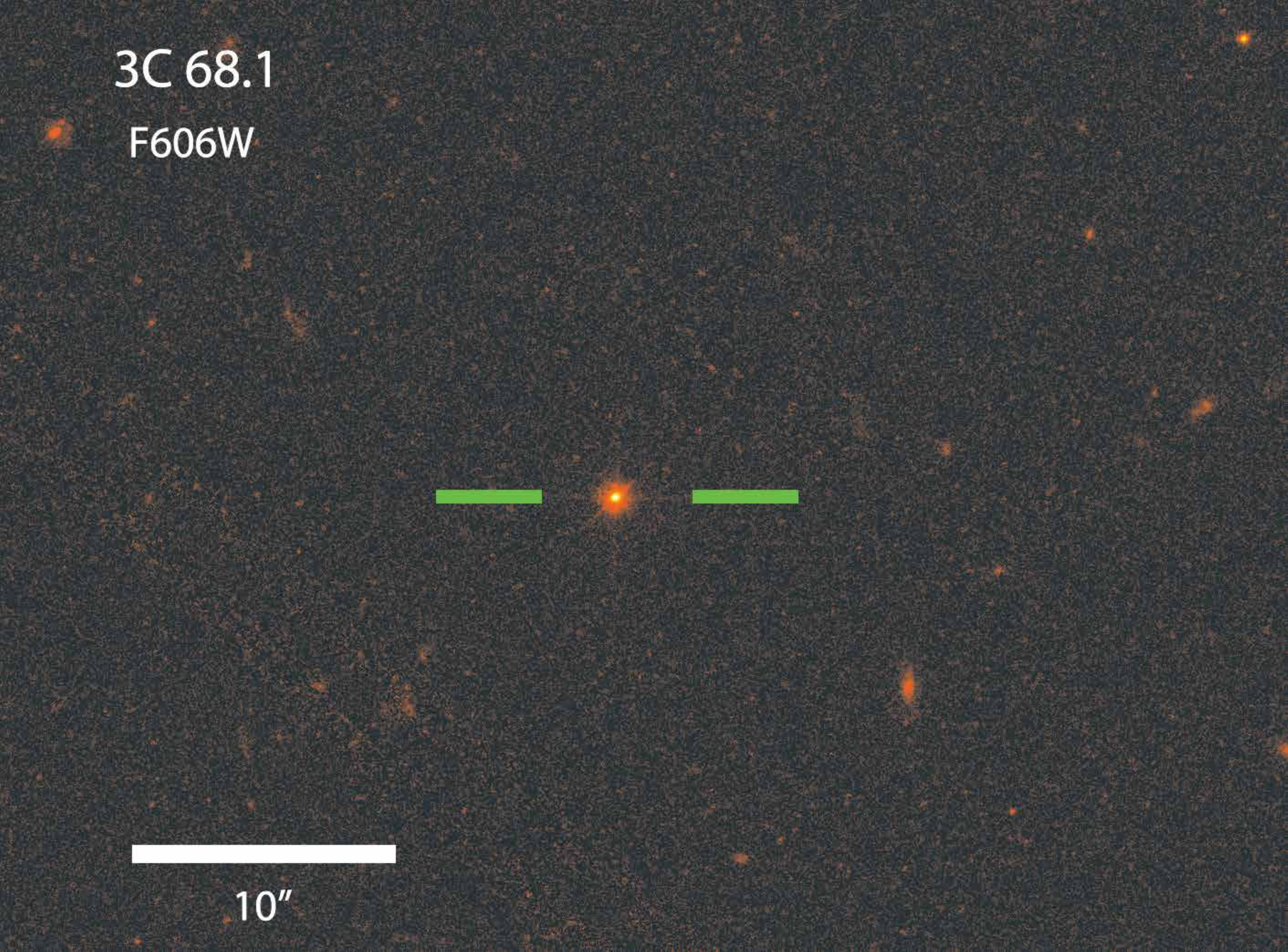}
\caption{Central $35\arcsec \times 50\arcsec$ of the UVIS image of QSO 3C 68.1. \redpen{The image has been rotated
so that North is up and East to the left. Green lines are placed on 
either side of the target to help identify its location.}}
\label{fig:3c68p1_uv}
\end{minipage}
\end{figure*}

\begin{figure*}[ht]
\begin{minipage}[b]{0.45\linewidth}
\centering
\includegraphics[width=\textwidth]{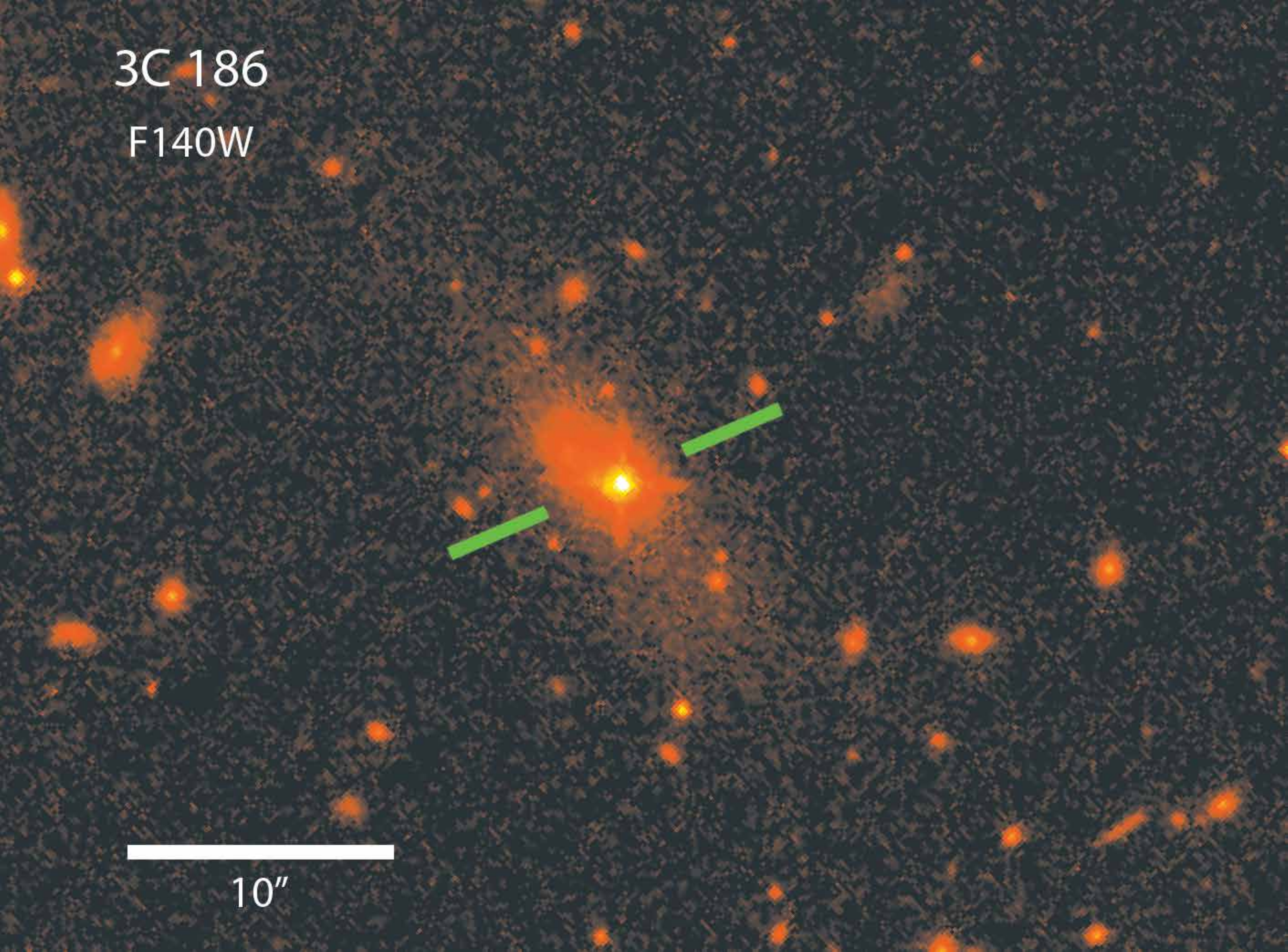}
\caption{Central $35\arcsec \times 50\arcsec$ of the IR image of QSO 3C 186. \redpen{The image has been rotated so 
that North is up and East to the left. Green lines are placed on 
either side of the target to help identify its location.}}
\label{fig:3c186_ir}
\end{minipage}
\hspace{0.5cm}
\begin{minipage}[b]{0.45\linewidth}
\centering
\includegraphics[width=\textwidth]{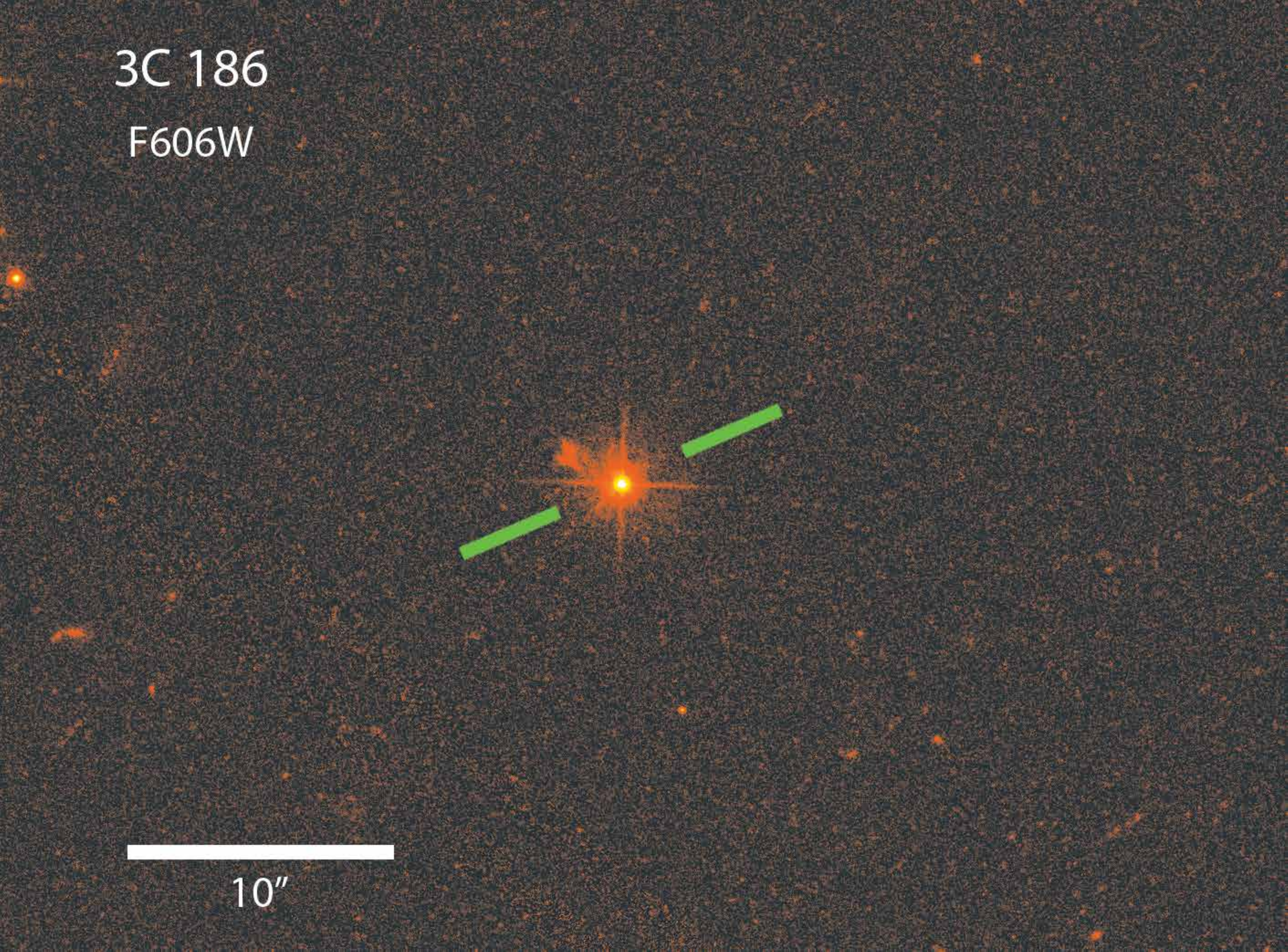}
\caption{Central $35\arcsec \times 50\arcsec$ of the UVIS image of QSO 3C 186. \redpen{The image has been rotated
so that North is up and East to the left. Green lines are placed on 
either side of the target to help identify its location.}}
\label{fig:3c186_uv}
\end{minipage}
\end{figure*}

\begin{figure*}[htp]
\begin{minipage}[b]{0.45\linewidth}
\centering
\includegraphics[width=\textwidth]{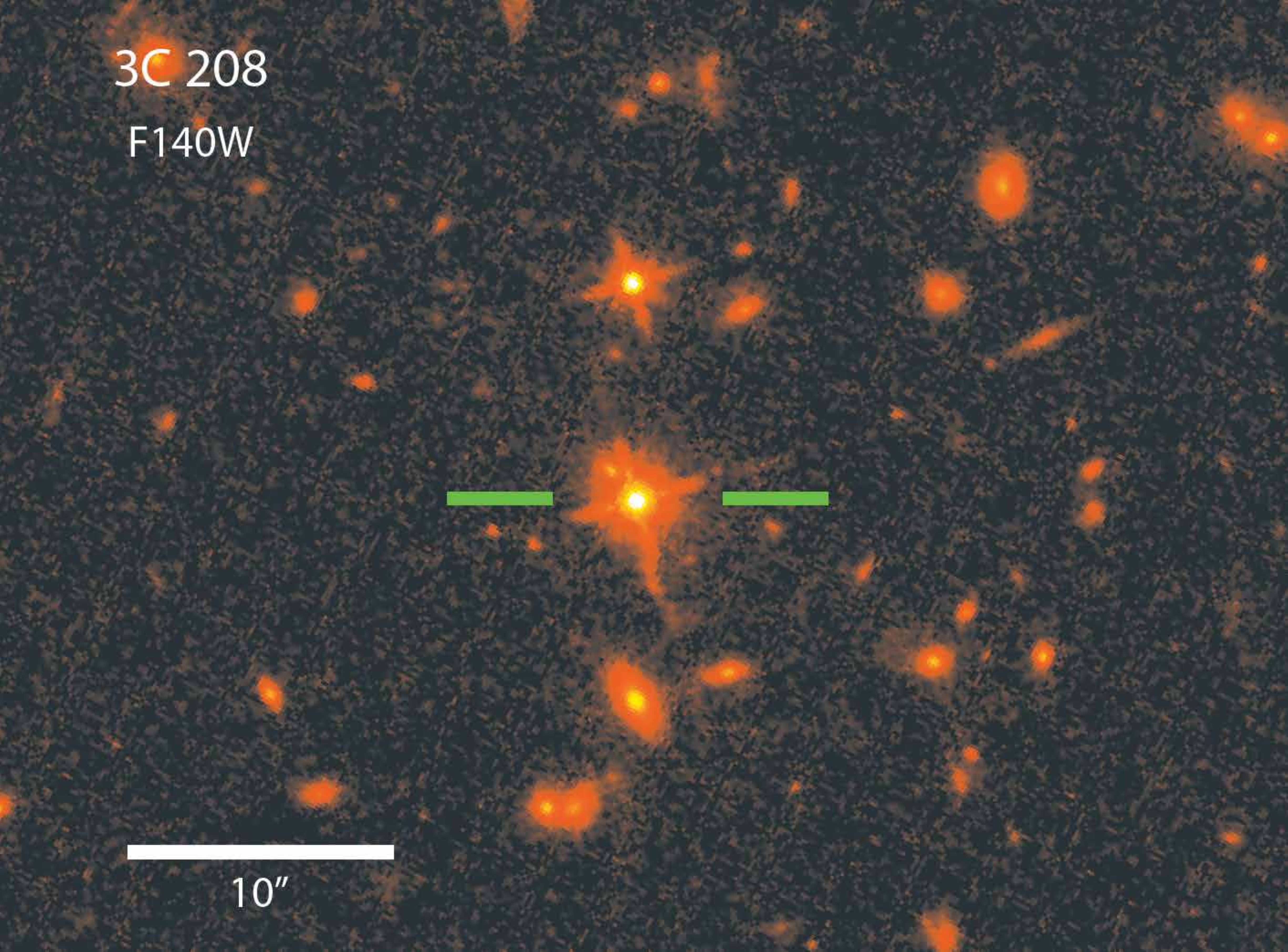}
\caption{Central $35\arcsec \times 50\arcsec$ of the IR image of QSO 3C 208. \redpen{The image has been rotated so 
that North is up and East to the left. Green lines are placed on 
either side of the target to help identify its location.}}
\label{fig:3c208_ir}
\end{minipage}
\hspace{0.5cm}
\begin{minipage}[b]{0.45\linewidth}
\centering
\includegraphics[width=\textwidth]{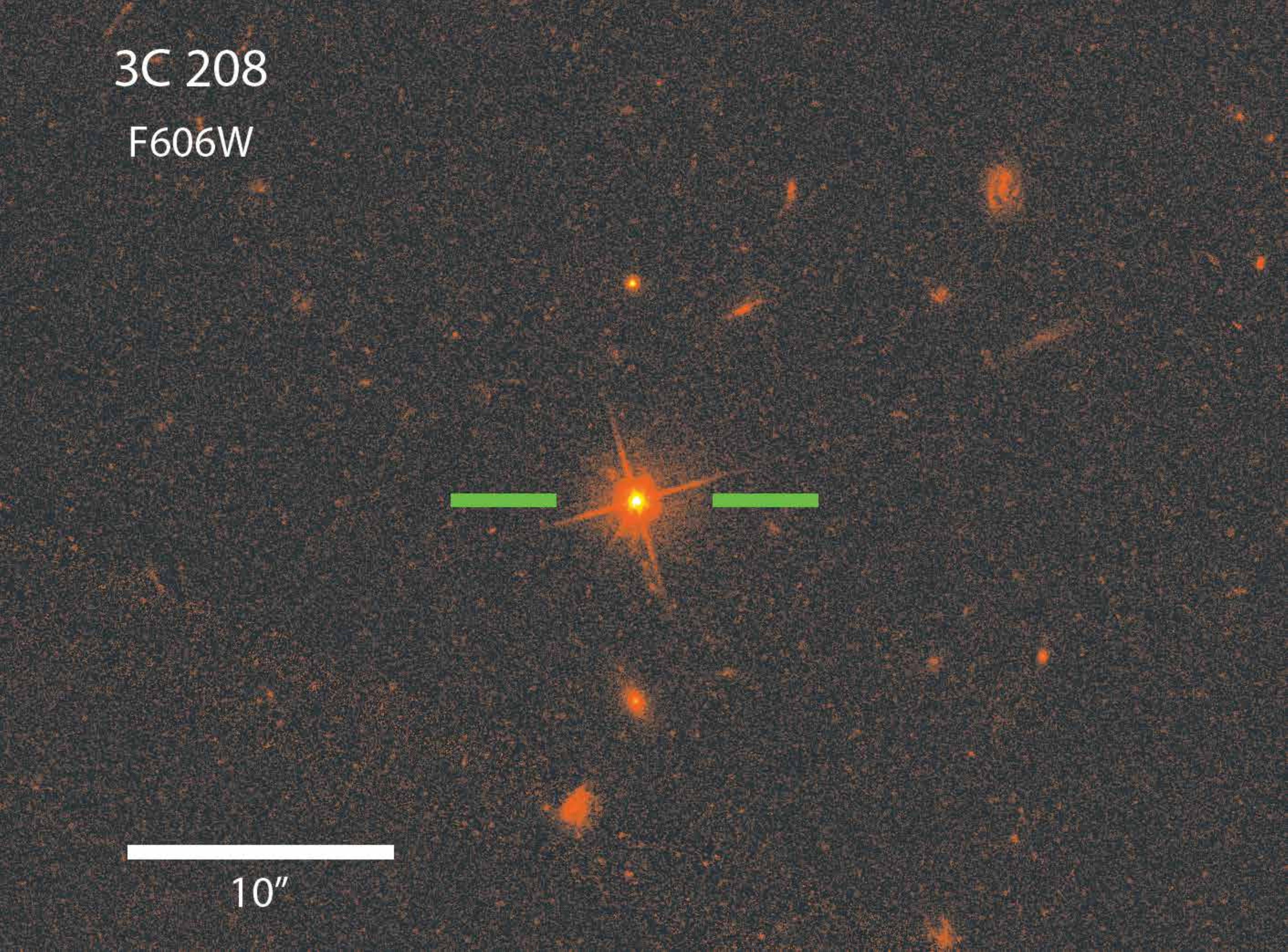}
\caption{Central $35\arcsec \times 50\arcsec$ of the UVIS image of QSO 3C 208. \redpen{The image has been rotated
so that North is up and East to the left. Green lines are placed on 
either side of the target to help identify its location.}}
\label{fig:3c208_uv}
\end{minipage}
\end{figure*}

\begin{figure*}[htp]
\begin{minipage}[b]{0.45\linewidth}
\centering
\includegraphics[width=\textwidth]{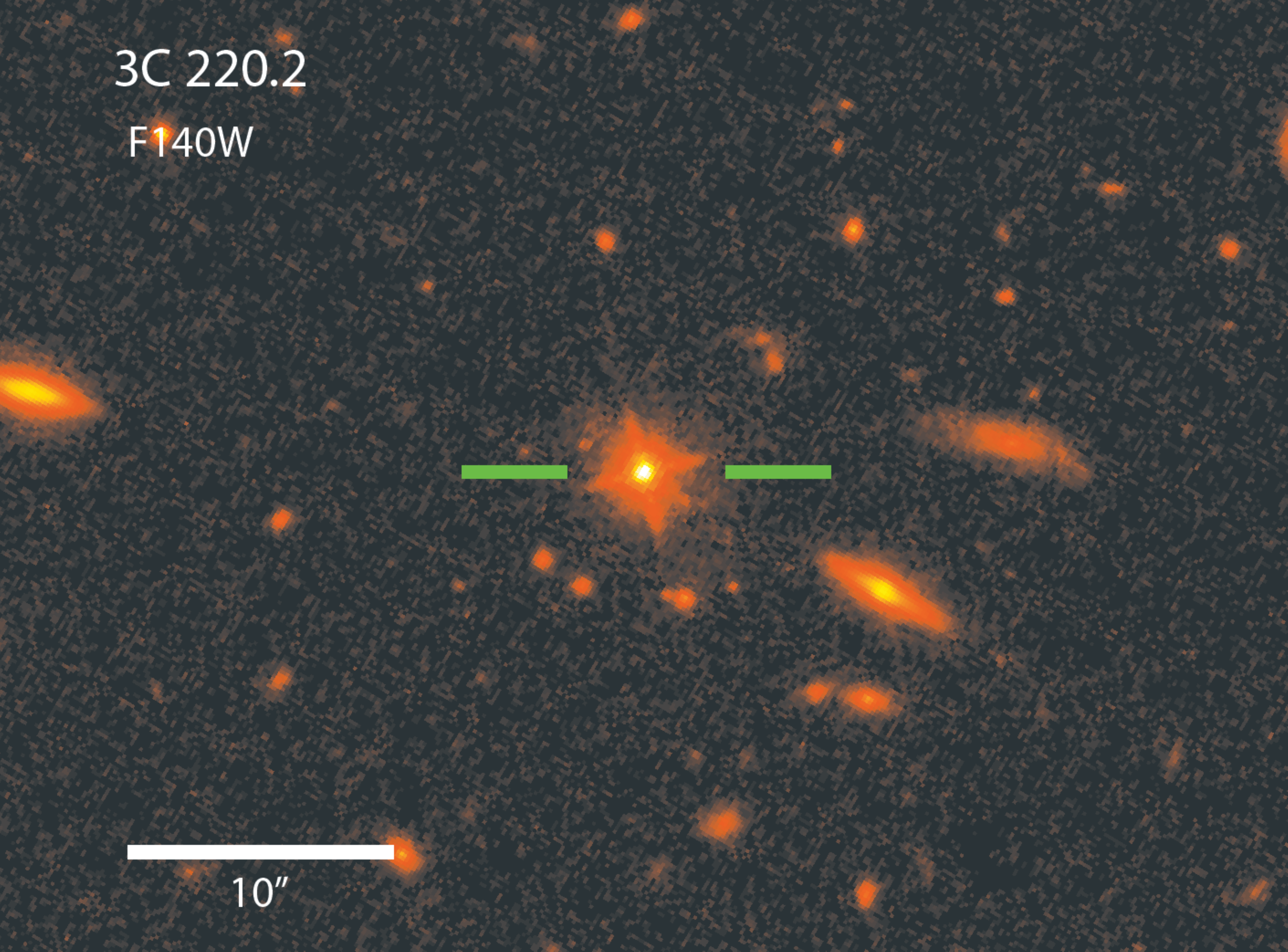}
\caption{Central $35\arcsec \times 50\arcsec$ of the IR image of QSO 3C 220.2. \redpen{The image has been rotated so 
that North is up and East to the left. Green lines are placed on 
either side of the target to help identify its location.}}
\label{fig:3c220p2_ir}
\end{minipage}
\hspace{0.5cm}
\begin{minipage}[b]{0.45\linewidth}
\centering
\includegraphics[width=\textwidth]{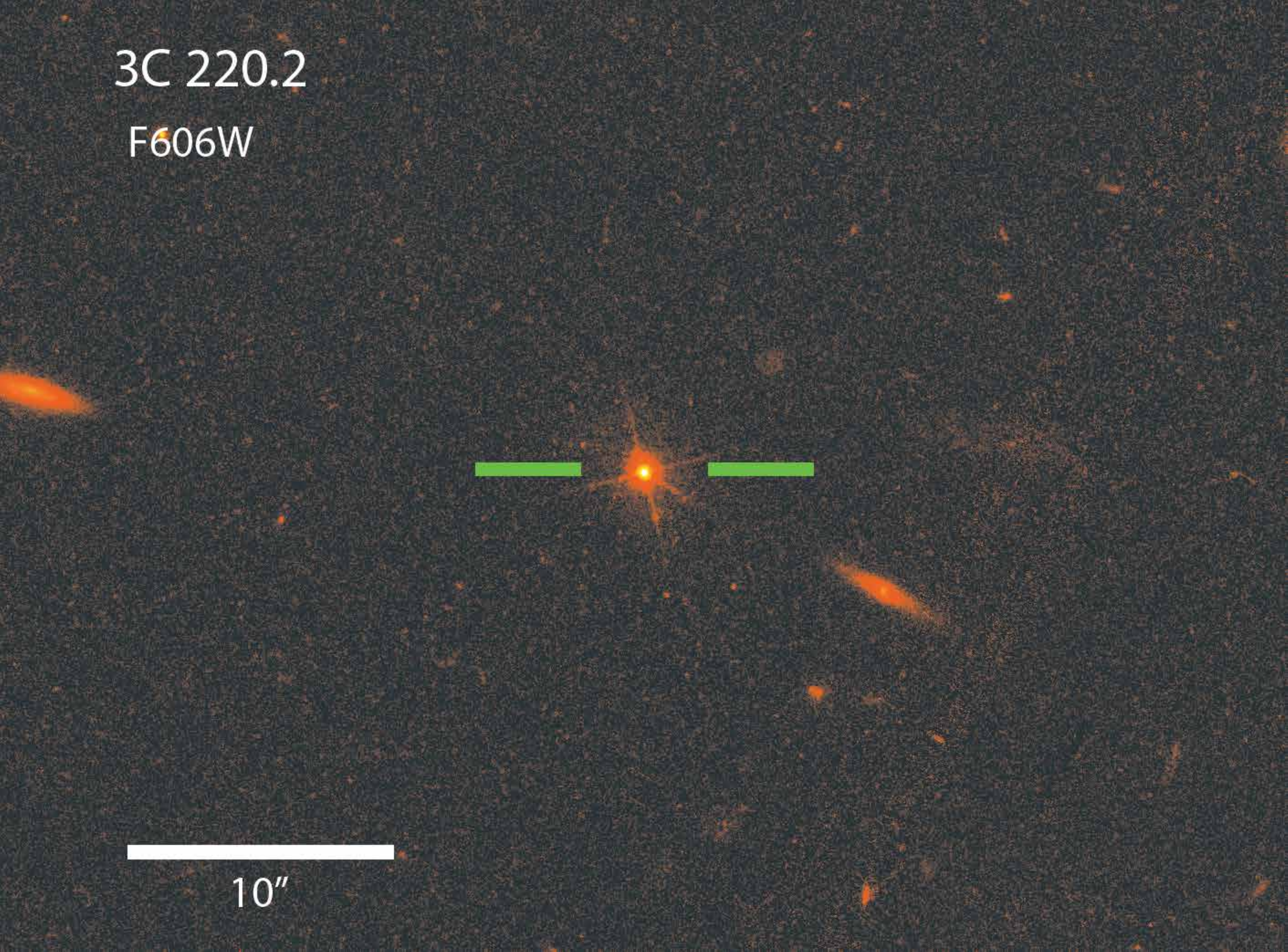}
\caption{Central $35\arcsec \times 50\arcsec$ of the UVIS image of QSO 3C 220.2. \redpen{The image has been rotated
so that North is up and East to the left. Green lines are placed on 
either side of the target to help identify its location.}}
\label{fig:3c220p2_uv}
\end{minipage}
\end{figure*}

\begin{figure*}[htp]
\begin{minipage}[b]{0.45\linewidth}
\centering
\includegraphics[width=\textwidth]{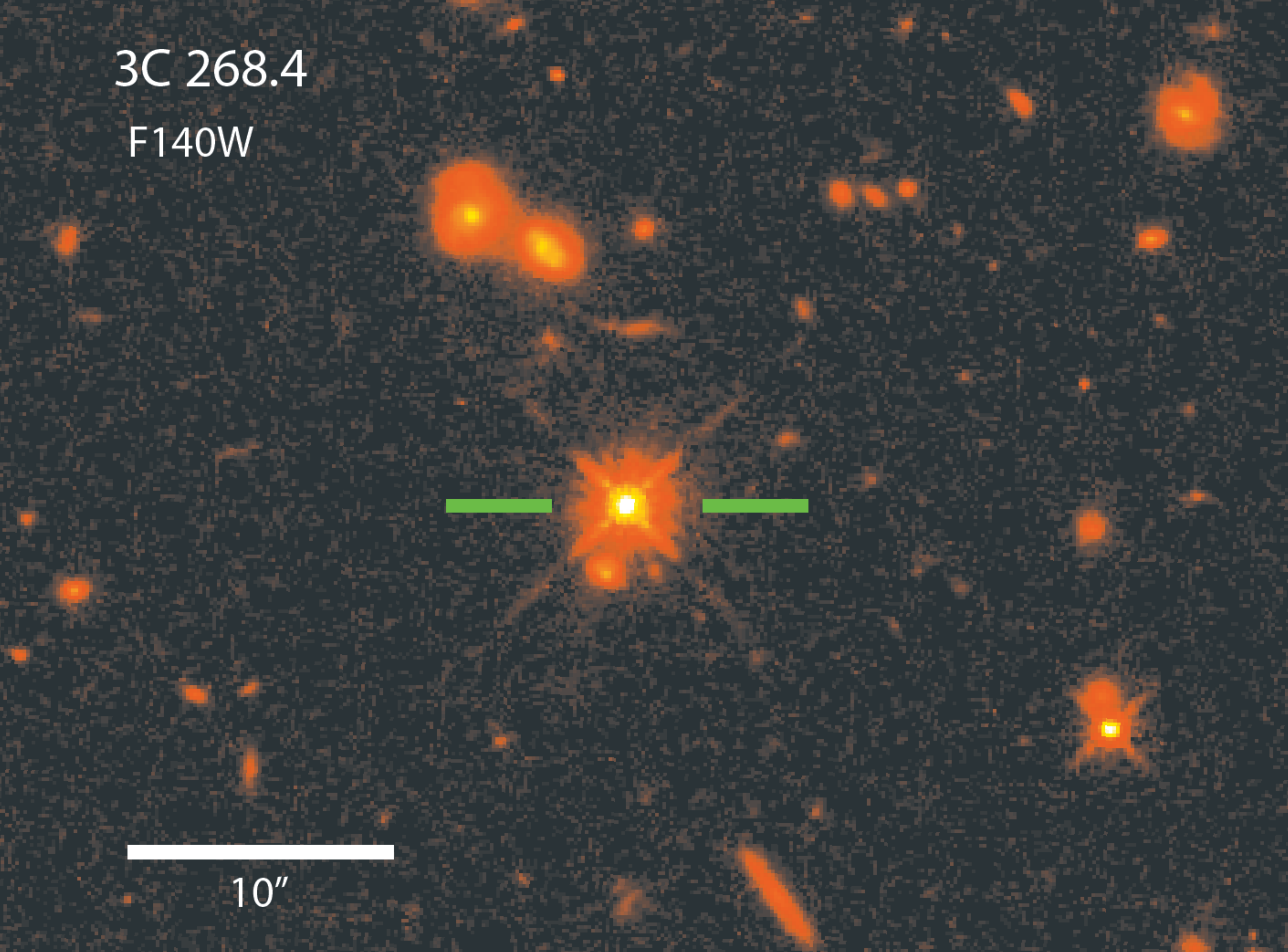}
\caption{Central $35\arcsec \times 50\arcsec$ of the IR image of QSO 3C 268.4. \redpen{The image has been rotated so 
that North is up and East to the left. Green lines are placed on 
either side of the target to help identify its location.}}
\label{fig:3c268p4_ir}
\end{minipage}
\hspace{0.5cm}
\begin{minipage}[b]{0.45\linewidth}
\centering
\includegraphics[width=\textwidth]{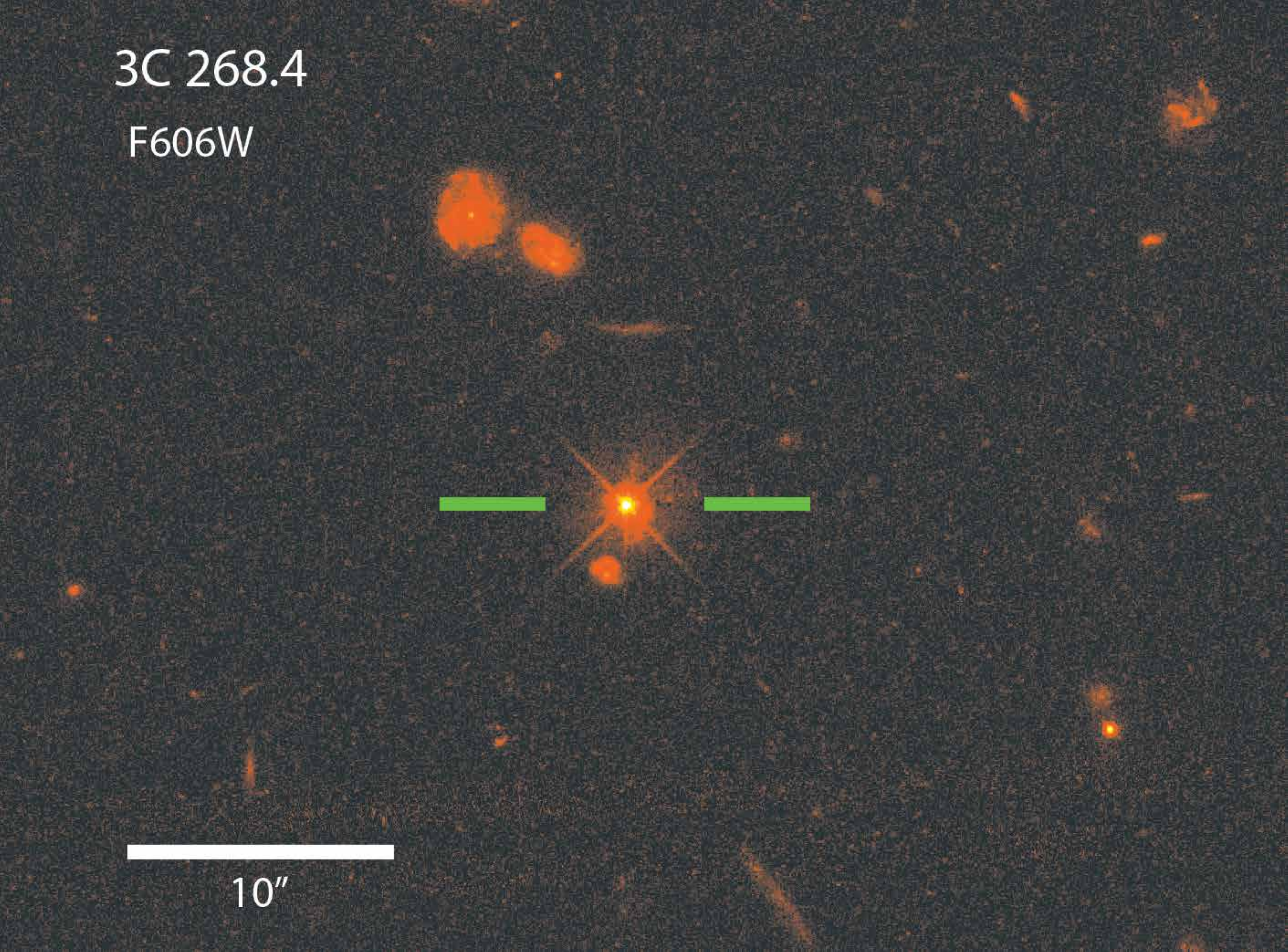}
\caption{Central $35\arcsec \times 50\arcsec$ of the UVIS image of QSO 3C 268.4. \redpen{The image has been rotated
so that North is up and East to the left. Green lines are placed on 
either side of the target to help identify its location.}}
\label{fig:3c268p4_uv}
\end{minipage}
\end{figure*}

\begin{figure*}[htp]
\begin{minipage}[b]{0.45\linewidth}
\centering
\includegraphics[width=\textwidth]{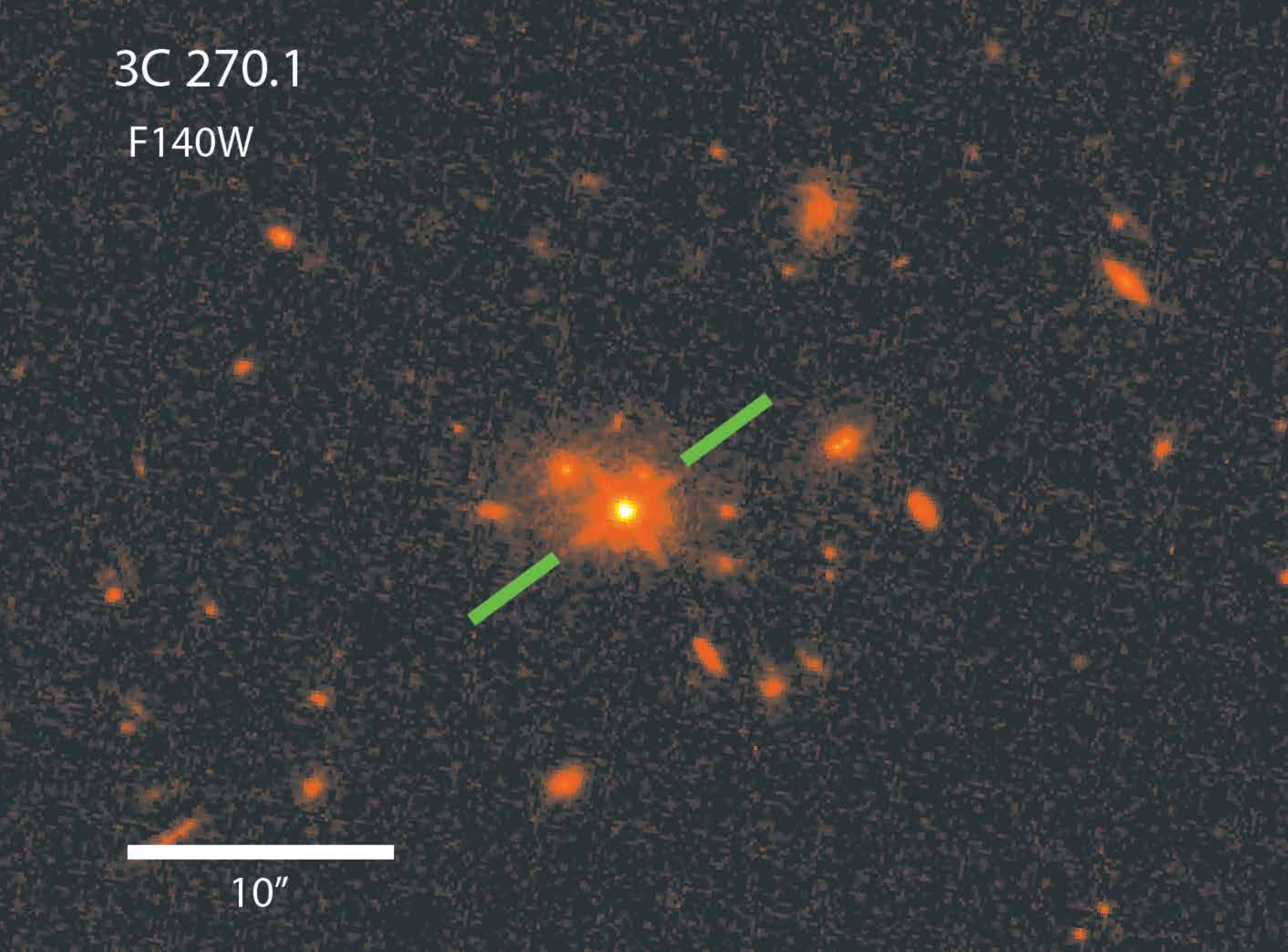}
\caption{Central $35\arcsec \times 50\arcsec$ of the IR image of QSO 3C 270.1. \redpen{The image has been rotated so 
that North is up and East to the left. Green lines are placed on 
either side of the target to help identify its location.}}
\label{fig:3c270_ir}
\end{minipage}
\hspace{0.5cm}
\begin{minipage}[b]{0.45\linewidth}
\centering
\includegraphics[width=\textwidth]{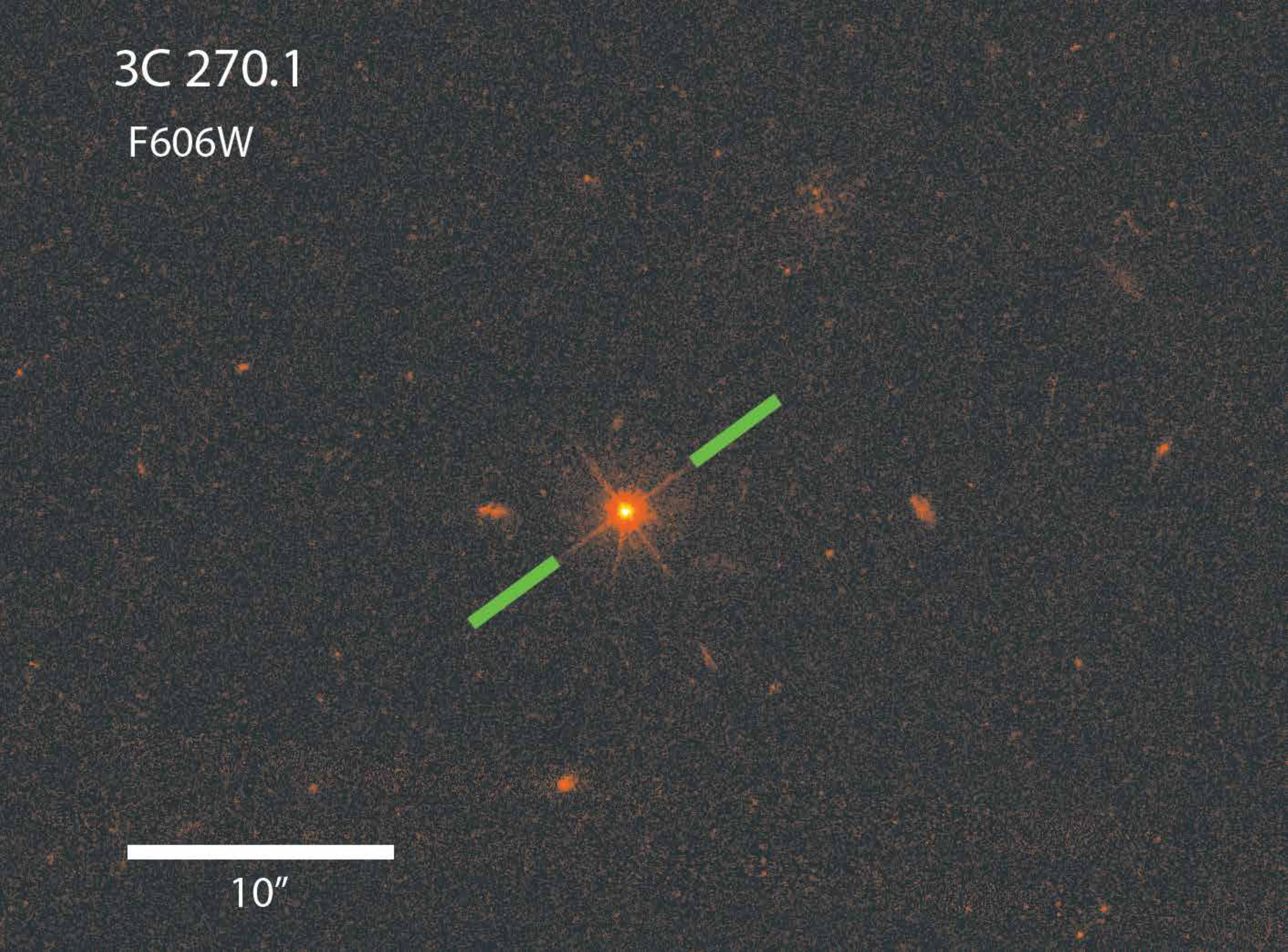}
\caption{Central $35\arcsec \times 50\arcsec$ of the UVIS image of QSO 3C 270.1. \redpen{The image has been rotated
so that North is up and East to the left. Green lines are placed on 
either side of the target to help identify its location.}}
\label{fig:3c270_uv}
\end{minipage}
\end{figure*}

\begin{figure*}[htp]
\begin{minipage}[b]{0.45\linewidth}
\centering
\includegraphics[width=\textwidth]{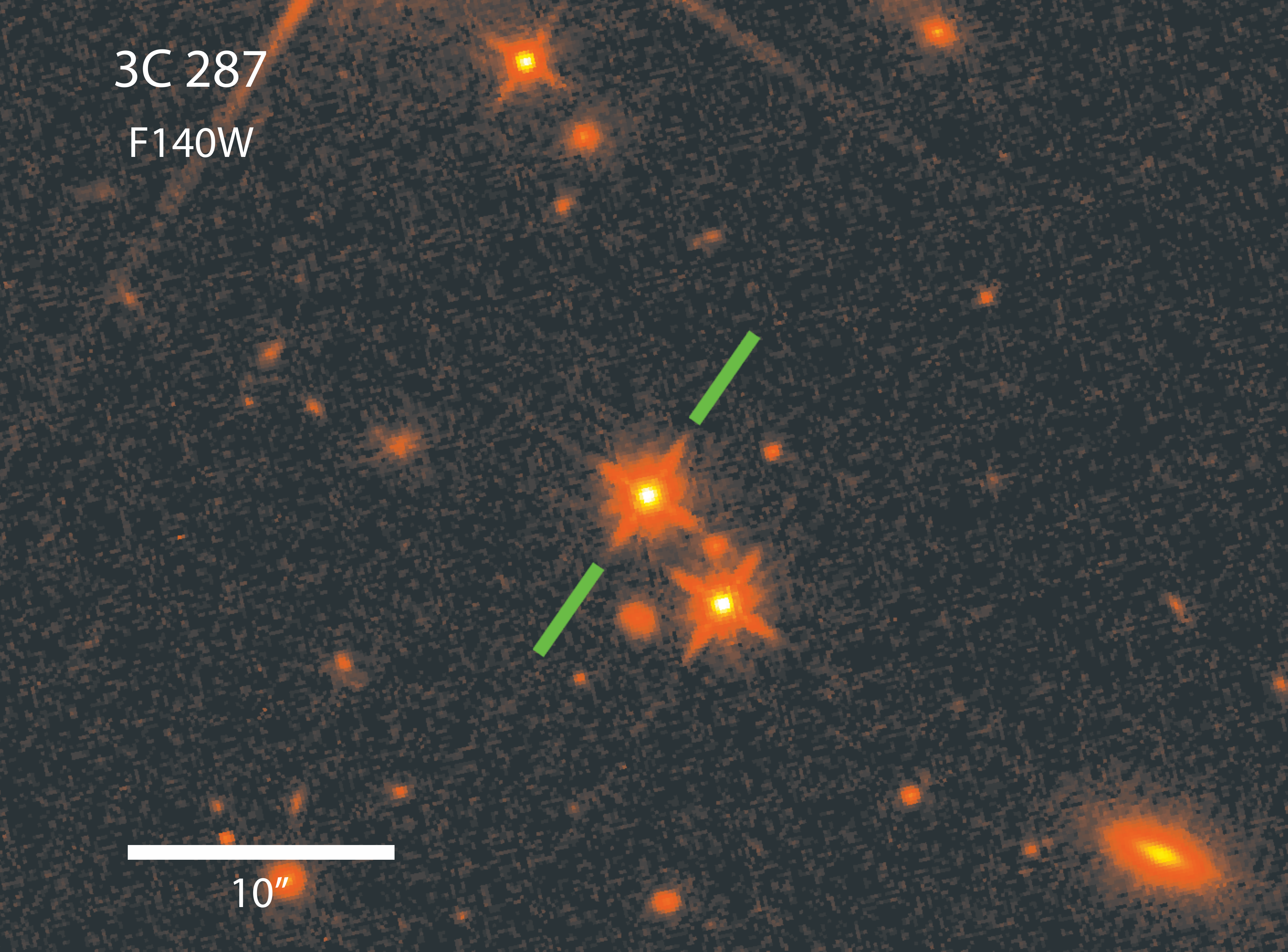}
\caption{Central $35\arcsec \times 50\arcsec$ of the IR image of QSO 3C 287. \redpen{The image has been rotated so 
that North is up and East to the left. Green lines are placed on 
either side of the target to help identify its location.}}
\label{fig:3c287_ir}
\end{minipage}
\hspace{0.5cm}
\begin{minipage}[b]{0.45\linewidth}
\centering
\includegraphics[width=\textwidth]{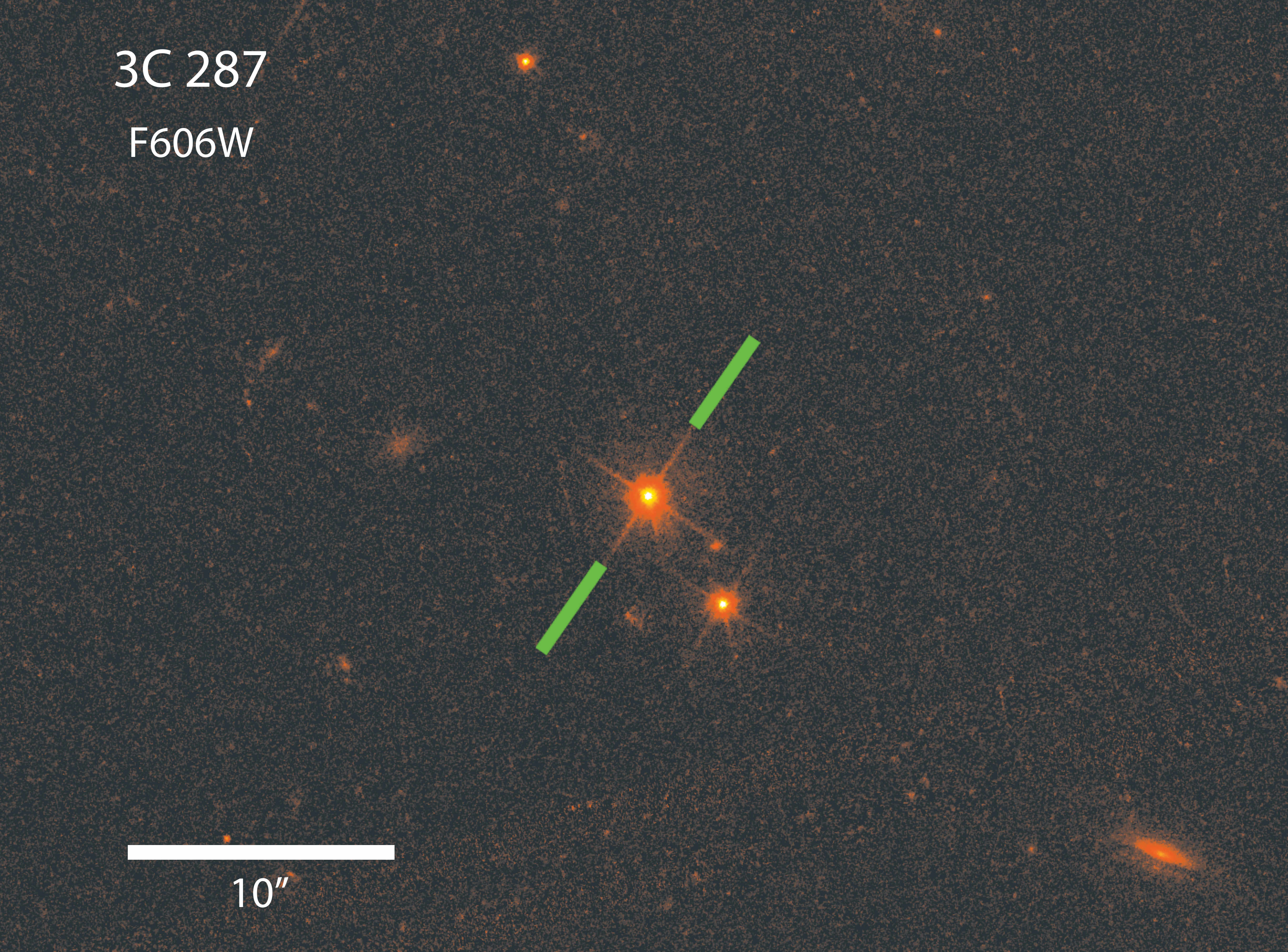}
\caption{Central $35\arcsec \times 50\arcsec$ of the UVIS image of QSO 3C 287. \redpen{The image has been rotated
so that North is up and East to the left. Green lines are placed on 
either side of the target to help identify its location.}}
\label{fig:3c287_uv}
\end{minipage}
\end{figure*}

\begin{figure*}[htp]
\begin{minipage}[b]{0.45\linewidth}
\centering
\includegraphics[width=\textwidth]{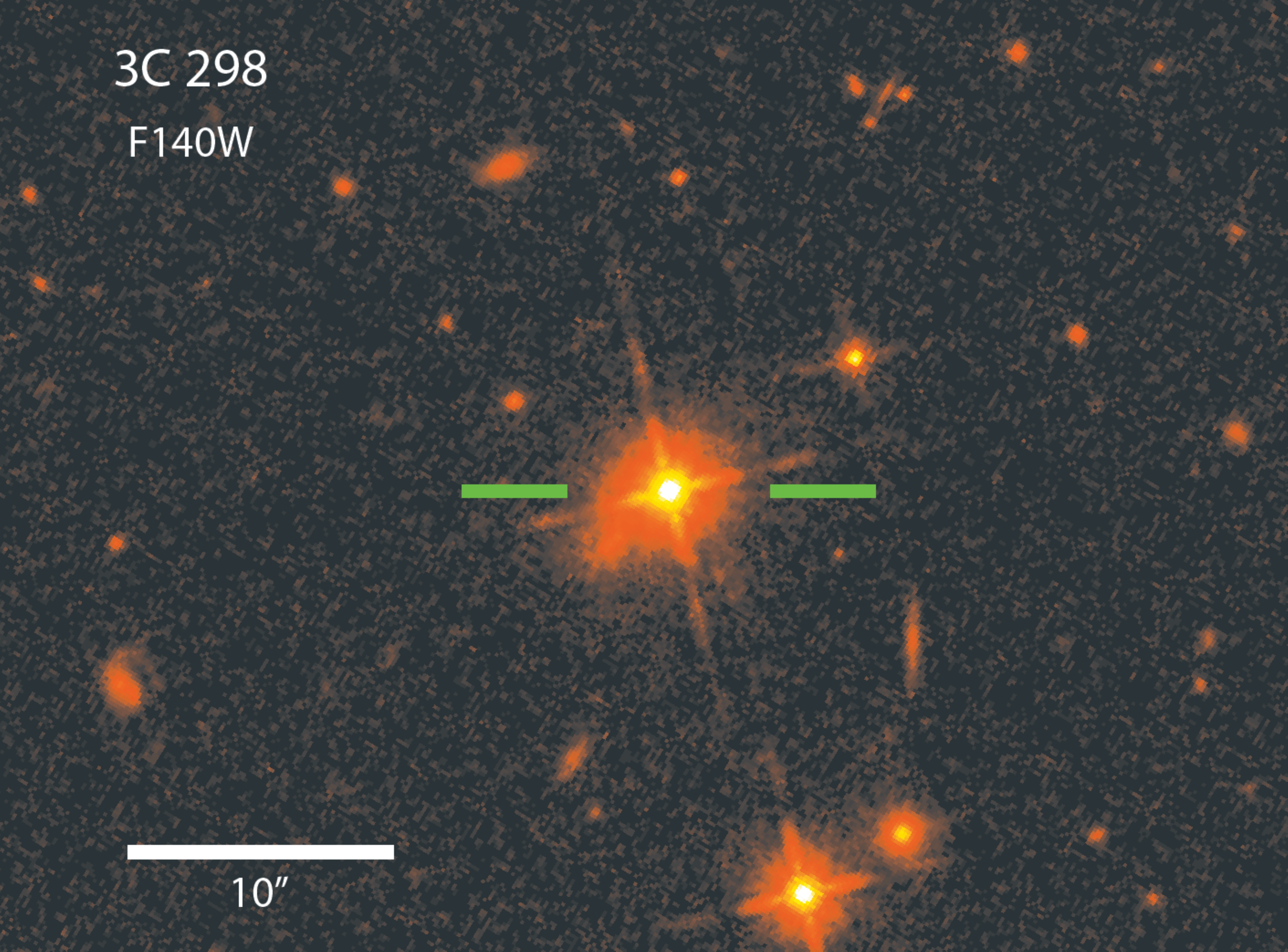}
\caption{Central $35\arcsec \times 50\arcsec$ of the IR image of QSO 3C 298. \redpen{The image has been rotated so 
that North is up and East to the left. Green lines are placed on 
either side of the target to help identify its location.}}
\label{fig:3c298ir}
\end{minipage}
\hspace{0.5cm}
\begin{minipage}[b]{0.45\linewidth}
\centering
\includegraphics[width=\textwidth]{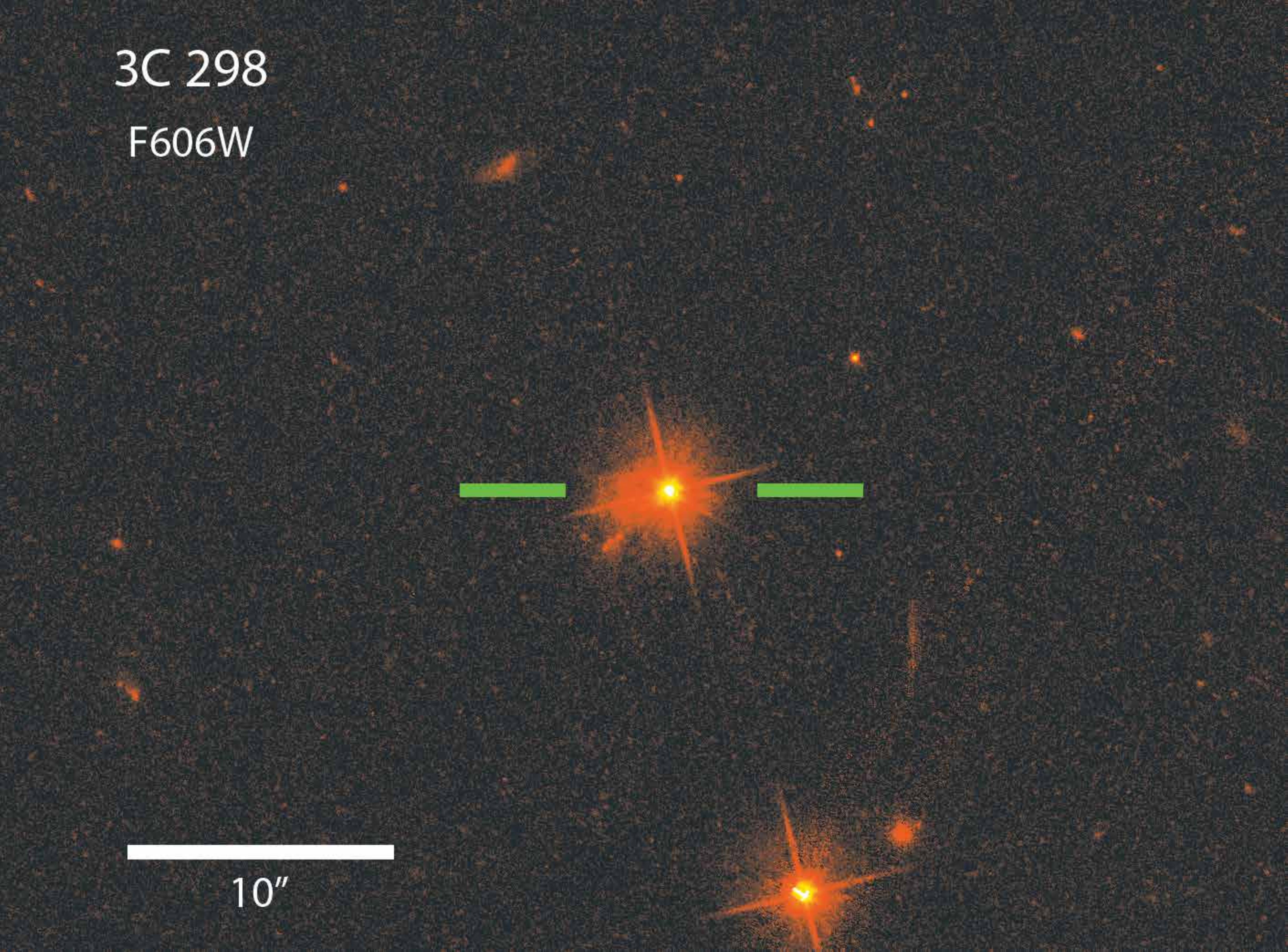}
\caption{Central $35\arcsec \times 50\arcsec$ of the UVIS image of QSO 3C 298. \redpen{The image has been rotated
so that North is up and East to the left. Green lines are placed on 
either side of the target to help identify its location.}}
\label{fig:3c298_uv}
\end{minipage}
\end{figure*}

\begin{figure*}[htp]
\begin{minipage}[b]{0.45\linewidth}
\centering
\includegraphics[width=\textwidth]{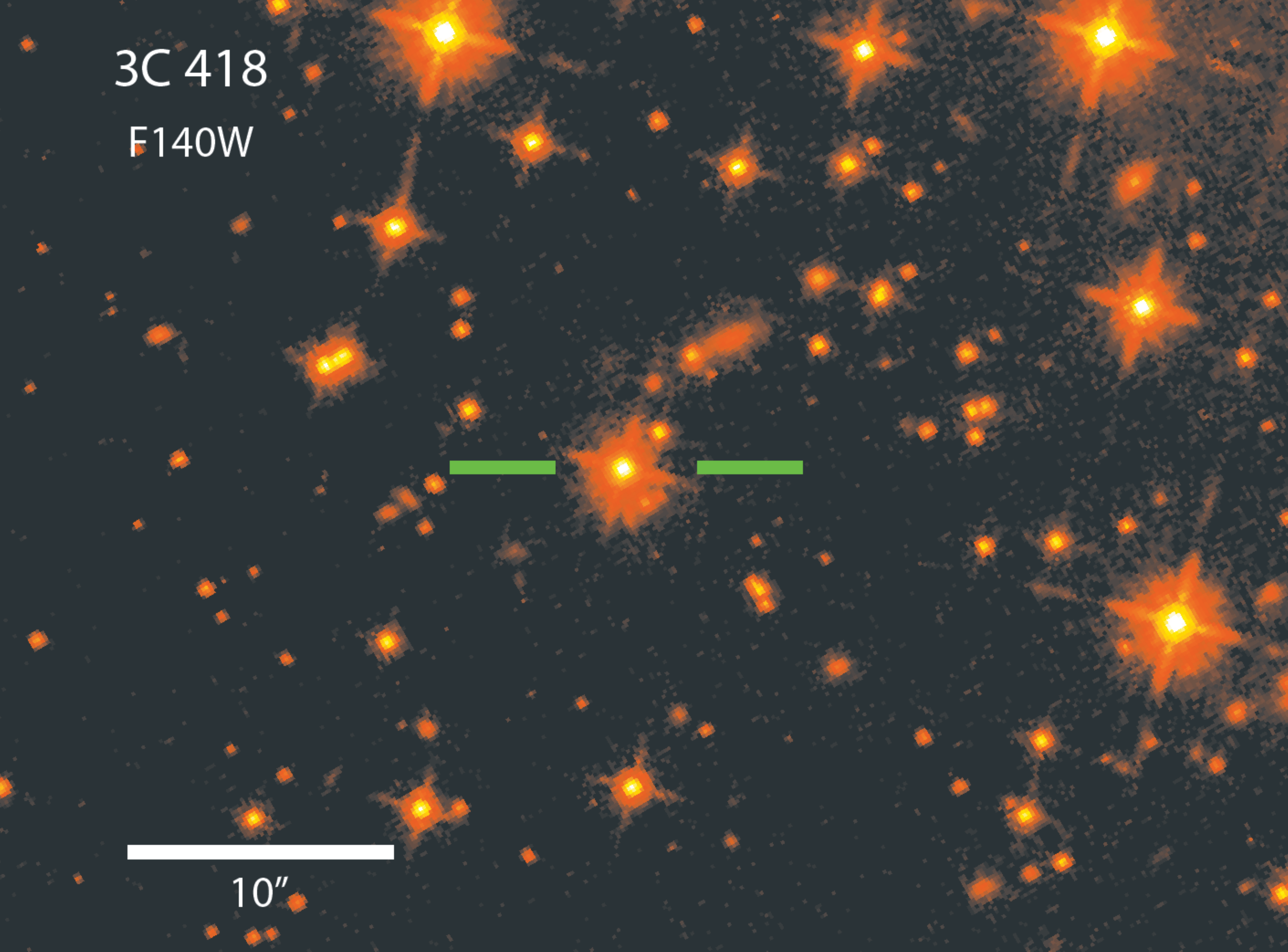}
\caption{Central $35\arcsec \times 50\arcsec$ of the IR image of QSO 3C 418. \redpen{The image has been rotated so 
that North is up and East to the left. Green lines are placed on 
either side of the target to help identify its location.}}
\label{fig:3c418_ir}
\end{minipage}
\hspace{0.5cm}
\begin{minipage}[b]{0.45\linewidth}
\centering
\includegraphics[width=\textwidth]{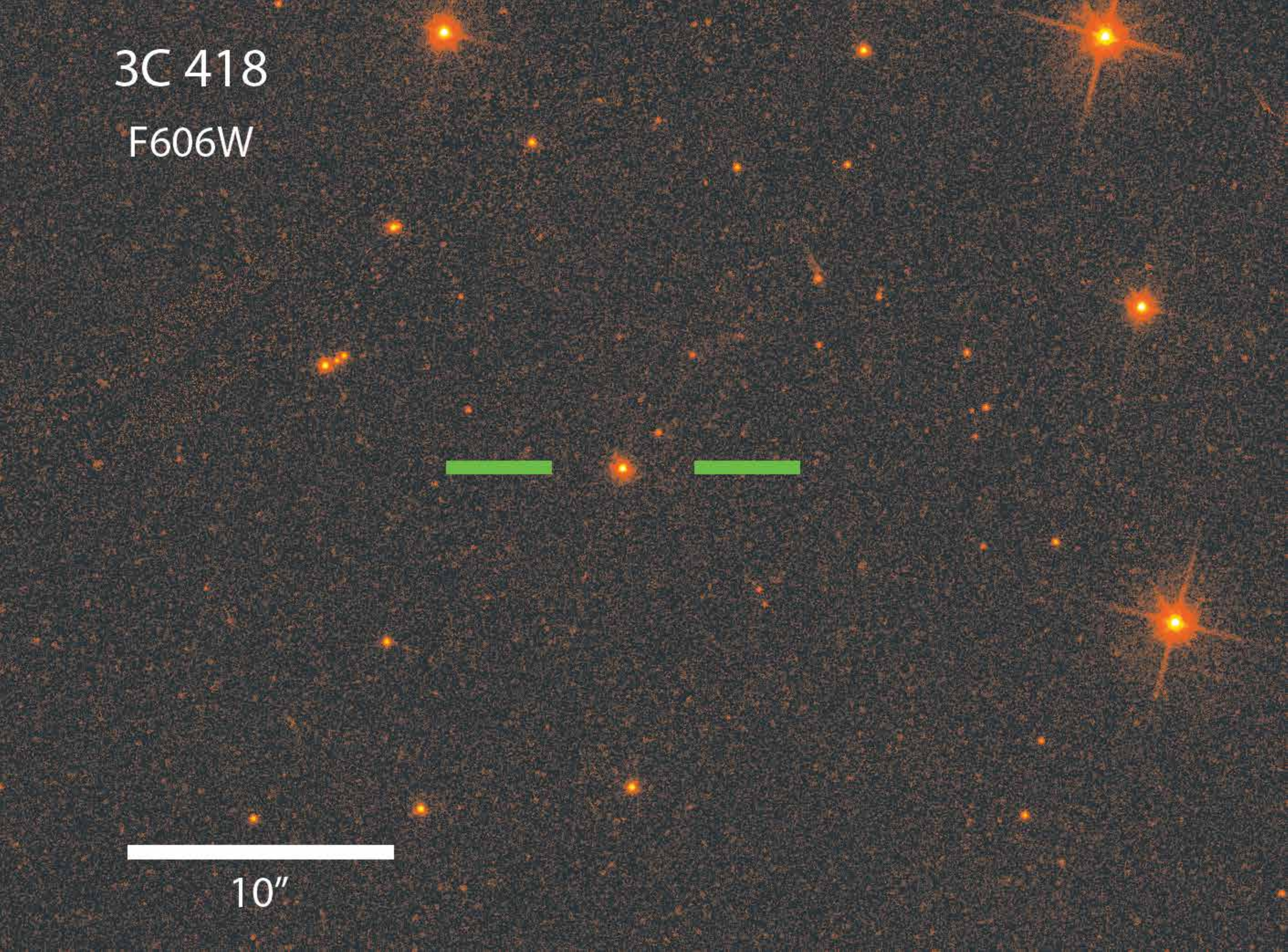}
\caption{Central $35\arcsec \times 50\arcsec$ of the UVIS image of QSO 3C 418. \redpen{The image has been rotated
so that North is up and East to the left. Green lines are placed on 
either side of the target to help identify its location.}}
\label{fig:3c418_uv}
\end{minipage}
\end{figure*}

\begin{figure*}[htp]
\begin{minipage}[b]{0.45\linewidth}
\centering
\includegraphics[width=\textwidth]{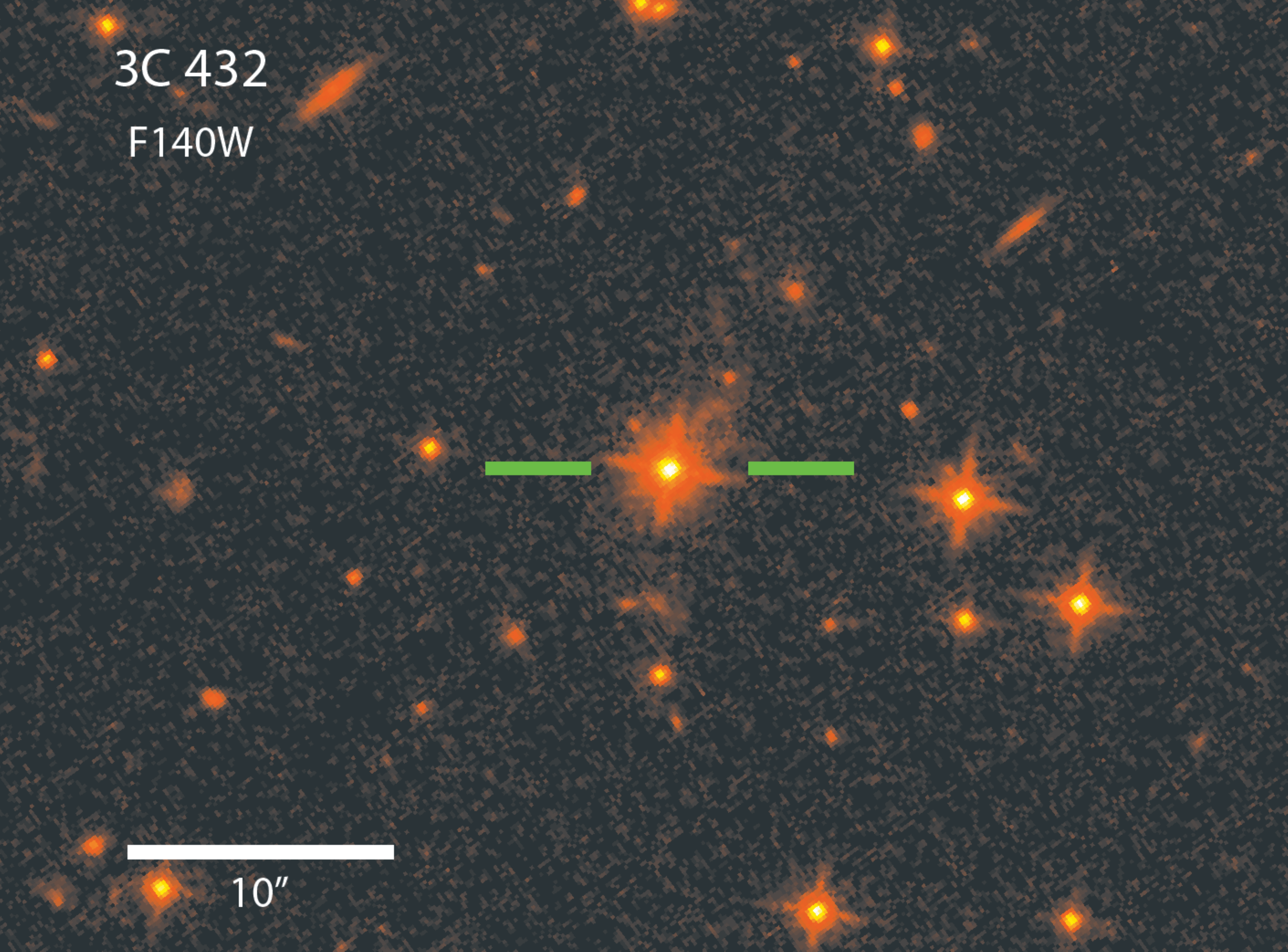}
\caption{Central $35\arcsec \times 50\arcsec$ of the IR image of QSO 3C 432. \redpen{The image has been rotated so 
that North is up and East to the left. Green lines are placed on 
either side of the target to help identify its location.}}
\label{fig:3c432_ir}
\end{minipage}
\hspace{0.5cm}
\begin{minipage}[b]{0.45\linewidth}
\centering
\includegraphics[width=\textwidth]{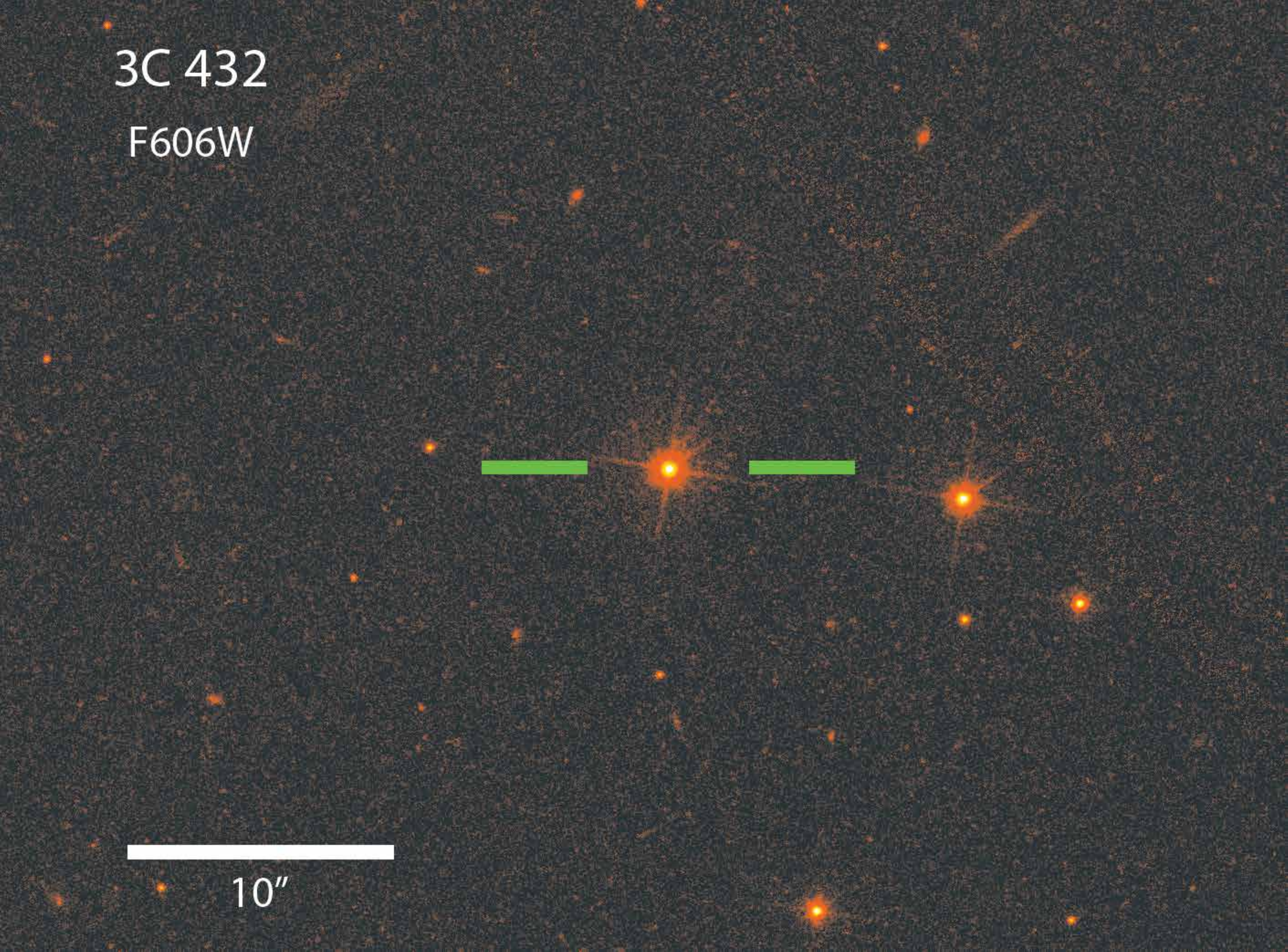}
\caption{Central $35\arcsec \times 50\arcsec$ of the UVIS image of QSO 3C 432. \redpen{The image has been rotated
so that North is up and East to the left. Green lines are placed on 
either side of the target to help identify its location.}}
\label{fig:3c432_uv}
\end{minipage}
\end{figure*}

\begin{figure*}[ht]
\centering
    \includegraphics[angle=90,width=\textwidth]{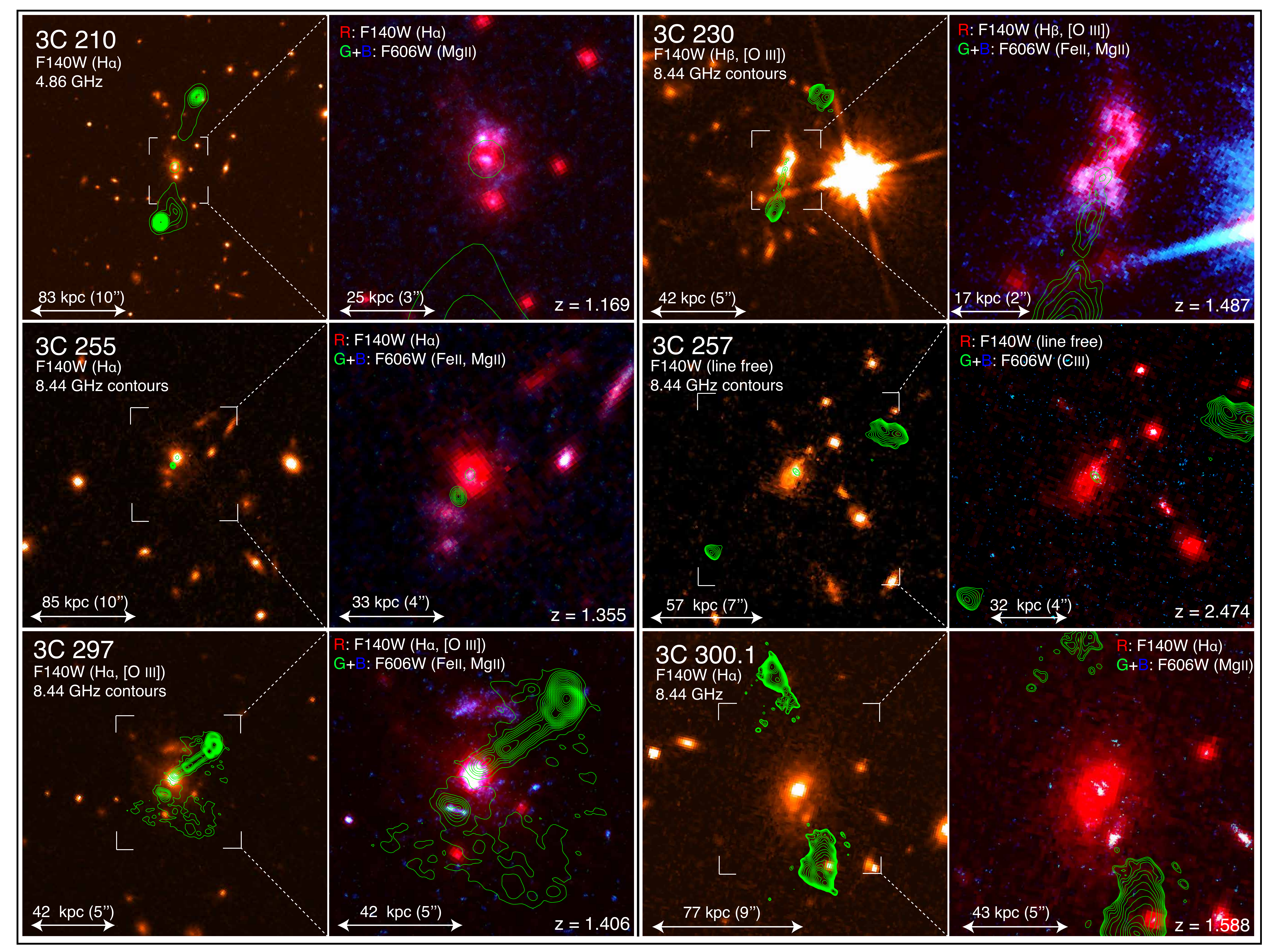}
    \caption{Radio galaxy optical and IR observations with overlain radio maps. \redpen{The images have been rotated 
so that North is up and East to the left.}}
    \label{fig:RG_6panel1}
\end{figure*}

\begin{figure*}[ht]
\centering
    \includegraphics[angle=90,width=\textwidth]{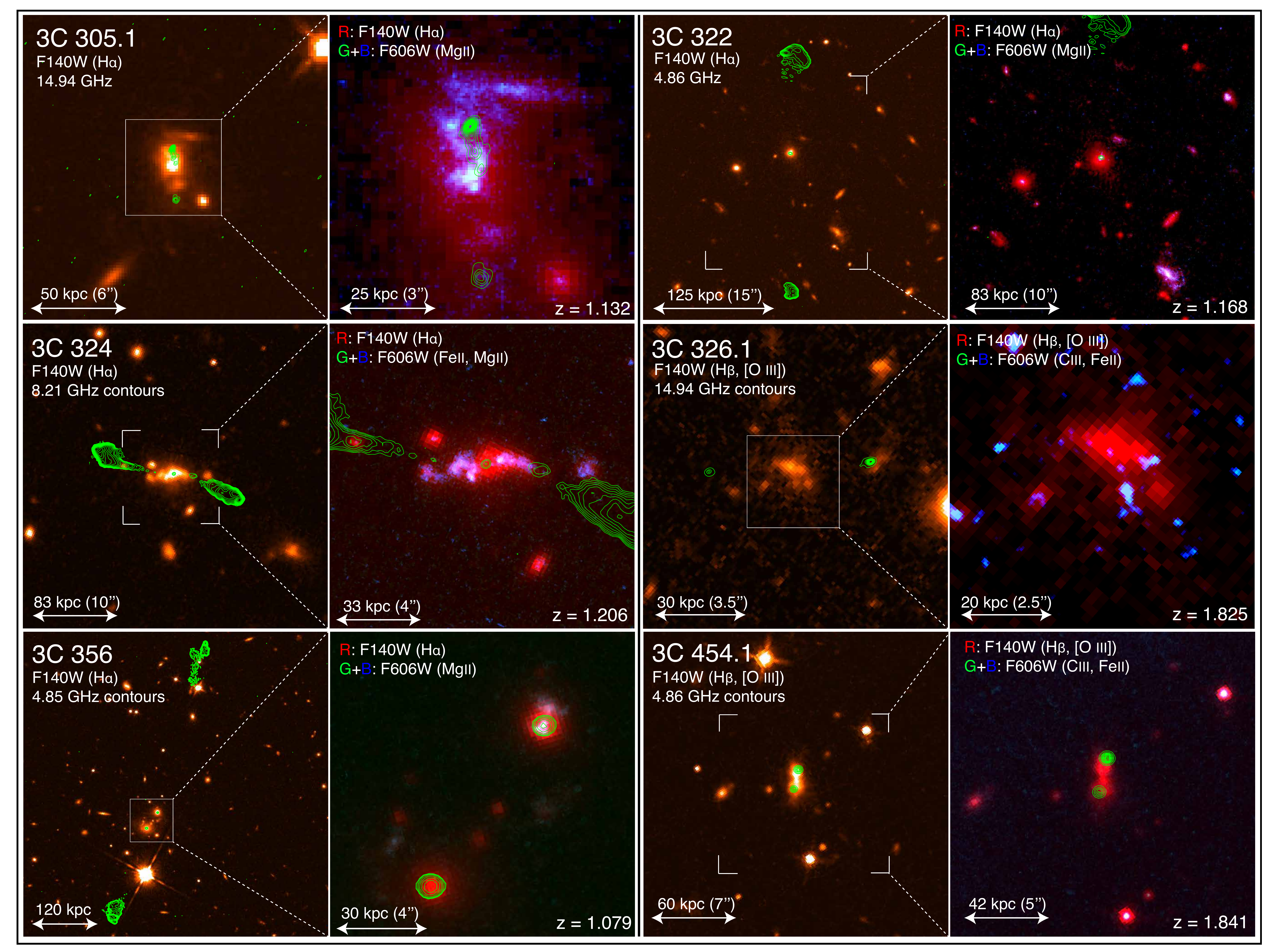}
    \caption{Radio galaxy optical and IR observations with overlain radio maps. \redpen{The images have been rotated 
so that North is up and East to the left.}}
    \label{fig:RG_6panel2}
\end{figure*}

\begin{figure*}[ht]
\centering
    \includegraphics[angle=90,width=\textwidth]{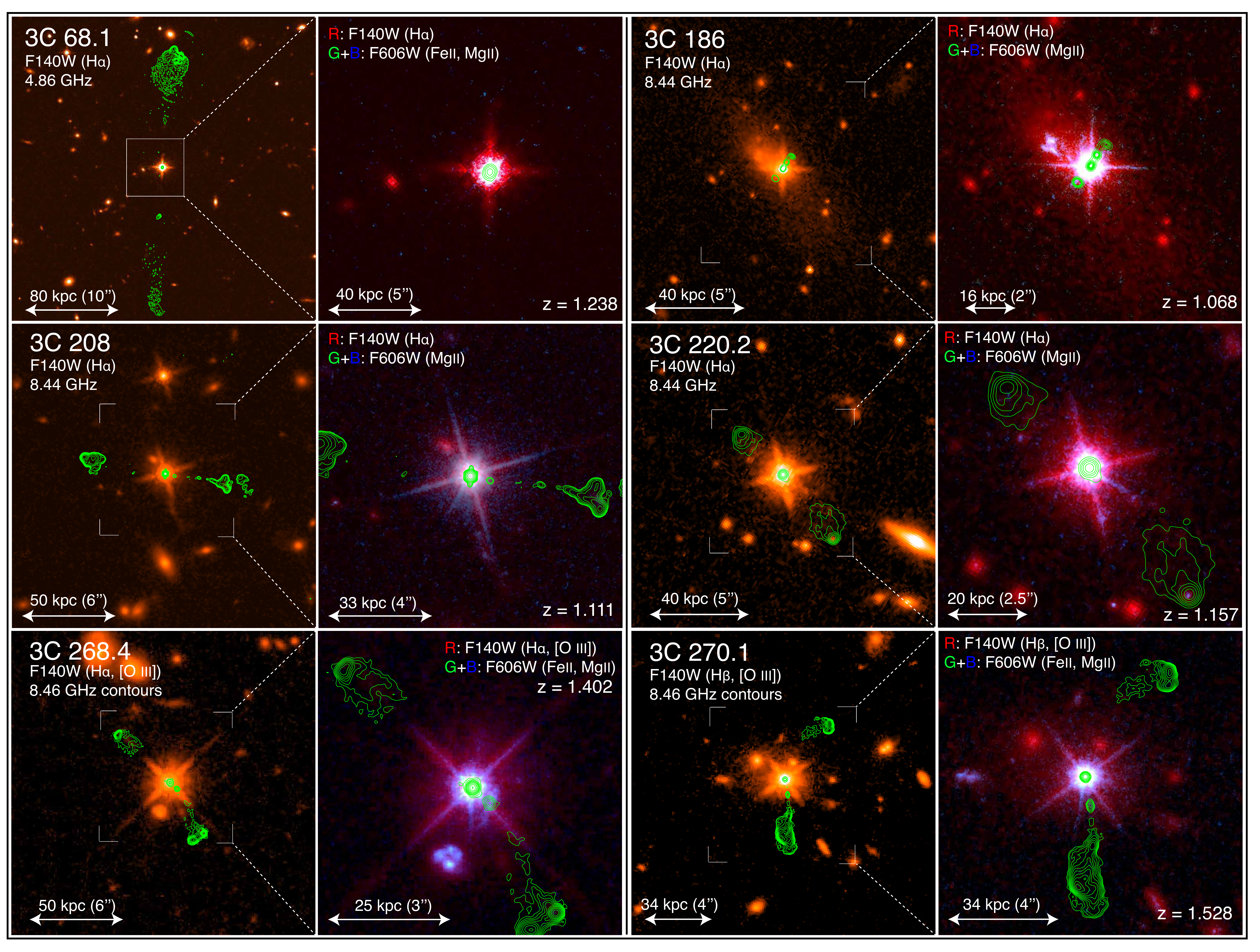}
    \caption{QSO optical and IR observations with overlain radio maps. \redpen{The images have been rotated 
so that North is up and East to the left.}}
    \label{fig:QSO_6panel1}
\end{figure*}

\begin{figure*}[ht]
\centering
    \includegraphics[angle=90,scale=0.22]{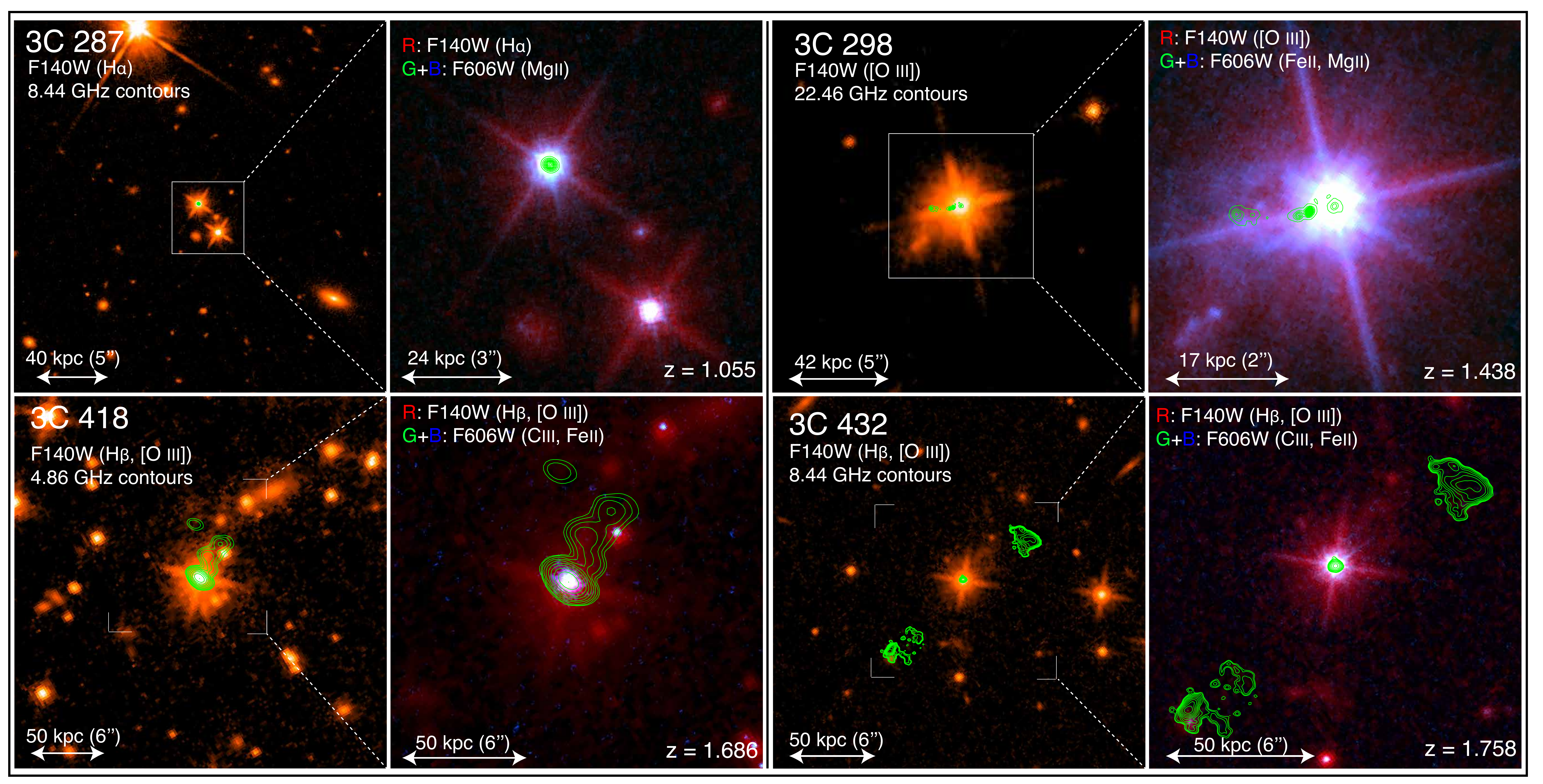}
    \caption{QSO optical and IR observations with overlain radio maps. \redpen{The images have been rotated 
so that North is up and East to the left.}}
    \label{fig:QSO_6panel2}
\end{figure*}

In this section we give qualitative descriptions of the objects and environments in both the visible/IR images
and the corresponding radio data, as well as notes on the alignment between the two. We look for the 
 alignment effect \citep{m08}, where the radio signal falls along the same line as the optical continuum emissions.
We note potential regions of active star formation based on the presence of concentrated areas of UVIS emission
in our observations. We also note that \citet{marco15} find that all RGs in this sample show signs of recent merger activity,
with the exception of 3C 230, where the host galaxy is not visible.
Figures \ref{fig:3c210_ir} through \ref{fig:3c432_uv} show the visible and IR 
observations for all targets, while figures \ref{fig:RG_6panel1} through \ref{fig:QSO_6panel2} show the enlarged central 
portions of the visible and IR observations with radio map overlays. 

The uncertainties in the world coordinate systems (WCS) of the VLA data are relatively large compared to the visible/IR
data. For sources containing a radio core, we shifted the radio data in order to align the radio core with the visible/IR core.
The magnitude of these shifts were typically less than 1$\arcsec$. From this we assume that the positional uncertainty for the
sources with no radio cores is also under 1$\arcsec$.

\subsection{Radio Galaxies}

\subsubsection{3C 210 (z=1.169)}
3C 210 is the eponymous member of the 3C 210 galaxy cluster \citep{stanford2002}. The cluster appears in our IR image as an 
overdense region of red galaxies.   The host of 3C210 is clearly undergoing major interactions with nearby
companions \citep{marco15}. Rest-frame 
UV emission, in the form of blobs most likely associated with star forming regions, is strongest in the 
northeastern-most part of the host galaxy. The galaxy immediately to the south of the target also shows a 
fainter compact UV blob. Diffuse UV emission is also present in those areas. The VLA data show a 
relatively weak radio core along with two strong lobes roughly 8$\arcsec$ to the north and 6$\arcsec$
to the south. The radio core is aligned with the optical core of the galaxy. 




\subsubsection{3C 230 (z=1.487)}

This object appears as a wide, double lobed radio galaxy. 
The very bright object in the center of the field is a foreground star. Just east of this star, 
a large number of small objects are visible in the IR image within a radius of 10$\arcsec$ from our 
target. The morphology in our optical and IR observations appears quite similar to 
that shown in the emission line images of \citep{s11}, and is reminiscent of the biconical structure
observed in the narrow line region of many Type II AGN. The size of the NLR as estimated from our images is 
$~24$kpc. The location of the core of galaxy is at the narrowest point (i.e. at the center) of the 
emission line region, but it is not clearly visible in our data. The radio data show a jet to the south 
and two lobes with compact hot spots. The radio jet is appears to be aligned with the conical structure
observed in the optical and IR images.

\subsubsection{3C 255 (z=1.355)}
Our IR image shows a collection of extended sources in the vicinity of 3C 255, which appears 
as the largest object in the central compact group. There are two objects to the southeast (at about 1.3$\arcsec$ 
and 2$\arcsec$ from the target), one to the west at 2.5$\arcsec$, and two elongated features to the northwest. 
Spectroscopic data from \citep{g90} suggests that all of the objects in the central compact group are members of a 
cluster. All of these objects, except for one of the elongated galaxies, appear in the optical image as well, 
suggesting that active star formation is ongoing.   

Radio maps indicate a relatively weak signal at the center of the host galaxy, as well as a brighter, compact 
radio source, possibly a hot spot, roughly 1$\arcsec$ to the southeast of the host galaxy.

\subsubsection{3C 257 (z=2.474)}
3C 257 is the most distant object in our sample and it is also the highest redshift object in the 3CR catalog. 
Its host galaxy appears in our IR image as a bright core surrounded by a tear-drop shaped region of emission which extends 
primarily south and east. Its morphology is clearly distorted, most likely as a result of an ongoing major merger. 
There is a 
small amount of signal in the optical image associated with the compact brighter area, which may indicate that star
formation is ongoing. This supports the relatively high star formation rate of $920^{+60}_{-50} M_\sun$ per year calculated using SED
fitting of Herschel data for 3c 257. \citep{podigachowski} 

The galaxy is surrounded by about 
10 other sources within a radius of $\sim 9\arcsec$ in the IR image. Most of these sources also appear in our optical 
image with very faint signals in their cores. 
The radio core is on the optical continuum core of the galaxy. In addition, the radio data show two lobes roughly 
7$\arcsec$ to the southeast and also to the northwest of the host.


\subsubsection{3C 297 (z=1.406)}
This target appears in our data with a complex morphology suggestive of ongoing merger activity \citep{marco15}. 
In the IR image, 3C 297 appears as a compact but 
resolved core surrounded by an area with a radius of more than 2$\arcsec$ of more diffuse emission.  
Directly north of the core by 2$\arcsec$, there is an arc-shaped area of emission. Approximately 1$\arcsec$ 
south of the target there is also a highly elongated source. 

In the optical image, the core of 3C 297 is resolved into two distinct areas of emission, with the eastern area 
much more extended than the western region. There is also faint emission associated with the extended 
arc-shaped area north of the target, as well as relatively strong emission from the highly elongated source 
south of the target, suggesting an active merger.

In addition to the sources described above, the IR image also shows 2 sources, approximately 2$\arcsec$ 
to the southwest and 2.5$\arcsec$ to the south of the target, which have no optical counterparts.

Radio maps show strong, wide-spread signal across the field. A lobe stretching from the galaxy more
than 5$\arcsec$ to the northwest displays emission along its entire length. Emission to the south spreads over
a much wider area, with a stronger concentration of signal centered on an elongated area approximately 
1$\arcsec$ from the galaxy. Also present in this area are two knots of emission in the optical image, although
with no radio core to aid in the alignment of the radio and visible/IR data, we cannot be completely 
sure of the relative positions of these sources.


\subsubsection{3C 300.1 (z=1.159)}
This galaxy appears within our IR image as an extended object containing a bright 
nucleus within an area of diffuse stellar emission. Also present within the extended envelope is a localized 
region of increased brightness to the southwest of the galaxy center. This is the only region of the 
galaxy other than the core that also appears within our optical image, suggesting an area of active star 
formation. The overall morphology is irregular most likely as a result of a recent merger event. 
Numerous extended sources are visible in the IR image around the central galaxy. Only a handful 
of these also show signal in the optical image. 

Radio maps show two bright, extended lobes to the north and south of the host galaxy by roughly 4$\arcsec$ 
and 2$\arcsec$ respectively, but no emission in the galaxy core at 8.4 GHz.

\subsubsection{3C 305.1 (z=1.132)}
This galaxy clearly appears in our IR image with a complex morphology suggestive of ongoing merger activity, and 
includes an extended, bright central area surrounded by more diffuse signal. Approximately 1.5$\arcsec$ 
north of the brightest emission, there is a highly elongated galaxy running almost directly 
east-west.  In addition, there is a distinct, compact source located 2$\arcsec$ southwest of the center 
of the target.  

In the optical image, the region in the center of the galaxy appears with an irregular shape spanning the 
same area covered by the brightest area in the IR image, suggesting wide spread star formation in the nucleus.
SED-fitting of Herschel data suggests a star formation rate of $220^{+40}_{-20} M_\sun$ per year for 3C 305.1. 
\citep{podigachowski}   The 
elongated object to the north also displays significant optical signal. The compact source to the southwest 
shows no significant optical signal.

The radio map shows a bright hot spot just to the north of the galactic center. There is also radio emission
at the nucleus of the host galaxy. A second, weaker 
lobe is seen roughly 3$\arcsec$ south of the galactic center. The positions of the radio emission lobes
is aligned with the elongation direction of the galaxy in the visible/IR data.


\subsubsection{3C 322 (z=1.168)}
This galaxy appears in our IR image as an extended object with a core that is slightly elongated in the 
northwest/southeast direction. There are two additional sources visible within 2.5$\arcsec$ of the target. 
In the corresponding optical image, signal from the target is limited to the northwestern-most point of the core. 
Only one of the two additional sources shows a limited amount of optical emission.

Corresponding radio data reveal emission within the galactic core at a location within 1\arcsec of the optical 
emission core. In addition, two lobes are present directly north and south of the galaxy by 
roughly 10 $\arcsec$ and 15 $\arcsec$ respectively.


\subsubsection{3C 324 (z=1.206)}
The radio galaxy 3C 324 appears in our IR image as an extended object stretching primarily east-west, with 
a smooth appearance to the west, and a more irregular, blobby appearance on the eastern side. In addition, 
there are 3 sources within 3$\arcsec$ of the galactic center, including an area immediately to the west and 
in-line with the galaxy's semi-major axis, as well as two sources to the north and east. 

The bright galactic center in the IR image has almost no signal in the corresponding optical image. 
The blobby appearance of the eastern half of the galaxy in the IR image is also present in the optical image, 
with a semi-circle of bright emission just to the east of the dark galactic center. The western half of the 
galaxy appears in the optical image with the brightest signal just to the west of the galactic center and a 
decreasing brightness to the west as the distance from the galactic center increases. This morphology is 
suggestive of enhanced star formation possibly associated with an ongoing merger. Using Herschel data, 
\citet{podigachowski} calculate a star formation rate of $180^{+40}_{-30} M_{\sun}$ per year for 3C 324.
In addition, the three nearby 
sources all show at least some signal in the optical image.

The radio map shows a compact core aligned with the bright (in the IR image) galactic center. Two extended 
lobes and two knotty jets are roughly aligned with the optical emission, stretching from just outside the 
galactic core to nearly 10$\arcsec$ northeast and southwest. One of the knots within the the eastern jet 
appears to be aligned with a source in the IR image. However, uncertainty on the position of the nucleus
makes this alignment unsure. Approximately 4$\arcsec$ to the west of the radio core, the jet bends to the
south. At nearly the same location as the bend, we see a galaxy in our IR and optical images. It is possible
that this galaxy is forcing the jet to bend and causing the western lobe out of alignment with the jet.

\subsubsection{3C 326.1 (z=1.825)}

This galaxy appears in our IR image as a relatively dim, elongated 
source stretching primarily east-west. The target is also surrounded by many small irregularly-shaped 
sources, several of which are within 1$\arcsec$ of the galactic center and nearly overlapping with the 
galactic edges, suggesting possible merger activity. Outside of these closest sources, there are 
numerous dim sources primarily north of the target, out to about 8$\arcsec$ from the target.

In our optical image, the target itself shows no significant emission. Of the small surrounding sources, 
several show a small amount of signal, including one of the sources closest to the target.

The corresponding radio map shows two compact lobes east and west of the galaxy by 3 $\arcsec$. 
The lobe to the west shows significantly stronger emission than the lobe to the east. The lobes are roughly
colinear with the elongation direction of the galaxy in the IR image. The signal in the IR image is
most likely contaminated by [OIII] emission line flux.



\subsubsection{3C 356 (z=1.079)}
The radio source 3C 356 appears in our IR image associated with two main bright elliptical galaxies lying on a 
southeast-northwest line, separated by 5$\arcsec$. The northern source appears brighter and 
more compact than the southern source. There are also 5-6 smaller distinct sources between 
the two main sources. 

In contrast, our optical image shows emission in the form of a bi-conical structure from the northern of 
the two main sources. There 
is also faint signal in the optical image associated with two of the more extended surrounding 
objects, as well as the compact object located between and equidistant from the two main sources.

This is clearly a merging system, with active star formation occurring in at least some of the sources 
connecting the two main galaxies \citep{marco15}.

The corresponding radio map shows two compact radio sources aligned with both of the bright IR 
components. In addition there are two lobes, roughly 30 $\arcsec$ to the north and 15 $\arcsec$ 
to the south of the bright IR components. The lobes are well aligned with the two main IR image 
sources. The northern source is the host of the FR II radio galaxy. \citep{cimatti97}


\subsubsection{3C 454.1 (z=1.841)}
3C 454.1, a Compact Steep Spectrum (CSS) source (e.g \citet{odea98}) showing signs of recent merger activity \citep{marco15}, 
appears in our IR image as a double-lobed object oriented north-south. The northern lobe 
is significantly brighter than the southern lobe. In addition, there are 4 distinct objects within 
about 3.5$\arcsec$ of the target. The optical image shows essentially no emission from any of these objects, despite a calculated star
formation rate of $750^{+40}_{-70} M_{\sun}$ per year \citep{podigachowski}. This may imply that dust obscuration in this object is substantial. 
 
The radio maps associated with the target show compact radio sources that are well aligned with the bright 
cores in the IR image.


\subsection{QSOs}

\subsubsection{3C 68.1 (z=1.238)}
This QSO appears in our IR image as a bright compact object. There are only two other sources 
within 8$\arcsec$ of the target.  These are both smaller, dimmer sources 3.75$\arcsec$ and 5.5$\arcsec$ to 
the east. Neither of these extra sources is visible in the corresponding optical image. 
The radio map of the area shows emission in the core of the quasar, as well as two lobes roughly 10$\arcsec$ 
to the north and 15 $\arcsec$ to the south.

Our photometry show that 3C 68.1 is significantly redder than all of our other QSO targets. This is most 
likely due a combination of a high inclination and a dusty environment. This allows dust-reddened light 
from the nucleus as well as reddened scattered light to contribute to the overall observed signal. 
\citep{brotherton98}

\subsubsection{3C 186 (z=1.069)}

The target, a well-studied CSS source \citep{odea98} appears in our IR image in 
an overdense region which is most likely a cluster of galaxies \citep{semi2008}. 
The host galaxy is clearly visible as a roughly elliptical area of diffuse emission extending to the 
northeast and southwest. 


A blob of relatively bright emission, about 1$\arcsec$ in size, has its center located roughly 
2$\arcsec$ northeast of the target. Our optical image reveals a well-defined area of signal in the 
same area, suggesting active star formation. Very few of the galaxies surrounding the target 
have appreciable signal in the optical image, supporting the assertion that they are cluster members at the 
redshift of the QSO. 

Radio data show emission coincident with the core of the QSO, as well as a jet extending from the QSO
to the northwest for roughly 2$\arcsec$. The signal in this jet is concentrated in two distinct knots. There
is also an area of emission visible to the southeast of the QSO core.

\subsubsection{3C 208 (z=1.112)}

In our IR image, 3C 208 appears as a compact object in our data, with 25-30 
sources visible within 15$\arcsec$. The closest source is just over 1$\arcsec$ northeast of 
the target. Most of these surrounding objects show a limited amount of signal in our optical image. 
There is corresponding radio emission coming from the core of the QSO, as well as lobes stretching roughly
6$\arcsec$ east and west.

\subsubsection{3C 220.2 (z=1.157)}
3C 220.2 appears in our IR and optical images as a compact source with a few immediate neighbors, located 
within a group of large spiral galaxies. In the IR there are 5 sources to the south and one to the northwest, 
all within 6$\arcsec$. There is little to no signal from these sources in our optical image.

Radio data show signal centered on the QSO, as well as two lobes to the northeast and southwest. There is 
evidence of a hot spot in the southwestern lobe. The
peak of the radio signal in the hot spot is coincident with an area visible in our optical 
image.

\subsubsection{3C 268.4 (z=1.402)}
Our IR image of 3C 268.4 reveals a bright QSO with a handful of small, dim sources, 
within about 12$\arcsec$. In addition, the nearest neighbor 
to the target is a relatively bright, compact object roughly 2.5$\arcsec$ to the south. In our optical image, 
this nearest neighbor is resolved into a clumpy mass with significant signal, suggesting active star formation.

Many of the other nearby dim sources also exhibit some flux in the optical. Finally, the optical image also contains 
an oblong source located only 0.8$\arcsec$ to the southwest of the target, suggesting a possible merger. If 
there is corresponding flux from this object in our IR image, it is not visible due to its close proximity to 
the target and the large pixel size in the IR channel. 

Radio emission is also visible coming from this adjacent oblong source. The QSO itself shows significant signal 
in the radio. There are also two radio lobes, roughly 3$\arcsec$ to the northeast and southwest.


\subsubsection{3C 270.1 (z=1.528)}
3C 270.1 appears as a bright unresolved source surrounded by many dimmer sources 
of various morphologies. There are at least 5 sources within 4$\arcsec$ of the target. 
Of these, only one is detected in our optical image. Most of the sources in the region surrounding the object are 
very red. 

The radio morphology is not symmetric with respect to the center of the QSO. In addition to radio signal present at the core of 
the QSO itself, we see radio lobes to the south and 
the northwest. Also visible is a jet connecting the radio core and the southern lobe. No jet is visible on the northern
side of the core.  The radio morphology is reminiscent of that of wide angle tailed (WAT) sources \citep{gomez97}, which typically inhabit 
clusters of galaxies.

\subsubsection{3C 287 (z=1.055)}
Our IR image of 3C 287 reveals the bright QSO with bright compact neighbor 
4$\arcsec$ to the southwest. In addition, four sources of varying brightnesses are present along a 
line separating the target from the bright neighboring source. Of these 4 sources, 2 show some signal 
in our optical image. The associated radio signal is limited to a compact core on the optical core of the QSO.




\subsubsection{3C 298 (z=1.438)}
In our IR image, 3C 298 appears as a bright, asymmetrical source with more signal to the east appearing in a 
conical shape reminiscent of the narrow line region seen in 3C 230.

In addition, there is a small area of increased flux 2.5$\arcsec$ to the southeast of the galaxy center in our IR
image.  This area is elongated and points radially away from the center of the QSO.  In the optical image, with its 
better resolution, this small source appears as three knots of emission. These sources could be regions of intense 
star formation at the edge of the NLR. 

Radio emission is visible in the core of the QSO, as well as two small areas 0.5$\arcsec$ and 1.5$\arcsec$
to the east; probably a knotty jet and a hot spot. The jet and hot spot appear roughly in the center of the conical area
of emission in the IR and optical images. Note that the jet is at a different orientation than the three knots of optical 
emission to the southeast of the QSO.

\subsubsection{3C 418 (z=1.686)}

The QSO 3C 418 appears in our IR image as a bright target located within a dense field of objects. 
In addition to the main QSO, there appears to be a distinct source located 1.5$\arcsec$ to the southwest, 
as well as a small and a larger extended source 3$\arcsec$ and 6$\arcsec$ to the northwest, respectively. 
To the east, there is also a pair of extended sources approximately 8$\arcsec$ from the target. While 
3C 418 appears in our optical image as a compact source, only the source 3$\arcsec$ to the northwest shows
any signal in the optical image.

Radio emission is concentrated in the core of the QSO, with a small jet extending approximately 6$\arcsec$
to the northwest.


\subsubsection{3C 432 (z=1.785)}

3C 432 appears in our IR data as an unresolved source within a field of objects. There is some
dim extended emission, potentially a narrow line region, to the northwest of the QSO. There are 
also several distinct sources in the
same direction as the extended emission, within 7.5$\arcsec$ of the QSO. In addition a small source 
is located 2$\arcsec$ to the northeast, and two extended sources are 4.9$\arcsec$ to the south of the target.

In our optical image, 3C 432 appears as a compact, isolated source. None of the nearby sources described above 
show optical signal. 

Radio emission is present in the core of the QSO, as well as in two lobes roughly 6-8$\arcsec$
to the northwest and southeast. These lobes are aligned with the potential NLR-emission seen in our IR image.

\section{Concluding Remarks}
We have presented qualitative descriptions along with photometric analyses of new WFC3 visible and near-IR
observations of a set of 22 1$>z>$2.5 radio galaxies and QSOs from the 3CR catalog. In addition, archival
radio data have been combined with the HST observations in order to provide a more complete picture
of the targets and their local environments.


Our photometric analysis has shown that RGs are generally more red than the QSOs. We also 
find that for QSOs, increasing IR channel flux is correlated with increased redness.

Emission line analysis indicates that the contamination from emission line signal is limited for both the QSOs and the radio galaxies. We find contamination rates of less than 2.6\% from each of the lines for which we have direct flux measurements. However, we do not estimate contributions from the potentially bright H$\alpha$ and [OIII] lines in the rest frame optical data.

For most of our radio galaxies, we see the alignment effect, where the lobes in the radio data are
colinear with the major axis of the galaxy in the visible/IR image. However, emission line contamination
means that for most sources, we are unable to distinguish between alignment with the optical continuum versus
alignment with the emission line gas.

We see two RGs (3C 322 and 3C 356) with compact
nuclei coincident (to within WCS uncertainties) with the radio core. Extended narrow line regions appear
visible in 2 RGs (3C 230 and 3C 305.1) and 5 QSOs (3C 186, 3C 268.4, 3C 270.1, 3C 298, and 3C 432). 
In addition we observe clumps of UV emission, most likely regions of active star formation,
in 4 of the RGs (3C 210, 3C 305.1, 3C 324, and 3C 326) and two of the QSOs (3C 186 and 3C 268.4). One of 
our sources, 3C 356, appears to be a double AGN. Finally, 3 RGs (3C 230, 3C 255, and 3C 305.1) and 2 QSOs 
(3C 220.2 and 3C 298) show evidence of hot spots in their radio maps.

\acknowledgments

The authors would like to thank the referee for a careful reading of our manuscript and many helpful comments, which enabled us to greatly improve the final document. J.P.K. and B.H. acknowledge support from $HST$-GO-13023.005-A. This research has made use of the NASA/IPAC Extragalactic Database (NED) which is operated by the Jet Propulsion Laboratory, California Institute of Technology, under contract with the National Aeronautics and Space Administration. This work is based on observations made with the NASA/ESA HST, obtained from the Data Archive at the Space Telescope Science Institute, which is operated by the Association of Universities for Research in Astronomy, Inc., under NASA contract NAS 5-26555. This work is based in part on observations taken by the VLA, operated by U.S. National Radio Astronomy Observatory which is a facility of the National Science Foundation, operated under cooperative agreement by Associated Universities, Inc.

{\it Facilities:} \facility{HST (WFC3)}, \facility{VLA}

\begin{figure*}[h]
\epsscale{0.6}
\plotone{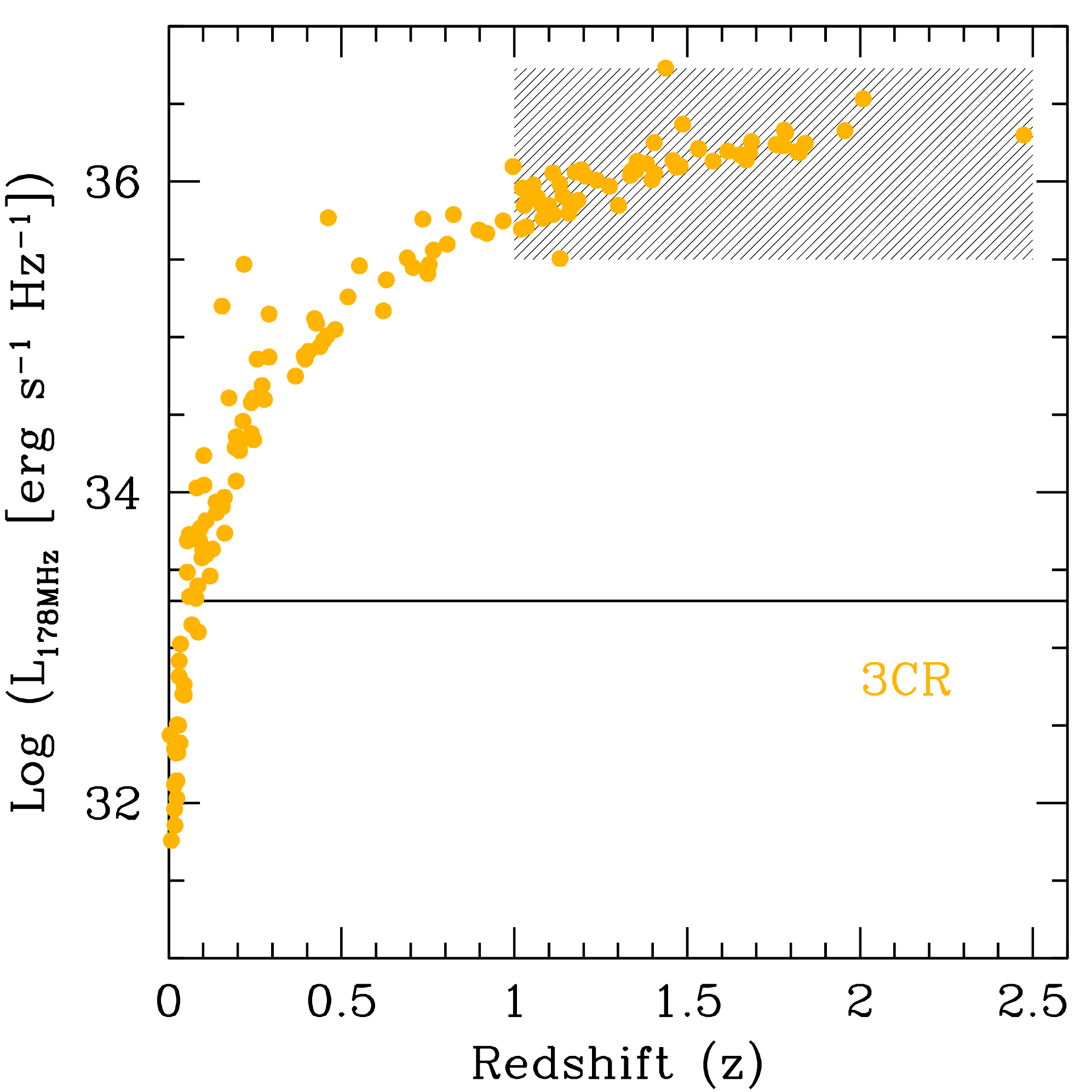}
\caption{The distribution in radio power (at 178MHz) and redshift of the entire 3CR catalog. The shaded area corresponds to the sample
of our HST SNAPSHOT program described in this paper. \label{rad_v_z}}
\end{figure*}

\clearpage

\end{document}